\documentclass[ejsv2,noshowframe,noinfoline,preprint]{imsart}

\RequirePackage[numbers]{natbib}
\RequirePackage[colorlinks,citecolor=blue,urlcolor=blue]{hyperref}
\RequirePackage{graphicx}

\usepackage{algorithm}
\usepackage{enumitem}
\usepackage{algpseudocode}
\usepackage{tikz}

\startlocaldefs
\theoremstyle{plain}

\newtheorem{theorem}{Theorem}[section]
\newtheorem{lemma}[theorem]{Lemma}
\newtheorem{remark}[theorem]{Remark}
\newtheorem{assumption}[theorem]{Assumption}
\theoremstyle{definition}
\newtheorem{definition}[theorem]{Definition}

\theoremstyle{remark}

\endlocaldefs

\newcommand{\be}{\begin{equation}}
\newcommand{\ee}{\end{equation}}
\newcommand{\bea}{\begin{eqnarray}}
\newcommand{\eea}{\end{eqnarray}}

\newcommand{\bft}{\boldsymbol{t}}
\newcommand{\bfpi}{\boldsymbol{\pi}}
\newcommand{\psitilde}{\tilde{\psi}}
\newcommand{\phitilde}{\tilde{\phi}}

\newcommand{\Otilde}{\widetilde{O}}
\newcommand{\Ftilde}{\widetilde{F}}
\newcommand{\calM}{\mathcal{M}}
\newcommand{\E}{\mathbb{E}}
\newcommand{\reals}{\mathbb{R}}

\newcommand{\that}{\hat{t}}

\newcommand{\Ltilde}{\widetilde{L}}

\begin{document}
\begin{frontmatter}
\title{Permutation Recovery on Manifold Data via Spectral Seriation}
\runtitle{Permutation Recovery via Spectral Seriation}

\begin{aug}
\author[A]{\fnms{Yuehaw}~\snm{Khoo}\ead[label=e1]{ykhoo@uchicago.edu}},
\author[B]{\fnms{Xin T.}~\snm{Tong}\ead[label=e2]{xin.t.tong@nus.edu.sg}},
\author[C]{\fnms{Wanjie}~\snm{Wang}\ead[label=e3]{wanjie.wang@nus.edu.sg}}\and
\author[A]{\fnms{Yuguan}~\snm{Wang}\ead[label=e4]{yuguanw@uchicago.edu}}
\address[A]{Department of Statistics, University of Chicago\printead[presep={,\ }]{e1,e4}}

\address[B]{Department of Mathematics, National University of Singapore\printead[presep={,\ }]{e2}}

\address[C]{Department of Statistics and Data Science, National University of Singapore\printead[presep={,\ }]{e3}}

\runauthor{Y. Khoo et al.}
\end{aug}

\begin{abstract}
Data points in many scientific experiments originate from an ordered structure, yet this ordering is often unavailable. 
We consider noisy data points  with the correct ordering to be recovered. 
The underlying structure naturally places the data on a 1-dimensional manifold. Because eigenfunctions of 1-dimensional manifold Laplacian are trigonometric functions, and the manifold Laplacian can be approximated by the graph data Laplacian, the data ordering can be recovered by inverting the data Laplacian eigenvectors. We propose two spectral algorithms, one for the periodic structure (closed loop) and one for the non-periodic structure (open curve). We have derived the uniform error bound for the algorithms, which is composed of two parts: the discretization error between the manifold eigenfunctions and the noiseless graph Laplacian eigenvectors, and the eigenvectors error caused by data noise. In numerical studies, our spectral seriation algorithms outperform other manifold learning methods. The superior performance of our algorithms is demonstrated further on a biomolecule data example.
\end{abstract}



\end{frontmatter}

\section{Introduction}\label{sec:intro}
Temporal progression systems arise naturally in many scientific settings. Consider a system $\{X(t); t \in [0, 2\pi]\}$, where each state $X(t) \in \reals^d$ is indexed by a single latent parameter $t$. The parameter $t$ may represent physical time, a rotation angle, an oscillatory phase, or a latent developmental stage in biological processes \citep{singer2011three, moscovich2020cryo}.
For example, if $X(t)$ denotes the image of a moving robot at time $t$, then $\{X(t); t \in [0, 2\pi]\}$ describes its full trajectory. Regardless of its interpretation, the mapping $t \rightarrow X(t)$ traces out a one-dimensional progression in a $d$-dimensional feature space and naturally forms a one-dimensional manifold.

In practice, the underlying trajectory $X(t)$ cannot be observed. Instead, we only have observations at discrete latent time points $t_i$. 
Let $Z_i$ denote the $i$-th observed data point, where $Z_i = X(t_i) + e_i$ with $e_i \in \reals^d$ being the observation noise, $i \in [N]$. 
Because the latent time points $t_i$ are unavailable, the central goal is to recover $t_i$ and, consequently, the ordering $\bfpi$ such that $t_{\pi(i)}$ increases with $i$. This task is commonly referred to as permutation recovery.

Such latent temporal systems appear broadly in scientific applications. In cryo-electron microscopy, for example, each particle image corresponds to an unknown conformational state of a molecule undergoing continuous motion, and determining its latent position along this trajectory is essential for accurate reconstruction \citep{moscovich2020cryo,cossio2022extracting,vant2022exploring}. Permutation recovery also plays an important role in embryo development studies, where samples collected at different growth stages must be aligned along a developmental timeline. Determining the developmental stage $t_i$ for each embryo is critical for understanding embryo development \citep{jaeger2004dynamic, dsilva2015temporal}. 
Similar ordering problems are also present in bioinformatics, paleontology, archaeology, and many other fields \citep{brown2016measurement, shennan1997quantifying, o1999seriation}.

The most direct seriation approach is to exhaustively search over all permutations, as first proposed in \cite{arabie1996overview}, with subsequent efforts to accelerate computation \citep{hubert2001combinatorial, brusco2005branch, hahsler2008getting}. Yet even the most recent work \cite{hahsler2008getting} reports feasible sample sizes of about 40. To address this challenge, another line of research focuses on the subject-to-subject similarity matrix rather than the raw data $Z_i$ \citep{giraud2023localization, issartel2024minimax}. Under the assumption that this similarity matrix has a monotone structure, \cite{atkins1998spectral} shows that the leading Laplacian eigenvector can recover the ordering, with several extensions \citep{fogel2014serialrank}.
In particular, when the noiseless matrix has monotone rows along the true ordering, \cite{ma2021optimalb} proposes an efficient estimator, with extensions to extreme values and double-sided ordered data \citep{ma2021optimal, cai2023matrix}. A more involved nonlinear monotone structure is discussed in \cite{rigollet2019uncoupled}. 
Recent works have extended these methods to complex structures like graphons \cite{janssen2022reconstruction,natik2026consistency}. However, all these methods rely on monotone or even linear dependence of matrix entries on the latent ordering. 
The only exception is \cite{recanati2018reconstructing} where a circular seriation is considered. But the related data Laplacian matrix is assumed to  symmetric and Toeplitz.
These restrictive assumption may fail in many complex nonlinear systems.

Our work considers a general temporal system $\{X(t); t \in [0, 2\pi]\}$ where $X(t)$ may be highly nonlinear. 
Define $\calM \subset \reals^d$ as the collection of all points traced out by $X(t)$. Then $X(t)$ is a mapping from $[0, 2\pi]$ to $\calM$. It indicates that $\calM$ is a one-dimensional manifold, irrespective of the complexity of $X(t)$. This viewpoint motivates the use of manifold theories to recover the latent labels $t_i$ and the permutation. 
Crucially, this one-dimensional structure comes from the temporal nature of this system and does not rely on any restrictive assumptions on $X(t)$. 

Viewed as a curve, the manifold $\calM$ can take two topological forms.
\begin{definition}
Consider the manifold $\calM$ formed by $\{X(t); 0 \leq t \leq 2\pi\}$. We call $\calM$ 
\begin{itemize}
    \item {\it an open curve (non-periodic)}, if $\calM$ has two disjoint endpoints.
    \item {\it a closed loop (periodic)}, if $\calM$ does not have endpoints and forms a closed cycle. 
    \end{itemize}
\end{definition}
Molecular conformations often exhibit periodic behavior with $X(0) = X(2\pi)$, and the corresponding $\calM$ is a closed loop \citep{Zelesko2019EarthmoverBasedML}. In contrast, embryo development and evolution sequences usually generate non-periodic data where $X(0) \neq X(2\pi)$. These two cases represent different topological structures, for which we propose different approaches and theoretical results. In this paper, we do not consider self-intersecting curves.


There is a beautiful mathematical property of a one-dimensional manifold: the eigenfunctions of its manifold Laplacian are simple trigonometric functions of $t$; see \citep{rosenberg1997laplacian}. This observation motivates our spectral seriation approach. Let $\hat{v} = (\hat{v}_1, \hat{v}_2, \cdots, \hat{v}_N)^T$ be an eigenvector of the graph Laplacian constructed from $Z$. If the corresponding manifold eigenfunction is, say, $\cos(t)$, then applying the inverse function $\arccos(\hat{v}_i)$ yields an estimate $\hat{t}_i$. 
These estimates consequently determine the permutation. This forms the basic intuition behind our method. For the two scenarios, the eigenfunctions take more involved shapes.

The consistency of $\hat{t}_i$ under spectral seriation requires controlling the distance between the manifold eigenfunction evaluated at $t_i$ and the corresponding eigenvector entry $\hat{v}_i$. While the convergence between the manifold Laplacian matrix and the graph Laplacian matrix has been established \citep{coifman2006diffusion, singer2006graph}, it is the convergence of their eigenvectors that is more delicate.
We first briefly review existing results.
For the closed-loop case, \cite{garcia2020error} proves the spectral convergence in the $\ell_2$ norm when $Z_i = X(t_i)$, and \cite{cheng2022eigen} extends this to settings with additive noise $e_i$. 
For the open curve case, 
\cite{peoples2021spectral} discusses the spectral convergence in the noiseless setting. However, we need eigenvector convergence in the $\ell_{\infty}$ norm. In this direction, the only known result is  \cite{dunson2021spectral}, which establishes $\ell_{\infty}$ convergence for the closed-loop case without noise. 
However, our work requires $\ell_{\infty}$ convergence for both the closed-loop and open-curve cases with additive noise.


We establish spectral convergence in the $\ell_{\infty}$ norm for both the open-curve and closed-loop cases, in the presence of high-dimensional noise $e_i$ and eigenvalue multiplicity. Our results reveal several notable phenomena:
\begin{itemize}
    \item for the closed-loop case, the $\ell_{\infty}$ error has the same order as the $\ell_2$ error;
    \item for the open-curve case, the $\ell_{\infty}$ error is significantly smaller than the $\ell_2$ error reported in \cite{peoples2021spectral} when both endpoints are excluded; 
    \item for the open-curve case with endpoints, we provide a hybrid approach to integrate a simple graphical-distance-based refinement and spectral seriation that improves the $\ell_{\infty}$ error.
\end{itemize}
It may seem counterintuitive that our bound for the $\ell_{\infty}$ error is better than the $\ell_2$ error bound in the open-curve case. The key is that large errors occur mostly near the boundary, and our analysis captures this phenomenon explicitly. 
We also would like to point out that our analysis framework is fundamentally different from the settings commonly assumed in spectral perturbation theory. In particular, the matrix we analyze does not contain a low-rank component with a nondegenerate eigenvalue gap, a typical assumption in works such as \cite{abbe2015community,cai2018rate,chen2021spectral,ding2023learning}.

Our work is also connected to the statistical literature on manifold learning, a constantly 
expanding area driven by the manifold hypothesis for modern data. Our idea is intuitive: to leverage manifold embeddings to recover the temporal labels and thereby obtain the permutation. While we focus on spectral embedding due to its computational efficiency, mathematical understanding, and rich analysis tools, other manifold embedding tools can also be implemented.  For example, our augmented Algorithm \ref{alg:opencurvepatch} incorporates ideas similar to Isomap \citep{tenenbaum2000global}. 
We refer the interested audience to \cite{whiteley2022statistical} for a comprehensive discussion of manifold learning methods.

\section{Methodology: seriation via spectral embedding}

\subsection{Data, notations, and manifold embedding}
Let $\{X(t); 0 \leq t \leq 2\pi\}$ denote the underlying system with a latent parameter $t$. 
For simplicity of notations, let $X_i = X(t_i)$ be the underlying truth at $t_i$.
We observe noisy data points $Z_i = X_i + e_i\in \reals^d$, $i \in [N]$, where $t_i$, $X_i$ and $e_i$ are all unobserved. 
The true timestamps $t_i$ induce a permutation ${\bfpi}: [N] \to [N]$ such that 
$t_{\pi(1)} < t_{\pi(2)} < \cdots < t_{\pi(N)}$. 
Accordingly, the reordered observations $\{Z_{\pi(i)}\}$ follow the correct temporal order.

Our goal is to recover the underlying temporal labels $\bft = (t_1, t_2, \cdots, t_N)$ and hence the permutation $\bfpi$. Recall that $\calM$ denotes the manifold on which the curve $X(t)$ lies. 
Denote $Z = (Z_1, \cdots, Z_N) \in \reals^{d \times N}$ and $X = \E(Z) = (X_1, \ldots, X_N) \in \reals^{d \times N}$ as the data matrix and the signal matrix, respectively. 
Although we work with the interval $[0, 2\pi]$ for convenience, all results extend directly to other finite intervals.


To propose an estimate $\hat{t}_i$, we seek a mapping $g:\reals^d\times \reals^{d \times N}\mapsto [0,2\pi]$ so that $t_i \approx \hat{t}_i = g(Z_i; Z)$. 
Determining such a mapping $g$ is challenging for several reasons. 
First, this is an unsupervised problem, and no training labels $t_i$ are available. 
Second, our goal is to recover the timestamp $t_i$ associated to each individual observation $Z_i$. Unlike classic parametric estimation which benefits from a large sample size, pointwise recovery of $t_i$ is inherently limited by the observation noise $e_i$, leading to an irreducible error that does not vanish as $N$ grows. These challenges are more severe when the feature dimension $d$ is large.

To address these challenges, we exploit the geometric structure of the data. Because $X(t)$ maps a one-dimensional interval $[0,2\pi]$ to $\reals^d$, the image $\calM$ is intrinsically one-dimensional, regardless of how complex $X(t)$ may appear. 
This observation motivates the use of manifold-embedding methods. Such methods are designed to capture nonlinear structure and to construct a low-dimensional representation of each data point $Z_i$ that reflects its position along the manifold. 
These approaches are grounded in the manifold hypothesis, where the high-dimensional data lies near a latent low-dimensional manifold \citep{whiteley2022statistical}. In our temporal system setting, this hypothesis is satisfied by construction.

We illustrate the pipeline of using manifold embedding for seriation, taking spectral embedding as an example.
Spectral embedding begins by quantifying local similarity between data points through a kernel function. Specifically, we use the Gaussian kernel with bandwidth $\sigma>0$:
\begin{equation}
\label{eqn:kernel}
    k(Z_i,Z_j; \sigma)=(\sqrt{2\pi}\sigma)^{-1}\exp(-(2\sigma^2)^{-1}\|x-y\|^2).
\end{equation}
Because $k(Z_i,Z_j;\sigma)$ decays rapidly when $\|Z_i - Z_j\|\gg \sigma$, this similarity score emphasizes local interactions. 
It should be noted that, even if data points are already sorted, the matrix with entries $k(Z_i,Z_j;\sigma)$ may not be monotone or Toeplitz. Therefore, methods and analysis from existing literature do not directly apply to this general setting.

Based on $k(Z_i,Z_j;\sigma)$, we form a similarity graph and define the corresponding normalized graph Laplacian $L^Z$, 
\begin{align}
\label{eqn: close normalization}
L^Z(i,j;\sigma)=1_{i=j}-k(Z_i, Z_j;\sigma)/\sqrt{d^Z(i)d^Z(j)}, \quad 
d^Z(i)=\sum\nolimits_{j\in[N]}k(Z_i,Z_j; \sigma),
\end{align}
where $d^Z(i)$ is the degree of node $i$. 

Let $\hat{v}_k$ be the eigenvector of $L^Z$ corresponding to the $k$-th smallest eigenvalue. Then the $K$-dimensional spectral embedding of the data point $Z_i$ is given by 
\[
V_i = (\hat{v}_2(i), \hat{v}_3(i), \cdots, \hat{v}_{K+1}(i)), \quad i \in [N]. 
\]
When $K = 1$, the embedding coordinate $V_i = \hat{v}_2(i)$ can be seen as a direct approximation of $t_i$. When $K > 1$, the embedding provides a low-dimensional representation from which an inverse mapping to $t$ can be constructed. This inversion step forms the basis of spectral seriation.

Although many other manifold-embedding methods exist, including isometric feature mapping (Isomap)\citep{tenenbaum2000global}, t-distributed stochastic neighbor embedding (t-SNE)\citep{maaten2008visualizing}, and uniform manifold approximation and projection (UMAP), we focus on the spectral embedding for several reasons. First, Isomap  has computation complexity at $\widetilde{O}(N^3)$ due to shortest-path computations; 
in comparison, spectral method in general incurs $\widetilde{O}(N^2)$ complexity.
Second, some underlying assumptions of these alternative methods may be violated in temporal systems. For example, Isomap assumes $\calM$ is geometrically convex, which fails in the closed-loop setting. 
Finally, the theoretical analysis for spectral embedding has a more solid foundation \citep{singer2006graph, peoples2021spectral, cheng2022eigen}. To the best of our knowledge, theoretical guarantees for Isomap are largely limited to graphical distance approximations \citep{bernstein2000graph}; and existing analysis for t-SNE focus on clustered data \citep{cai2022theoretical}. Given these perspectives, we adopt spectral embedding in our methodology.

Throughout the paper, for a vector $a$, we use $\|a\|$ and $\|a\|_{\infty}$ to denote its $\ell_2$ and $\ell_{\infty}$ norms, respectively. For a matrix $A$, $\|A\|$ and $\|A\|_F$ denote its spectral norm and Frobenious norm. 
For an integer $N$, we write $[N]=\{1,2,\ldots,N\}$. Denote $1_{A}$ as the indicator function of event $A$. 

\subsection{Spectral seriation for the ideal case}\label{subsec:oracle}

In this section, we formalize the spectral seriation algorithm. Although the intuition has been introduced earlier, an explicit choice of the embedding dimension $K$ and a concrete map from $V_i$ to $t_i$ are still required.

A key advantage of spectral embedding is that its asymptotic behavior is well understood, which provides useful guidance for algorithm design. Recall the normalized Laplacian $L^Z$ defined in \eqref{eqn: close normalization}. In \cite{singer2006graph},
it is shown that when $N\to \infty$ and the bandwidth $\sigma\to 0$ in the Gaussian kernel, the rescaled matrix $\sigma^{-2}L^Z$ converges to the negative manifold Laplacian $-\Delta_{\calM}$ in operator norm. In the one-dimensional setting, the eigenfunctions of $-\Delta_{\calM}$ have explicit analytic forms, which we summarize below. 
\begin{itemize}
    \item For the closed-loop case where $X(0) = X(2\pi)$, the system is periodic. 
The first eigenfunction is constant with eigenvalue 0, and the subsequent eigenfunctions take the form $\cos(kt)$ and $\sin(kt)$, with eigenvalues $k^2$ for integers $k \geq 1$. Therefore, for a point $X(t_i) \in \calM$, the two eigenfunctions with eigenvalue $1$ yield an ideal two-dimensional embedding
    \[
    V^{o}_i = (\cos(t_i), \sin(t_i)). 
    \]
    The inverse mapping from $V^o_i$ to $t_i$ is obtained by identifying the angle corresponding to $V^{o}_i$, i.e. mapping the point $V^o_i$ on the unit circle in $\reals^2$ to an angle in $[0, 2\pi]$. 
    \item For the open-curve case where $X(0) \neq X(2\pi)$, the system is non-periodic.
The first eigenfunction remains constant with eigenvalue 0, while the subsequent eigenfunctions are of the form $\cos(kt/2)$ and $\sin(kt/2)$, with eigenvalues $k^2/4$. In this case, a single eigenfunction corresponding to $1/4$ suffices to represent the temporal structure. The ideal embedding yields
    \[
    V^{o}_i = \cos(t_i/2). 
    \]
    The inverse map follows directly as $t_i = 2\arccos(V^{o}_i)$.
\end{itemize}

This analysis concludes two important facts. First, the appropriate embedding dimension $K$ depends on the topology of the underlying manifold. Specifically, one may take $K = 1$ for the open-curve case and $K = 2$ for the closed-loop case. 
In practice, one may also use larger $K$ since additional eigenvectors continue to encode information. But we only consider the minimal $K$ in this work for simplicity.
Second, in both settings, the inverse mapping from the ideal embedding $V^o_i$ to the temporal label $t_i$ can be explicitly identified. 
This elegant connection between $V_i^o$ and $t_i$ comes from the fact that the manifold $\calM$ is a 1-dimensional manifold with well-characterized spectral properties. It explains our choice of spectral embeddings and motivates the two spectral seriation algorithms in the following sections.

\subsection{Spectral seriation for the closed-loop case}\label{subsec:algclosed}
Consider the closed-loop case where $X(t)$ is periodic. 
Recall the normalized Laplacian $L^Z$ defined in \eqref{eqn: close normalization}. Denote $\lambda_k^Z$ to be the $k$-th smallest eigenvalue of $L^Z$, where $\lambda^Z_1 = 0$.

Our analysis of the ideal setting shows that the eigenvectors corresponding to $\lambda_2^Z$ and $\lambda_3^Z$ are sufficient for temporal label and permutation recovery. 
We denote by $F^Z_+$ and $F^Z_- \in \reals^N$ the eigenvectors associated with these two eigenvalues. 
The spectral embedding $V_i$ for $Z_i$ is
\[
V_i = (F^Z_+(i),F^Z_-(i)), \qquad i \in [N].
\]

\indent
In the ideal setting, $F^Z_+(i)$ and $F^Z_-(i)$ converge to $\cos(t_i)$ and $\sin(t_i)$, respectively. However, this convergence holds only up to a rotation and rescaling. 
Because $\cos(t)$ and $\sin(t)$ are associated with the same eigenvalue $1$, any rotation of them yields another valid eigenbasis. Therefore, the convergence exists quotient a rotation, i.e., there exist constants $c > 0$, $0 < \theta < 2\pi$, so that for all $i \in [N]$, 
\[
V_i = (F^Z_+(i),F^Z_-(i))\approx c(\cos(t_i+\theta),\sin(t_i+\theta))\quad  i\in [N]. 
\]
Let $\hat{V}_i$ denote the normalized version of $V_i$ with unit norm. 
The temporal label can then be recovered as $\hat{t}_i = g(\hat{V}_i(1), \hat{V}_i(2))$, where $g$ maps a point on the unit circle on $\reals^2$ to its corresponding angular coordinate. 
The algorithm can be found in Algorithm \ref{alg:closedcurveuniform}.

It can be found that the temporal labels $t_i$ are identifiable only up to a global rotation. Moreover, since eigenvectors are defined up to a sign, replacing $F_-^Z$ by $-F_-^Z$ yields $\tilde{t}_i = 2\pi - t_i$. Therefore, $t_i$ is determined up to a reflection. 
As a result, the estimated permutation $\bfpi$ is identifiable only up to a cyclic shift and a possible reversal. 
Such non-identifiability is intrinsic in periodic systems.

\begin{algorithm}
\caption{Temporal label and permutation recovery for periodic data.}
\label{alg:closedcurveuniform}
\begin{algorithmic}[1]
    \Require Data $Z$; Gaussian kernel bandwidth $\sigma$
    \State Compute the graph Laplacian $L^Z$, where
    \[
        L^Z(i,j) = \mathbf{1}_{i=j} - k(Z_i, Z_j;\, \sigma) \big/ \sqrt{d^Z(i)\, d^Z(j)}.
    \]
    \State Find the second and third smallest eigenvalues and their eigenvectors $F^Z_+$ and $F^Z_-$ of $L^Z$.
    \State Find $\hat{t}_i \in [0, 2\pi]$ such that
    \[
        \cos(\hat{t}_i) = \frac{F^Z_+(i)}{\sqrt{(F^Z_+(i))^2 + (F^Z_-(i))^2}}, \qquad
        \sin(\hat{t}_i) = \frac{F^Z_-(i)}{\sqrt{(F^Z_+(i))^2 + (F^Z_-(i))^2}}.
    \]
    \State Find a permutation $\hat{\pi}$ of $\hat{\bft}_{[N]}$ such that $\hat{\bft}_{\hat{\pi}(i)}$ is increasing in $i$.
    \Ensure Estimated time stamps $\hat{\bft}$ and permutation $\hat{\bfpi}$
\end{algorithmic}
\end{algorithm}


\begin{remark}
The algorithm involves a tuning parameter $\sigma$, the bandwidth of the Gaussian kernel. By Theorem \ref{thm:closed}, we suggest choosing $\sigma=\max\{N^{-1/7},\epsilon^{1/4}\}$, where $\epsilon $ represents the approximate magnitude of $Z_i - \E(Z_i)$. In our numerical analysis, we adopt a data-driven selection of $\sigma$ in \cite{4623181}; see Appendix \ref{sec:appA} for details. 
\end{remark}

\subsection{Spectral seriation for the open-curve case}\label{sec:openss}

Consider the open-curve case where $\calM$ has two endpoints. According to our analysis in the ideal setting, a single eigenfunction $\cos(t/2)$ suffices to recover $\bft$ and $\bfpi$. Denote $F^Z$ as the eigenvector of the graph Laplacian $L^Z$, corresponding to its second smallest eigenvalue $\lambda_2^Z$. 
If $F^Z$ converges to the limiting eigenfunction, then there exists a constant $c$, so that
\begin{equation}
F^Z(i)\approx c \cos(t_i/2), \quad i\in [N]. 
\end{equation}
The estimated temporal label follows that $\hat{t}_i = 2\arccos(c^{-1}F^Z(i))$.

Unlike the closed-loop case, spectral convergence in the open-curve setting is more delicate. This difficulty arises because $\calM=\{X(t); t \in [0,2\pi]\}$ is not a closed manifold but a manifold with two boundary points $\{X(0), X(2\pi)\}$. So the matrix convergence results of \cite{singer2006graph}, which apply to manifolds without boundary, may no longer hold. \cite{peoples2021spectral} proposes an alternative normalization of the graph Laplacian based on the same Gaussian kernel, under which spectral convergence can be established in the noiseless case. 
We employ this modified definition of $L^Z$, in which the difference from the closed-loop case lies solely in the normalization term,
\begin{equation}
\label{eqn: open normalization}
L^Z(i,j)= 1_{i=j}-k(Z_{i},Z_{j};\sigma)(\frac{1}{2d^Z(i)}+\frac{1}{2d^Z(j)}).    
\end{equation}
In Theorem \ref{thm:openeigen}, we establish the spectral convergence in the presence of noise $e_i$.

Compared with the closed-loop case, the open-curve case does not suffer from rotational ambiguity. However, the scaling constant 
$c$ must be identified. Our theorems show that $F^Z(i)$ converges to $\pm N^{-1/2}\cos(t_i)$, which indicates $c = \pm N^{-1/2}$. The sign of $c$ corresponds to a possible reflection of the temporal order. Therefore, we have the estimate $\hat{t}_i = 2\arccos(N^{1/2}F^Z(i))$. The recovered temporal labels $\hat{\bft}$ and permutation $\hat{\bfpi}$ are up to a reflection. It gives us the spectral seriation algorithm for the open curve case, in Algorithm \ref{alg:opencurveuniform}.

\begin{algorithm}
\caption{Temporal label and permutation recovery for non-periodic data.}
\label{alg:opencurveuniform}
\begin{algorithmic}[1]
    \Require Data $Z$; Gaussian kernel bandwidth $\sigma$
    \State Compute the graph Laplacian $L^Z$, where
    \[
        L^Z(i,j) = \mathbf{1}_{i=j} - k(Z_i, Z_j;\, \sigma) \big/ \sqrt{d^Z(i)\, d^Z(j)}.
    \]
    \State Let $F^Z$ be the eigenvector of $L^Z$ corresponding to the second smallest eigenvalue.
    \State Estimate $\hat{t}_i = 2\arccos\!\left(\sqrt{N}\, F^Z(i)\right)$.
    \State Find a permutation $\hat{{\bfpi}}$ of $\hat{t}_{[N]}$ such that $\hat{t}_{\hat{\pi}(i)}$ is increasing in $i$.
    \Ensure Estimated time stamps $\hat{t}_{[N]} \in [0, 2\pi]$ and ranking $\hat{{\bfpi}}$
\end{algorithmic}
\end{algorithm}


\begin{remark}
    For non-periodic data, Theorem \ref{thm:open} suggests choosing the kernel bandwidth as $\sigma=\max\{N^{-1/14},\epsilon^{2/7}\}$ with $\epsilon$ being the noise level. In practice, we adopt a data-driven choice of $\sigma$, following the procedure in the Appendix.
\end{remark}

\subsection{Hybrid permutation recovery for the open-curve case with endpoints}\label{subsec:hybrid}

The spectral seriation algorithm in Section \ref{sec:openss} performs well for interior points of an open curve, but its accuracy drops quickly near the endpoints $t \in \{0, 2\pi\}$. 
This behaviour is not surprising. The method relies on the eigenfunction $\cos(t/2)$ to recover $t_i$ for $X_i$, whose derivative is $-\sin(t/2)/2$. As $t \to 0$ or $t \to 2\pi$, this derivative approaches 0. It means separate $t_i$ values are mapped to nearly identical eigenvector entries $F^Z(i)$, making them hard to distinguish.
As a result, even mild noise can lead to large estimation errors at the boundary points.

To solve this problem, we exploit the relationship between the geodesic distance on the manifold and the graphical distance in a similarity graph. Consider a trajectory $X(t): [0, 2\pi] \to \calM$ as a curve connecting two endpoints $X(0)$ and $X(2\pi)$ on $\calM$. If the geodesic distance $d_{\calM}(Z_i, X(0)) = t_i$ were available, then the temporal ordering of $Z_i$ would follow directly from this distance. In practice, the geodesic distance is unknown and must be approximated. We therefore estimate it using graph distances in a weighted graph constructed from the data. Let the vertex set consist of the data points, and an edge $(Z_i, Z_j)$ exists if and only if $\|Z_i - Z_j \| \leq \sigma_0$ for a threshold $\sigma_0 > 0$. The graphical distance $d_G(Z_i, Z_j)$ between two data points is then defined as the length of the shortest path connecting them, which can be found using the Dijkstra algorithm. When $\calM$ is isometrically embedded in $\reals^d$, local Euclidean distances provide approximations of geodesic distances, and the graph distances capture the global geometry of the curve. The temporal ordering can be achieved by ordering $d_G(Z, Z_i)$ for a given $Z$ close to $X(0)$.

To obtain an overall accurate permutation, we combine spectral seriation and graph-distance-based ordering in a hybrid procedure. We first identify the endpoint by selecting the data point with the smallest degree $i^* = \arg\min_i d^Z(i)$. 
This is because the endpoints have less neighbors, which induces a smaller degree. 
Based on the graphical distance to $i^*$, we separate the points near the boundary and interior points. 
Next, we apply the spectral seriation method in Algorithm \ref{alg:opencurveuniform} to the full dataset and then restrict the resulting ordering $\hat{\bfpi}$ to the interior points. 
For points near the boundary, we instead use the ordering induced by graph distances. These two partial rankings are merged to obtain the final ranking. Details are in Algorithm \ref{alg:opencurvepatch}.

\begin{algorithm}
\caption{Hybrid permutation recovery for open curve.}
\label{alg:opencurvepatch}
\begin{algorithmic}[1]
    \Require Data $Z_{[N]}$; bandwidth $\sigma$, threshold $\sigma_0$, proportion $\delta$
    \State Graph distance sorting:
    \begin{enumerate}[label=(\alph*)]
        \item Find the endpoint $i^* = \arg\min_{i}\, d^Z(i)$.
        \item Build a weighted graph $G$ on $N$ data points, where $G(i,j) = \|Z_i - Z_j\|$ if $\|Z_i - Z_j\| \leq \sigma_0$, and $G(i,j) = 0$ otherwise.
        \item Find the graph distance between $i^*$ and all $i \in [N]$ via Dijkstra's algorithm, denoted $d_G(i, i^*)$.
        \item Let $\hat{{\pi}}^G$ be the ranking obtained by sorting $d_G(i, i^*)$.
        \item Set $\hat{t}^G_i = 2\pi N^{-1} \hat{\pi}^G_i$.
    \end{enumerate}
    \State Apply the spectral seriation method in Algorithm~\ref{alg:opencurveuniform} to obtain $\hat{t}$ and $\hat{{\bfpi}}$.
    \State Combine the two rankings:
    \begin{enumerate}[label=(\alph*)]
        \item Let $i_1$ and $i_2$ be the $\delta N$-th and $(1-\delta)N$-th data points according to $\hat{{\pi}}^G$.
        \item If $\hat{t}_{i_1} > \hat{t}_{i_2}$, set $\hat{t}_i = 2\pi - \hat{t}_i$ and $\hat{\pi}_i = N + 1 - \hat{\pi}_i$ for all $i$.
        \item For all $i$ with $\hat{\pi}^G_i \notin [\delta N,\, (1-\delta)N]$, set $\hat{t}_i = \hat{t}^G_i$.
    \end{enumerate}
    \State Sort $\hat{{\bfpi}}$ according to $\hat{t}_i$.
    \Ensure Estimated time stamps $\hat{t}_{[N]}$ and permutation $\hat{{\bfpi}}$
\end{algorithmic}
\end{algorithm}


The graph distance idea was employed in Isomap \citep{tenenbaum2000global}. 
However, while Isomap focuses on constructing a low-dimensional embedding and requires computing all-pairs shortest paths, our Algorithm \ref{alg:opencurvepatch} is designed for the permutation recovery. Therefore, we only need Dijkstra algorithm for one run to find the distance $d_G(i,i^*)$ from the identified endpoint $i^*$ to all $i$. It leads to a substantially lower computational cost.

\section{Theoretical guarantee}
\subsection{Model assumptions on the temporal system}
To introduce the theoretical results of spectral seriation algorithms, we first set up the assumptions on the temporal system $\{X(t)\}_{t \in [0,2\pi]}$ and the noise. Recall that $\calM$ is the manifold formed by $\{X(t)\}_{t \in [0,2\pi]}$ and the observed data points are $Z_i = X(t_i) + e_i$, $i \in [N]$. 
\begin{assumption}
\label{aspt:manifold}
$\calM$ is a one-dimensional manifold isometrically embedded in $\reals^d$. In particular, there is a constant $L_{\calM} > 0$, so that for any two points $x, y \in \calM$ and $d_{\calM}(x,y)$ denoting the distance between $x$ and $y$ on $\calM$, there is $\|x-y\|\leq d_{\calM}(x,y)\leq \|x-y\|(1+L_{\calM}\|x-y\|^2)$.
\end{assumption}
This assumption requires the system $X(t)$ to be a regular isometric embedding. It ensures that the geodesic distance on $\calM$ can be captured by the Euclidean distance in $\reals^d$, with $L_{\calM}$ controlling the curvature of this embedding.
This is a common 
assumption for manifold embedding methods \citep{garcia2020error,cheng2022eigen}. Notably, the local isometry is necessary for the identifiability of temporal labels $t_i$. Otherwise, we may recover $\tilde{t} = h(t)$ for any monotone function $h$, and the temporal system $X(h^{-1}(\tilde{t}))$ is the same.

\begin{assumption}
\label{aspt:uniform}
The time points $t_i$ follow the uniform distribution on $[0, 2\pi]$ for $i \in [N]$. 
\end{assumption}
This uniform assumption guarantees that the whole curve is covered in the observations.
For non-uniformly distributed time points, \cite{cheng2022eigen} and \cite{peoples2021spectral} suggest a graph Laplacian based on density correction. In practice, our spectral seriation algorithms could adopt this generalized graph Laplacian for non-uniform $t_i$, while the theoretical analysis becomes more involved.
\begin{assumption}
\label{aspt:noise}
There is a constant $\epsilon > 0$, so that $\sup_{i \in [N]}\|e_i\|\leq \epsilon$.
\end{assumption}

Under this assumption, the noise is uniformly bounded by a constant $\epsilon$, which plays an important role in accuracy analysis. Assumptions of similar form can be found in manifold learning literature \citep{garcia2020error,ding2023learning}. 
 In more challenging settings where $d \to \infty$, one may need to denoise the data so that the noise level $\epsilon$ remains constant. A brief discussion on data denoising algorithms and theoretical results can be found in \cite{whiteley2022statistical}, \cite{tong2025uniform}, and 
our appendix.

These assumptions allow us to analyze the estimation error. The estimation error consists of two components, which lie in approximating the manifold Laplacian of $\calM$ by the empirical graph Laplacian $L^Z$. 
The first arises from discretization: the map $X(t)$ is continuous, but only $N$ time points are observed. Even without noise, the graph Laplacian based on $X_i$ is a discrete approximation, which becomes accurate only as the sampling becomes arbitrarily dense, i.e., $N \to \infty$. 
The second arises from noise: estimation of $\bft$ and $\bfpi$ requires pointwise accuracy, instead of an accurate estimate of model parameters. Therefore, the observation noise $e_i$ introduces an inherent error that vanishes only when the noise level $\epsilon \to 0$.
These two conditions address different aspects of the problem and cannot replace one another.


In the following context, $L^Z$ denotes the graph Laplacian constructed on the data $Z$ by Algorithms \ref{alg:closedcurveuniform} and \ref{alg:opencurveuniform}, depending on the scenario. $F^Z_j \in \reals^{N}$ denotes the eigenvector of $L^Z$ corresponding to the $j$-th smallest eigenvalue.  We say $a_N =\Otilde(b_N)$ or $a_N \lesssim b_N$, if there are universal constants $C$ and $m$ so that 
$a_N\leq C (\log N)^m b_N$ for $N \to \infty$. 


\subsection{Convergence guarantees for the closed-loop case}\label{subsec:closederror}
When $\calM$ is a closed loop, there are no natural starting or ending points. As discussed in Section \ref{subsec:algclosed}, $\bft$ is identifiable up to a rotation and a rotation-reflection. To take this issue into account, we denote $[a]_{2\pi}: \reals \to [-\pi, \pi)$ as the representative of $a$ within the interval $[-\pi, \pi)$ under modulo $2\pi$. 
For two time sequences $\bft,\hat{\bft}\in \reals^N$, the error is the minimal $\ell_\infty$ distance between $\bft$ and $\hat\bft$ under all possible rotational matching or reflexive rotational matching, that is, 
\begin{equation}
\label{eqn: closed errors}
Err(\hat\bft, \bft)= \min_{r=\pm 1, \theta\in [0,2\pi]}
\max_{i\in[N]} \left|[rt_i+\theta-\hat{t}_i]_{2\pi}\right|.
\end{equation}
Here $\theta$ denotes the additional rotational angle, $r=1$ leads to a matching without reflection, and $r=-1$ leads to a matching with an additional reflection. 

For the permutation $\bfpi$, the rotation effect is captured by a shift $n$. The reflection effect in ranking is denoted as $\bfpi^r$ where $\pi^r(i) = N - \pi(i)$. To account for the modular nature of the ranking, we allow an additional integer offset $j$. We define the error measure as
\begin{equation}\label{eqn: closed ranking errors}
Err(\hat\bfpi, \bfpi)=\min_{\tilde{\bfpi} \in \{\bfpi, \bfpi^r\},n \in [N]}
\max_{i\in[N]} \min_{j\in \{0,\pm 1\}}\{|\tilde{\pi}(i)+n+jN-\hat\pi(i)|\}/N.
\end{equation}

Both definitions concern the maximum error in the time sequence or the ranking, namely, the $\ell_{\infty}$ error.  Bounding this error is both challenging and crucial, as it implies an overall consistent limit on estimation accuracy. 

To prove the effectiveness of Algorithm \ref{alg:closedcurveuniform}, we first show the convergence of the Fiedler eigenvectors to the manifold eigenfunctions. 
\begin{theorem}
\label{thm:closedeigen}
[Convergence of eigenvectors]
Suppose $\calM$ is a closed loop and Assumptions \ref{aspt:manifold} -- \ref{aspt:noise} are satisfied with a constant $\epsilon > 0$. 
Consider Algorithm \ref{alg:closedcurveuniform} with a Gaussian kernel bandwidth of $\sigma > 0$.
Let $F^Z_k$ be the $k$-th smallest eigenvector of $L^Z$, $k = 2, 3$, then  with high probability $1 - O(N^{-1})$, there is a constant $c > 0$, so that 
\[
\min_{r = \pm1, \theta \in [0,2\pi]}\max_{i \in [N]}|\sqrt{N}F^Z_k(i)-c\cos((k-1)(rt_i+\theta)) |=\Otilde(\sigma^{-2} \epsilon +\sigma^{-3/2} N^{-1/2}+\sigma^2),
\]
where $r$ and $\theta$ are the rotational and reflective parameters.
\end{theorem}

\begin{theorem}
\label{thm:closed}
Under the same assumptions as in Theorem \ref{thm:closedeigen}, further suppose that the sample size $N$ and the Gaussian kernel bandwidth $\sigma$ satisfy that $N\sigma>1$ and $\sigma<1/L_{\calM}$, then with high probability $1 - O(N^{-1})$, the estimates $\hat{\bft}(Z)$ and $\hat{\bfpi}(Z)$ by Algorithm \ref{alg:closedcurveuniform} satisfy that 
\begin{eqnarray*}
Err(\hat{\bft},\bft)\leq \Otilde(\sigma^{-2} \epsilon +\sigma^{-3/2} N^{-1/2}+\sigma^2),\\
Err(\hat{\bfpi},\bfpi)\leq \Otilde(\sigma^{-2} \epsilon +\sigma^{-3/2} N^{-1/2}+\sigma^2).
\end{eqnarray*}

\end{theorem}
These results establish the error bound of Algorithm \ref{alg:closedcurveuniform} even when confronted with high-dimensional noise such that $|Z_i - X(t_i)| \leq \epsilon d^{-1/2}$. The error bound depends on the noise level $\epsilon$, sample size $N$, and the Gaussian kernel bandwidth $\sigma$ employed in constructing $L^Z$. 
The effects of dimensionality $d$ have been removed by the denoising step \citep{tong2025uniform}, leading to the notable phenomenon that the error bound does not depend on $d$. 

\begin{remark}
By Theorems \ref{thm:closedeigen} and \ref{thm:closed}, it is essential for the bandwidth $\sigma$ to meet the conditions $\sigma > \epsilon^{1/2}$ and $\sigma > N^{-1/3}$ to effectively counter data noise. 
When $N$ and $\epsilon$ are fixed, we choose $\sigma=\max\{N^{-1/7},\epsilon^{1/4}\}$, which yields an error of order $\max\{N^{-2/7},\epsilon^{1/2}\}$. 
\end{remark}

\begin{remark}
We illustrate our error bound under two special cases.
When there is no noise, that is, $\epsilon = 0$, the ideal bandwidth is $\sigma = N^{-1/7}$ with $Err(\hat\bft, \bft) = \tilde{O}(N^{-2/7})$. This error rate coincides with the upper bound of the $\ell_2$ error established in \cite{garcia2020error} for the noiseless scenario. Remarkably, our approach achieves a uniform error bound that is the same with the $\ell_2$ error bound.
When $\epsilon$ is fixed and the sample size $N\to \infty$, the optimal (for minimizing the upper bound) bandwidth $\sigma=\epsilon^{1/4}$ 
leads to $Err(\hat\bft, \bft) =  \epsilon^{1/2}$. It can be seen as the ``essential error" induced by noises. 
\end{remark}

\subsection{Convergence guarantees for the open-curve case}\label{subsec:openerror}

For an open curve $\mathcal{M}$ with two separate endpoints, rotation is no longer a concern. Consequently, the time sequence $\bft$ can be identified up to a reflection. Hence, we define $Err(\hat{\bft}, \bft)$ and $Err(\hat\bfpi, \bfpi)$ similar to \eqref{eqn: closed errors} and \eqref{eqn: closed ranking errors}, where the only difference is that we remove $\theta$ and the shift $n$, respectively. 

In Section \ref{subsec:hybrid}, we mentioned that our spectral seriation approach succeeds on the interior region but fails at endpoints. Therefore, we further define the error on the interior of the interval defined by a parameter $\delta$, i.e., $int_{\delta} = [\delta, 2\pi - \delta]$. Taking the reflection effect into account, the error is defined as follows:
\begin{align}
\label{eqn: open errors}
Err_{\delta}(\hat\bft, \bft)= \min\bigl\{
 \max_{i: \delta<t_i<2\pi-\delta}|t_i-\hat{t}_i|, \max_{i: \delta<t_i<2\pi-\delta}|2\pi-t_i-\hat{t}_i|\bigr\}.
\end{align}

When it comes to the ranking $\pi$, the ranks under consideration are $S_{\delta}(\pi) = \{i: N\delta \leq \pi(i) \leq N(1-\delta)\}$. The error of $\hat{\bfpi}$ is defined as:
\begin{align}
\label{eqn: open rank errors}
Err_{\delta}(\bfpi,\bfpi')= N^{-1}\min\bigl\{
 \max_{i \in S_{\delta}(\pi)}|\pi(i)-\pi'(i)|,\max_{i \in S_{\delta}(\pi)}|N-\pi(i)-\pi'(i)|\bigr\}.
\end{align}

We first discuss the estimation error of spectral seriation on the interval $[\delta, 2\pi - \delta]$. For the open-curve case with boundaries, the convergence rate is slower. 
\begin{theorem}
\label{thm:openeigen}
Suppose $\calM$ is an open curve and Assumptions \ref{aspt:manifold}--\ref{aspt:noise} hold. 
Let $F^Z_k$ be the $k$-th smallest eigenvector of $L^Z$ for a fixed integer $k \in [N]$ by Algorithm \ref{alg:opencurveuniform}. Recall that the eigenfunction of the manifold Laplacian is $\cos(kt/2)$ with respect to the $k+1$-th smallest eigenvalue. Then with high probability  at least $1-O(N^{-2})$, for $\delta > \sigma^2+N^{-1/2}\sigma^{-3/2}(\log N)^{1/2}$,
\[
\min_{r\in \pm 1}\max_{\delta \leq t_i \leq 2\pi - \delta}|\sqrt{N}F^Z_{k+1}(i)- r\cos(kt_i/2) |=\Otilde (\sigma+\sigma^{-5/2}N^{-1/4}+{\sigma^{-5/2}}{\epsilon}).
\]
\end{theorem}

Following the eigen-analysis, we have the estimation error in $\hat{\bft}$ and $\hat{\bfpi}$ in Algorithm \ref{alg:opencurveuniform}.
\begin{theorem}
\label{thm:open}
Under the same assumptions with Theorem \ref{thm:openeigen}, further suppose that the sample size $N$ and the Gaussian kernel bandwidth $\sigma$ satisfy that $N\sigma>1$ and $\sigma<1/L_{\calM}$, then with high probability $1 - O(N^{-2})$, for $\delta > \sigma^2+N^{-1/2}\sigma^{-3/2}(\log N)^{1/2}$, the estimates $\hat{\bft}(Z)$ and $\hat{\bfpi}(Z)$ by Algorithm \ref{alg:opencurveuniform} satisfy that 
\begin{eqnarray*}
Err_{\delta}(\hat\bft,\bft)\leq \Otilde (\sigma+\sigma^{-5/2}N^{-1/4}+{\sigma^{-5/2}}{\epsilon}),\\
Err_{\delta}(\hat\bfpi,\bfpi)\leq \Otilde (\sigma+\sigma^{-5/2}N^{-1/4}+{\sigma^{-5/2}}{\epsilon}).
\end{eqnarray*}
\end{theorem}

When the sample size $N$ and the noise level $\epsilon$ are fixed, the optimal Gaussian kernel bandwidth is $\sigma=\max\{N^{-1/14},\epsilon^{2/7}\}$, which yields an error of order $Err_{\delta}(\hat\bfpi,\bfpi) = \tilde{O}(\max\{N^{-1/14},\epsilon^{2/7}\})$. 
In the special case where $\epsilon = 0$, the optimal bandwidth $\sigma = N^{-1/14}$ leads to an error at $\tilde{O}(N^{-1/14})$, indicating the discretization error. In the special case where $N \to \infty$, choosing $\sigma = \epsilon^{2/7}$ yields $Err_{\delta}(\hat\bfpi,\bfpi) = \tilde{O}(\epsilon^{2/7})$, implying the noise effect. In both cases, the open-curve setting demonstrates a slower convergence rate than the closed-loop case. 

Surprisingly, for the noiseless case $\epsilon = 0$, our convergence rate is better than the $\ell_2$ error upper bound established in \cite{peoples2021spectral}. 
Although the $\ell_{\infty}$ norm provides point-wise control and the $\ell_2$ norm benefits from averaging, our rate is $\Otilde(N^{-1/14})$ whereas their rate is $\tilde{O}(N^{-1/28})$. 
This counterintuitive outcome stems from our error measurement \eqref{eqn: open errors}, which excludes the endpoints. Errors near the endpoints dominate the global error if they are included.

While the spectral method performs well on the interior, accurate ranking of the endpoints is often of interest. In Section \ref{subsec:hybrid}, we propose Algorithm \ref{alg:opencurvepatch} to handle this issue. Below, we show that this improved algorithm extends the error bounds in Theorem \ref{thm:open} from $int_\delta$ to the entire curve.
\begin{theorem}
\label{thm:openpatched}
Under the same assumptions with Theorem \ref{thm:open}, further suppose that the graph distance threshold $\sigma_0$ satisfies $(10L_{\calM})^{-1/2}>\sigma_0>\min\{20\epsilon, ({10N^{-1}\log N})^{1/2}\}$, then with high probability $1 - O(N^{-2})$, the estimates $\hat{\bft}(Z)$ and $\hat{\bfpi}(Z)$ by Algorithm \ref{alg:opencurvepatch} satisfy that 
\begin{eqnarray*}
Err(\hat\bft,\bft)\leq \Otilde (\sigma_0+\sigma+\sigma^{-5/2}N^{-1/4}+{\sigma^{-5/2}}{\epsilon}),\\
Err(\hat\bfpi,\bfpi)\leq \Otilde (\sigma_0+\sigma+\sigma^{-5/2}N^{-1/4}+{\sigma^{-5/2}}{\epsilon}).
\end{eqnarray*}
\end{theorem}
The extra $\sigma_0$ has an order at $\min\{N^{-1/2}, \epsilon\}$, which is smaller than the original error rate.

\subsection{Getting the $\ell_{\infty}$ bound: a quick guide}\label{subsec: infinity error}
The analysis on $\ell_{\infty}$ spectral perturbation bounds is gaining interest within the statistics community. 
The Davis--Kahan theorem \citep{davis1970rotation} is useful in the $\ell_2$ eigenvector perturbation analysis. But for certain statistical procedures, the uniform error control is required \citep{tong2025uniform, hu2024network}. Current studies leverage leave-one-out analysis to transform this $\ell_2$ perturbation bound into an $\ell_{\infty}$ bound \citep{abbe2020entrywise}. This transformation guarantees the strong consistency, which is more robust than the consistency results on the average accuracy.

In this section, we explain how to obtain an $\ell_{\infty}$ bound from an $\ell_2$ error bound. We start with the following simple reformulation for the $\ell_{\infty}$ error. 
\begin{lemma}
\label{lem:Xinfty}
Consider a matrix $L^Z$ and let $Q^Z = I - L^Z$. Let $Q^Z_i$ and $L^Z_i$ denote the $i$-th row of $Q^Z$ and $L^Z$, respectively. 
Let $F^Z$ be an eigenvector of $L^Z$ with eigenvalue $\lambda^Z$. 
Fixing an index $i$, suppose for a vector $v \in \reals^N$ and some constants $\lambda$, $\delta_1, \delta_2, \delta_3 > 0$, the followings hold
\[
|(Q^Z_i)'(F^Z-v)|\leq N^{-1/2}{\delta_1},\; |(L^Z_i)'v-\lambda v(i)|\leq N^{-1/2}{\delta_{2}},\; |\lambda-\lambda^Z| |v(i)|\leq N^{-1/2}{\delta_3}. 
\]
Then there is
$|F^Z(i)-v(i)|\leq  (\delta_{1}+\delta_{2}+\delta_3)/\{N^{1/2}(1-\lambda^Z)\}$.
\end{lemma}
\begin{proof}
$F^Z$ is the eigenvector of $L^Z$ with eigenvalue $\lambda^Z$, so 
$\lambda^ZF^Z=L^Z F^Z,$
which leads to
\begin{equation}
\label{tmp:linfty1}
\lambda^Z(F^Z - v) = \lambda^ZF^Z-\lambda^Z v = L^Z (F^Z-v)+L^Z v - \lambda^Z v
\end{equation}
Take the $i$-th row of \eqref{tmp:linfty1} and rearrange it,
$(1-\lambda^Z)\{v(i) - F^Z(i)\}=(Q^Z_i)'(v-F^Z) +\{(L^Z_i)' v-\lambda v(i)\}+(\lambda-\lambda^Z) v(i)$.
Dividing both sides by $1 - \lambda^Z$ and plugging in bounds from the conditions, the result is proved.
\end{proof}

Apply Lemma \ref{lem:Xinfty} to the graph Laplacian $L^Z$ defined in the algorithms, where the prescribed eigenvector $v$ is taken as the discretized eigenfunctions of the manifold Laplacian, i.e. $v(i)\propto \cos(t_i/2)$ for the open-curve case or $v(i)\propto \cos(t_i)$ for the closed-loop case, and the constant $\lambda$ is the eigenvalue corresponding to $v$. 
Our goal is to bound $\|F^Z - v\|_{\infty}$. By Lemma \ref{lem:Xinfty}, the bound is based on $\|Q^Z(F^Z-v)\|_{\infty}$, $ \|L^Zv-\lambda v\|_{\infty}$, and  $|\lambda-\lambda^Z| \|v\|_{\infty}$. 

Bounding the last two terms is not complicated, by existing results and the random matrix theory. For $|\lambda-\lambda^Z|$, 
\cite{peoples2021spectral} and \cite{cheng2022eigen} have set the upper bound for $|\lambda - \lambda^X|$ and $|\lambda^X - \lambda^Z|$ is bounded by random matrix theory. 
The term $ \|L^Zv-\lambda v\|_{\infty}$ evaluates the ``error" when fitting $\lambda$ and $v$ in the eigenvector equation respective to $L^Z$, $L^Z v=\lambda v$. 
Since $L^Z$ is a noisy and discretized version of the manifold Laplacian, of which $v$ is the true eigenvector, such an error can be well-bounded.

The challenging part is to bound the projection $\|Q^Z(F^Z-v)\|_{\infty}$. We first show that using the Cauchy--Schwarz inequality does not work. 
For an index $i$, the Cauchy-Schwarz inequality gives
$|(Q^Z_i)'(F^Z-v)|\leq \|Q^Z_i\|\|F^Z-v\|$, which means the product of $\max_i\|Q^Z_i\|$ and the $\ell_2$ error bound of $F^Z-v$ will induce an upper bound on the $\ell_{\infty}$ error. 
In detail, we show that $\|Q^Z_i\|=\tilde{O}((N\sigma)^{-1/2})$ in the Appendix. 
$\|F^Z - v\| \leq \|F^Z - F^X\| + \|F^X - v\|$, where the effects of noise $\|F^Z-F^X\|=\Otilde(\epsilon/\sigma^2)$ by the Davis-Kahan Theorem. 
The effects of discretization $\|F^X-v\|=\Otilde(\sqrt{\sigma})$ for the open curve case, by \cite{peoples2021spectral}. 
Introduce these terms into the Cauchy-Schwarz estimate gives 
$\|F^Z-v\|_\infty=\Otilde(N^{-1/2})$; this bound is at the same order with $\|v\|_{\infty}$ and cannot be used for study accuracy of $t^Z_{[N]}$. For the closed-loop case, this estimate gives a usable but weaker estimate than what we offer in theorems. 

Our tight bound exploits the fact that $F^Z$ is an eigenvector of $L^Z$ and $v$ is an approximate eigenvector of $L^Z$, corresponding to a small eigenvalue $\lambda^Z = O(\sigma^2)$. 
We relate the inner product $\langle Q_i^Z, F^Z - v\rangle$ to $L^Z$ by a discretized Stein's equation. 
Starting from a fixed index $i$,  suppose there is a vector $U_i \in \reals^N$ such that $L^Z U_i = Q^Z_i$, then the inner product 
\begin{equation}
\label{eqn:Czintuition}
\langle Q^Z_i, F^Z-v\rangle=\langle L^Z U_i,  F^Z-v\rangle=\langle  U_i,  L^Z(F^Z-v)\rangle\approx \lambda^Z\langle U_i, F^Z-v\rangle. 
\end{equation}
Since $\lambda^Z = O(\sigma^2)$ and $\|F^Z - v\|$ is bounded, a sharp uniform bound can be found if the $\|U_i\|$ is well-bounded for all $i \in [N]$. 

Now the problem is to bound $U_i$ where $L^ZU_i = Q_i^Z$ for $i \in [N]$. Suppose there exists a function $\psi_i$, so that $\psi_i(t_j) = U_i(j)$. $Q^Z_i$ is close to the kernel matrix of $Z$ up to a constant factor. It gives that $Q^Z_i(j) \approx \phi_i(t_j)$, where $\phi_i(t) = \sqrt{2\pi} \exp(-\sigma^{-2}\|t_i-t\|^2/2)$. 
Introduce these terms into the equation about $U_i$, then the goal is to find $\psi_i(t)$ so that 
\begin{equation}
\label{eqn:Poisson}
L^Z U_i \approx L^Z \psi_i(t_{[N]})\approx \psi_i(t)-\E_{\xi\sim \mathcal{N}(0,1)}[\psi_i(t+\sigma \xi)] = \phi_i(t). 
\end{equation}
It is the Stein's equation about $\psi_i(t)$ and $\phi_i(t)$. 
The solution $\psi_i(t)$ can be explicitly written down as an infinite summation. By bounding the explicit terms in the summation, we can bound the vector $\psi_i(t)$ and then $\|U_i\|$. 
The rigorous analysis can be found in the Appendix.

\section{Simulation}
We compare our spectral seriation methods with three manifold learning approaches for both the open-curve case and the closed-loop case. We explore the permutation recovery performance as the noise level, the feature dimension, and the sample size vary. 

Consider a simple curve embedded in $\reals^d$. For any $t \in [0, 2\pi]$, define a vector $v(t) \in \reals^{2}$. 
The underlying temporal system is the projection of this unit circle $X(t) = Uv(t)$, where $U \in \reals^{d \times 2}$ is a random embedding matrix satisfying $U'U = I$. Generate $N$ time points where $t_i \sim \text{Unif}(0,2\pi)$, and hence the data points
    $Z_i=Uv(t_i)+e_i$ where $e_i \sim \mathcal{N}(0, \tau^2 I)$, $i \in [N]$.
The noise level $\epsilon$ is affected by both the variance $\tau^2$ and the feature dimension $d$.

We apply our methods and three comparison methods to the data matrix $Z$. Due to the large dimension $d$, we first employ the standard principal component analysis method to reduce the dimensionality to at most 50; see Algorithm A.2 in Appendix. Then we apply the spectral seriation methods to the denoised data.  
The other methods are popular manifold learning methods, including: t-SNE \citep{maaten2008visualizing}, UMAP \citep{McInnes2018}, and Isomap \citep{tenenbaum2000global}. 
For these methods, we first embed $Z_i$ into a 1-dimensional space, and then recover the permutation by ordering these embeddings. 
For any recovered permutation $\hat\bfpi$, we define the permuted matrix as $X_{\hat\bfpi} \in \reals^{d \times N}$, whose columns are ordered according to the estimated temporal order.  
The relative error is then measured by 
\begin{equation}\label{eqn:dataerr}
    Err(\bfpi,\hat\bfpi)={\|X_{\bfpi}-X_{\hat\bfpi}\|_F}/{\| X_{\bfpi}\|_F}.
\end{equation}

\begin{figure}
    \centering
    \begin{tikzpicture}[scale=0.7]
    \node[inner sep=0] at (-6,0) {\includegraphics[width=0.3\textwidth]{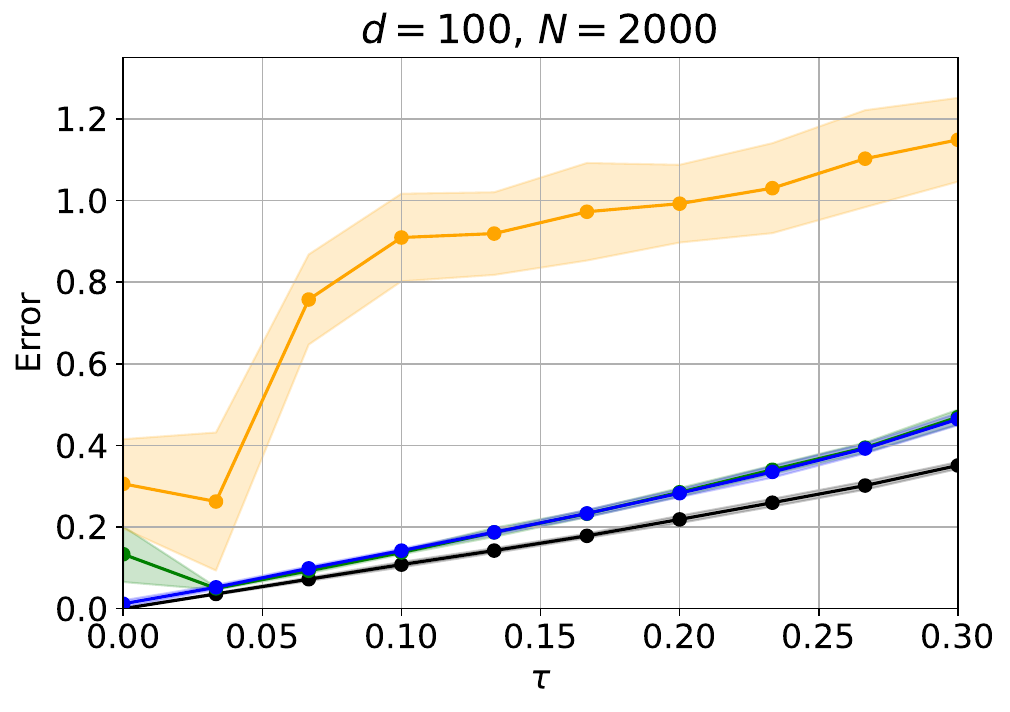}}; 
  \node[inner sep=0] at (0,0) 
    {\includegraphics[width=0.3\textwidth]{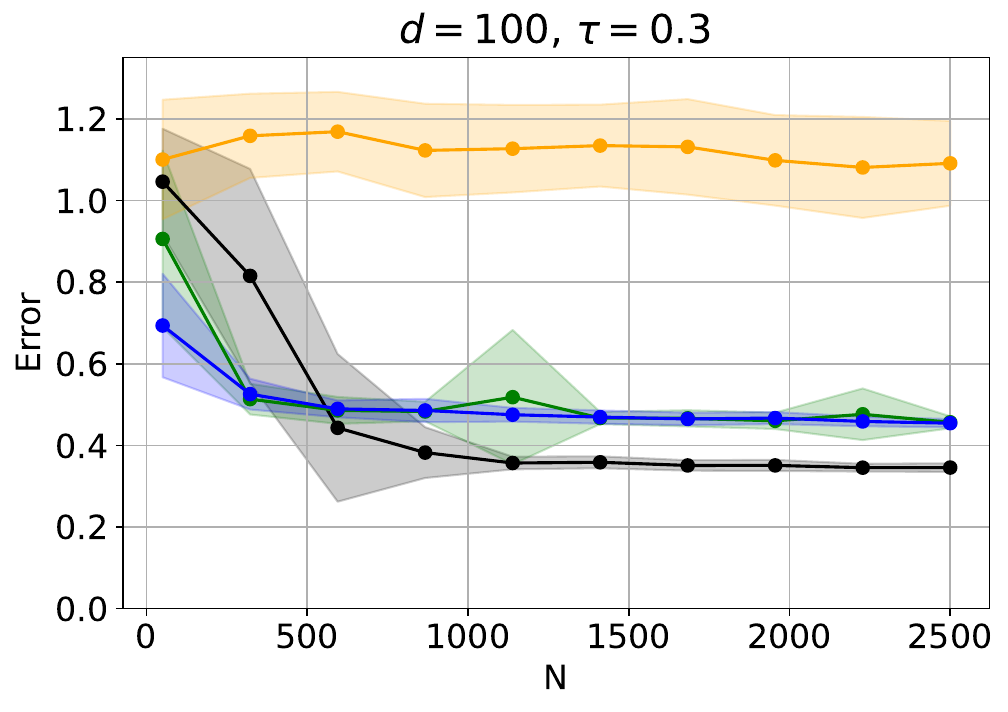}};
\node[inner sep=0] at (6,0) 
    {\includegraphics[width=0.3\textwidth]{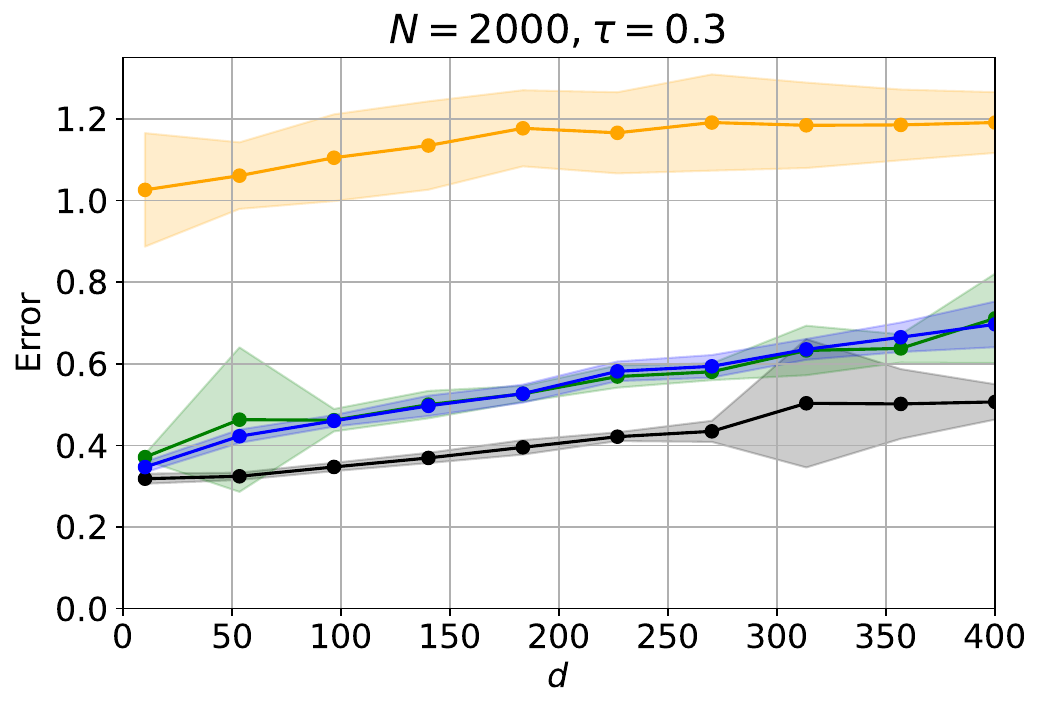}};
        \node[inner sep=0] at (-6,-5) {\includegraphics[width=0.3\textwidth]{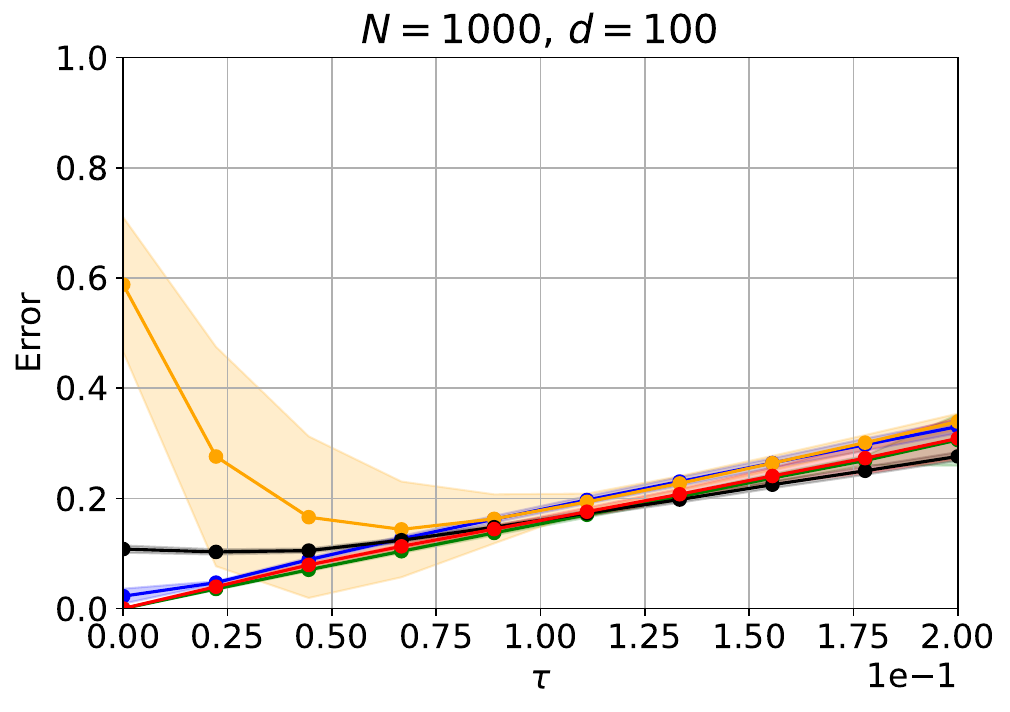}}; 
  \node[inner sep=0] at (0,-5) 
    {\includegraphics[width=0.3\textwidth]{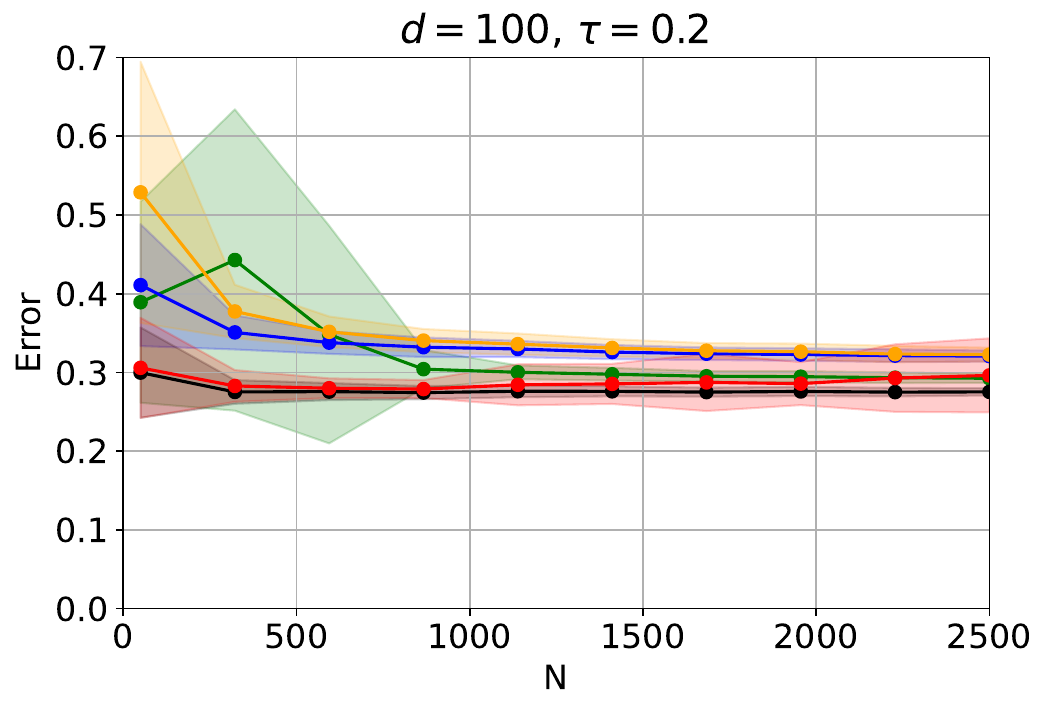}};
     \node[inner sep=0] at (6,-5) 
    {\includegraphics[width=0.3\textwidth]{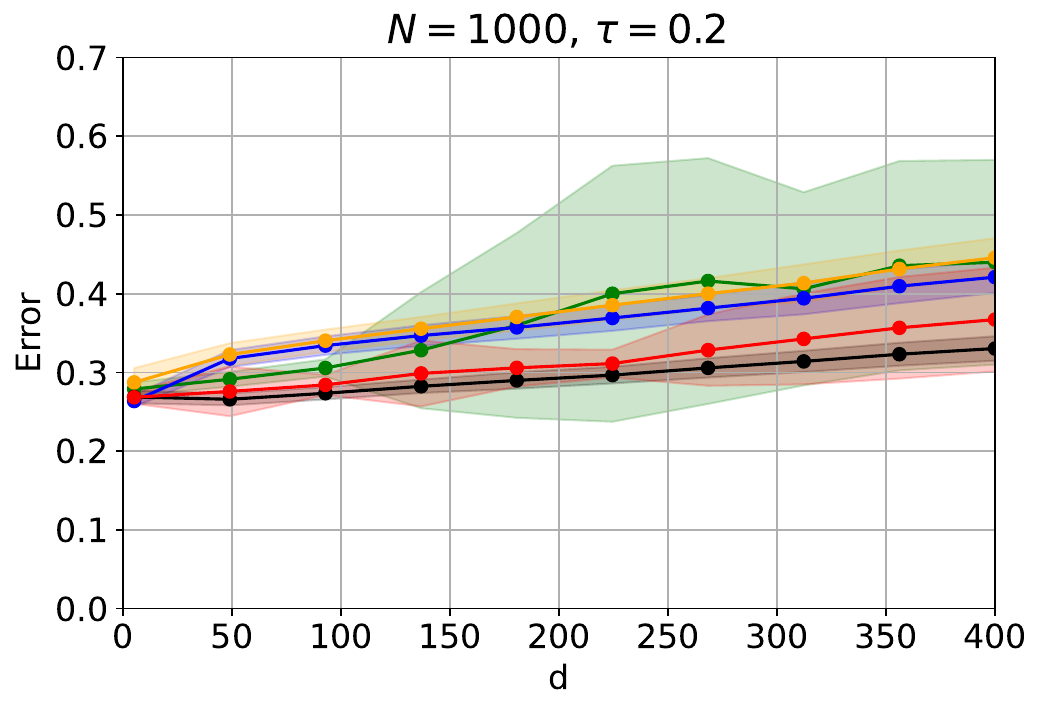}};
  \end{tikzpicture}
   \caption{Relative errors for the closed-loop case (top row) and the open-curve case (bottom row). The methods contain our spectral seriation method (black), our hybrid method (red, only for the open-curve case), t-SNE (green), Isomap (blue), and UMAP (yellow). The shaded area denotes $\pm 1$ standard deviation. }
    \label{fig:simulation}
\end{figure}

{\bf Experiment 1 (closed-loop case)}. For the closed-loop case, we take $v(t) = (\cos(t), \sin(t))'$, a unit circle in $\reals^2$. 
Draw time points as $t_i \sim \text{Unif}(0, 2\pi)$ so that the manifold is the whole circle. We consider three scenarios: (a) the noise parameter $\tau \in [0, 0.3]$, with a step size $\approx 0.033$; (b) the dimension $N \in [50, 2500]$, with a step size $\approx 272$; and (c) the sample size $d \in [10, 400]$ with a step size $\approx 43$. We take $N = 2000$, $d = 100$, $\tau = 0.3$ as the default values. Figure \ref{fig:simulation} summarizes the mean and standard deviation of the errors across 250 replicates.

Overall, all methods enjoy a lower relative error with a smaller noise variance $\tau$, a smaller feature dimension $d$, or a larger sample size $N$. However, even for large $N$, there is still an unavoidable error for all methods. This is the inherent error from noise. For most cases, our spectral seriation method outperforms all other manifold learning methods in terms of accuracy and stability, particularly when $N \geq 1000$.

\indent {\bf Experiment 2 (open-curve case)}. 
For the open-curve case, we consider the curve 
$v(t)=(\cos(t),\frac{1}{2}\sin(t))'$. Sample $t_i$ from $\text{Unif}[0,\pi]$.
Therefore, the underlying temporal system is an open curve with two endpoints. We consider three scenarios: (a) the noise parameter $\tau \in [0, 0.2]$, with a spacing $\approx 0.022$; (b) the sample size $N \in [50, 2500]$, with a spacing $\approx 272$; (c) the feature dimension $d$ from 10 to 400 with a step size $\approx 43$. We take the sample size $N = 1000$, $d = 100$ and $\tau = 0.2$ as the default setting. 
We evaluate all methods on the full dataset. Figure \ref{fig:simulation} summarizes the average and standard deviation of the errors across 250 replicates. 

As shown in Figure \ref{fig:simulation}, our hybrid method outperforms all the other methods in error rate and stability, across a wide range of noise levels, feature dimensions, and sample sizes. Our spectral seriation method is comparable with the hybrid method in most settings, indicating that it outperforms other manifold learning methods even when applied to the full interval. The only exception occurs in the very-low-noise regime, where the graphical distance methods are favored. In this case, our hybrid method achieves comparable performance to the competing approaches. Overall, our spectral seriation and hybrid methods consistently outperform other methods.

\section{Rotation of a real molecule}\label{sec:molecule}
We consider the density maps of a biological molecule undergoing a rotation. 
Since the molecule is an object in 3-dimensional space, the density map at a given rotation angle $t_i$ is a 3-dimensional tensor of size $16 \times 35 \times 16$ after downsampling. Vectorizing the voxel data yields a data point $Z_i \in \reals^d$, where $d = 8960$. The original dataset was published in~\cite{Zelesko2019EarthmoverBasedML}.

We consider the 3-dimensional rotation of the molecule around the $y$-axis. We first generate the rotation angles, which serves as temporal labels $t_i$, uniformly from $[0, 2\pi]$. This corresponds to a closed-loop scenario. At each sampled angle $t_i$, $X_i \in \reals^{8960}$ denotes the corresponding density map. 
Augmented by a Gaussian noise $e_i \sim \mathcal{N}(0, \tau^2 I)$, the observation is $Z_i = X_i + e_i$. 
While $d = 8960$ and the molecule is in a 3-dimensional space, the temporal system $X(t)$ forms a 1-dimensional manifold, since it is an embedding of $[0, 2\pi]$ into this high-dimensional space. Our spectral seriation method can be applied to this high-dimensional molecule data.

We consider the combinations of sample sizes $N \in \{100,1000,10000\}$ and  noise levels $\tau\in\{0.001, 0.01,0.02\}$. 
Due to the high dimensionality, we first apply a dimension reduction algorithm to $Z$ \citep{supp}. It reduces the ambient dimension $d = 8960$ to a comparatively low dimension $r \in [200,300]$, which is automatically chosen by the algorithm. Then we apply the spectral seriation method in Algorithm~\ref{alg:closedcurveuniform}. 

\begin{figure}
    \centering
    \begin{tikzpicture}[scale=0.7]
  \node[inner sep=0] at (-10,0) {\includegraphics[width=.21\textwidth]{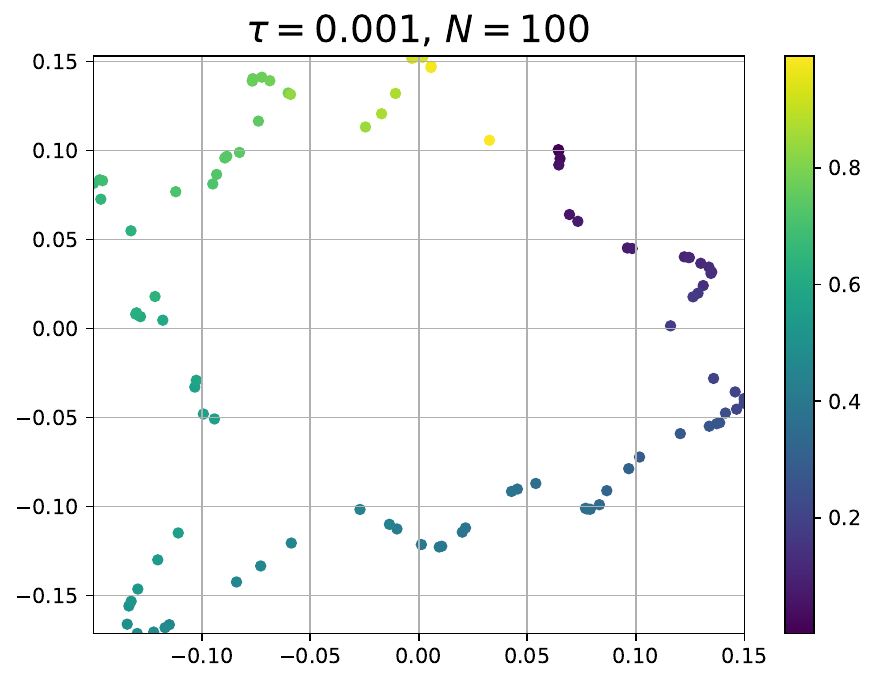}}; 
  \node[inner sep=0] at (-5,0) {\includegraphics[width=.21\textwidth]{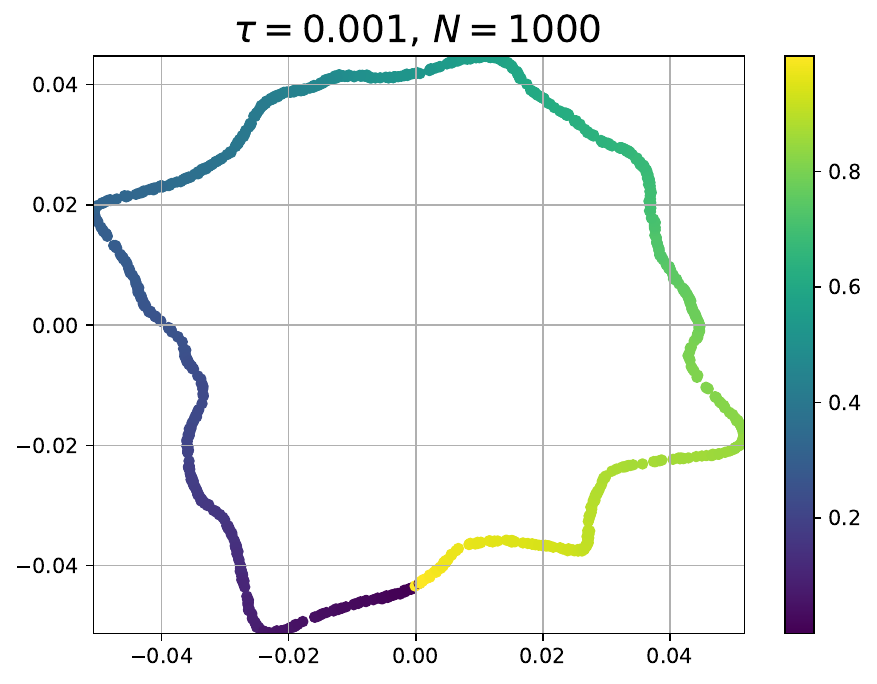}}; 
  \node[inner sep=0] at (0,0) {\includegraphics[width=.21\textwidth]{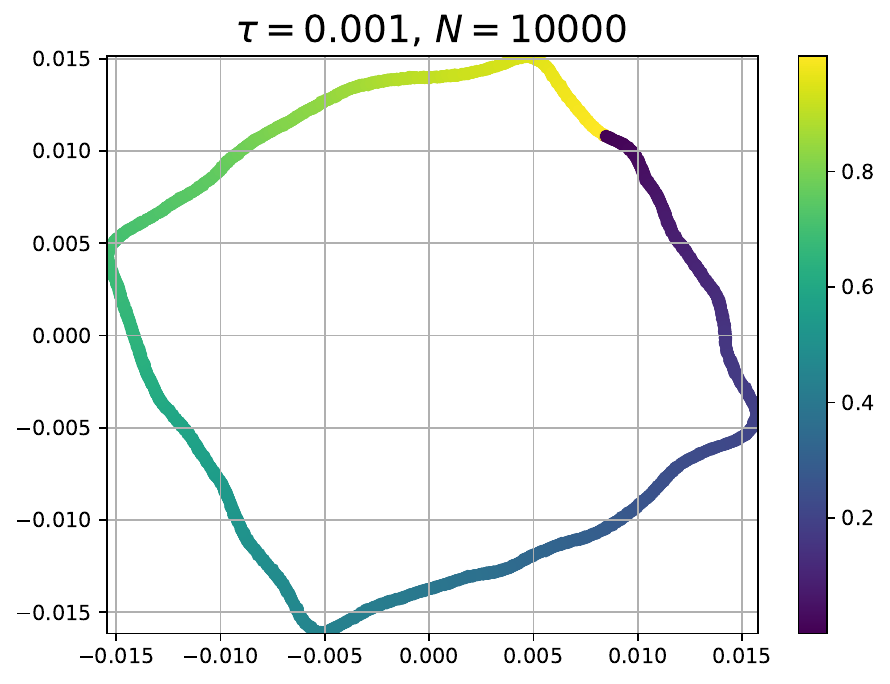}}; 
  \node[inner sep=0] at (-10,-4) {\includegraphics[width=.21\textwidth]{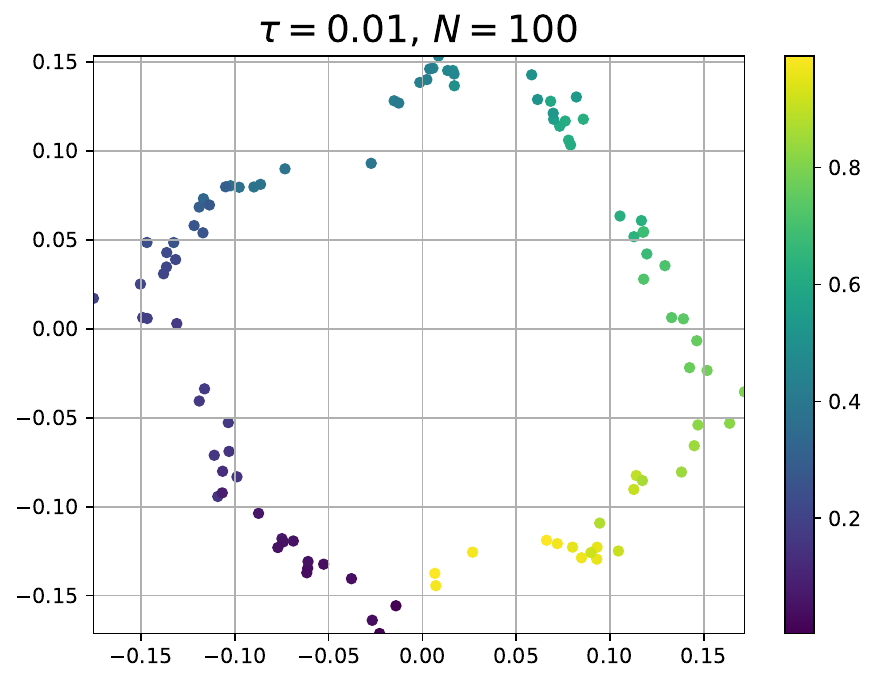}}; 
  \node[inner sep=0] at (-5,-4) {\includegraphics[width=.21\textwidth]{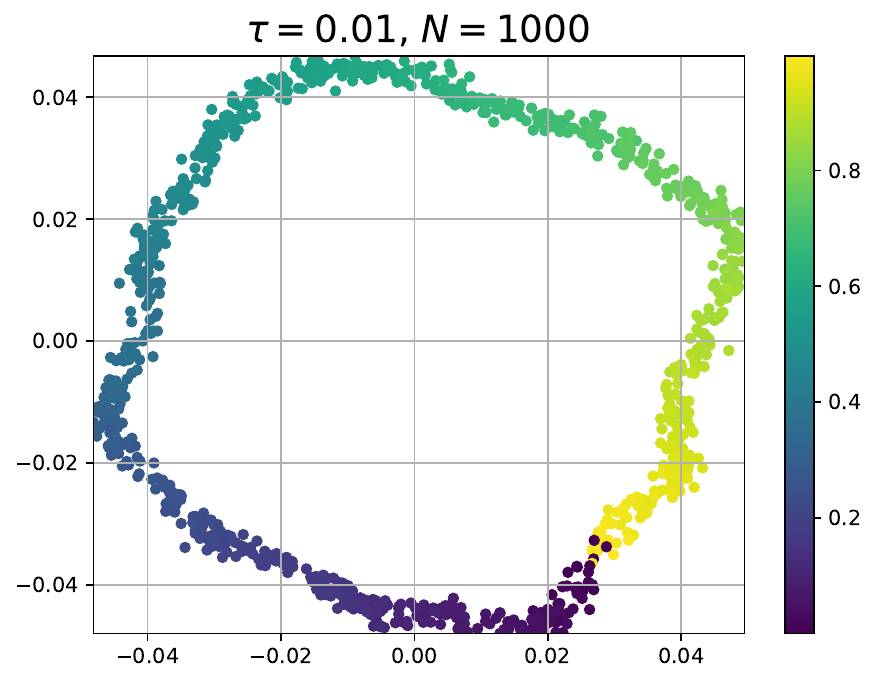}}; 
  \node[inner sep=0] at (0,-4) {\includegraphics[width=.21\textwidth]{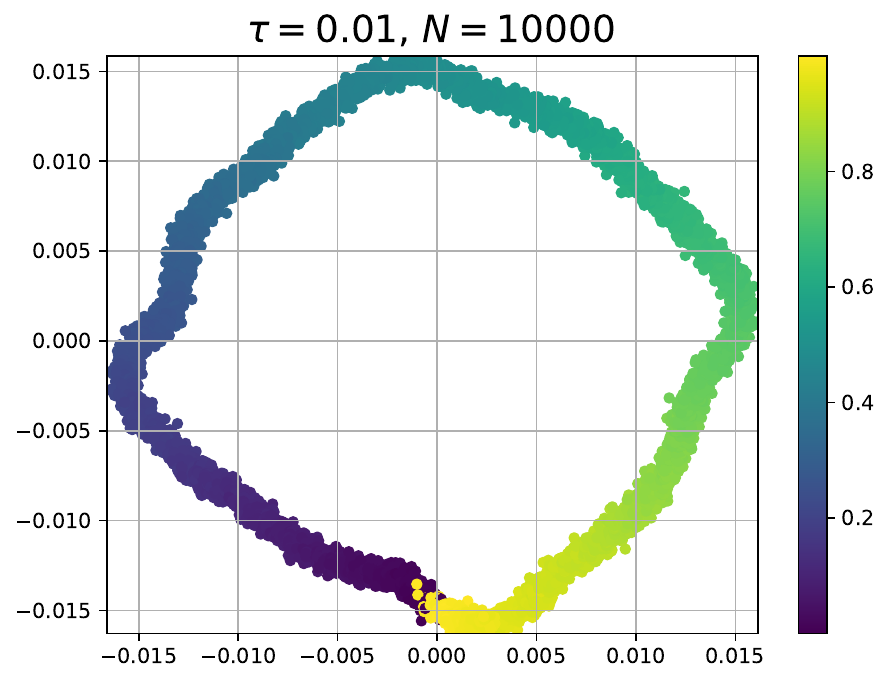}}; 
  \node[inner sep=0] at (-10,-8) {\includegraphics[width=.21\textwidth]{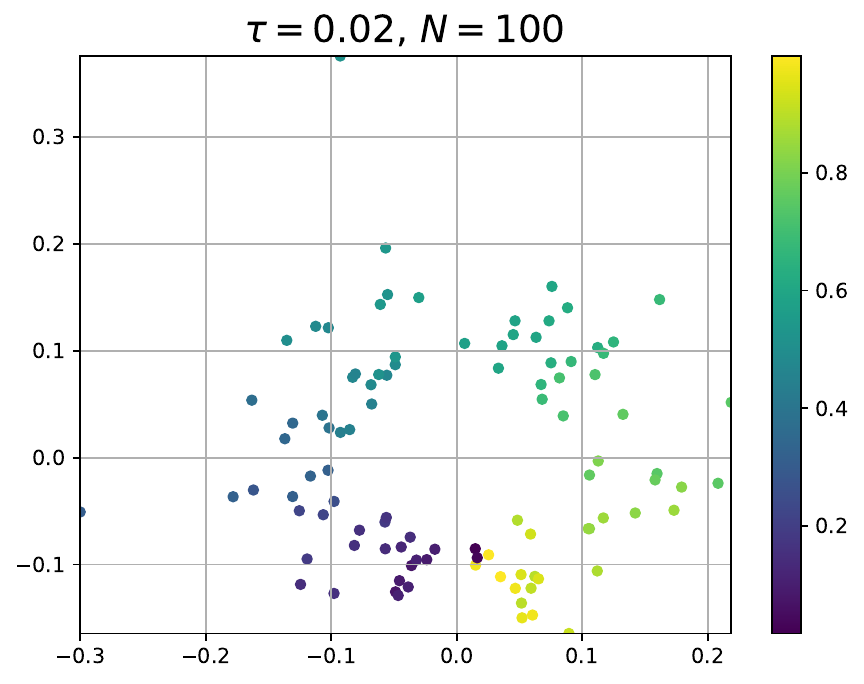}}; 
  \node[inner sep=0] at (-5,-8) {\includegraphics[width=.21\textwidth]{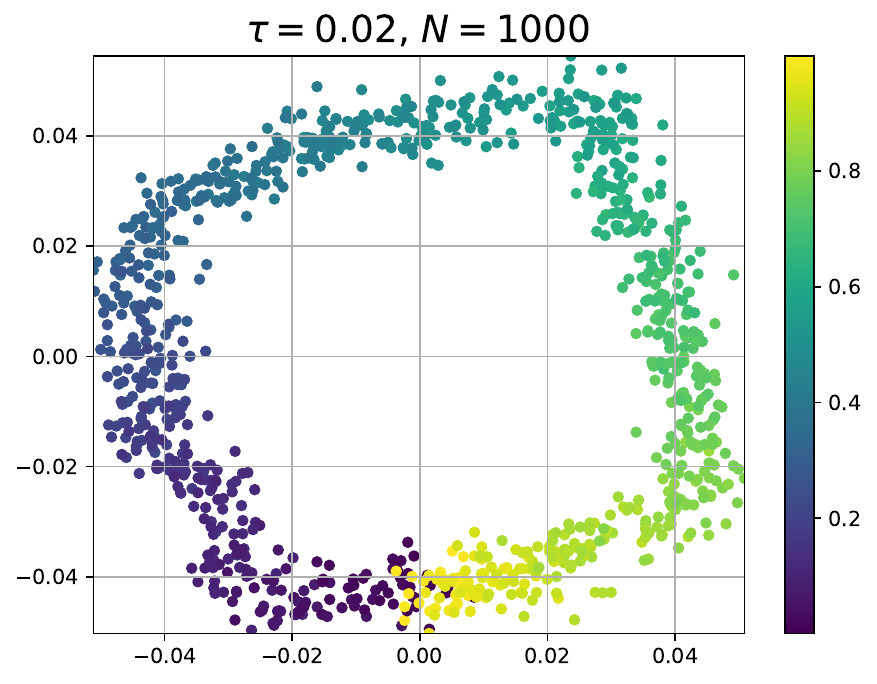}}; 
  \node[inner sep=0] at (0,-8) {\includegraphics[width=.21\textwidth]{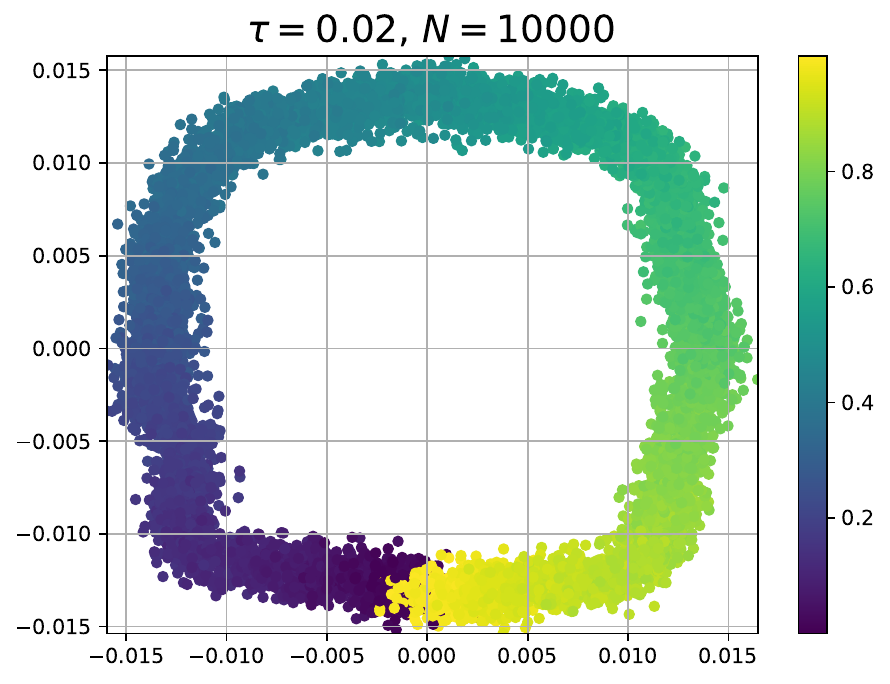}}; 
 \end{tikzpicture}
    \caption{The second and third eigenvectors of $L^Z$ from the molecular data.}
    \label{fig:fiedler_close}
\end{figure}

We first study the eigenvectors of the graph Laplacian versus the true temporal labels $t_i$.
For each combination of $(N, \tau)$,  the scatter plots of $(F^Z_+(i), F^Z_-(i))$ are presented in Figure~\ref{fig:fiedler_close}, with the color indicating $t_i$. 
These plots show a circle that captures the rotation angle $t_i$, illustrating the intuition of using $F^Z_+$ and $F^Z_-$ to recover the angles. As $N$ increases, the recovered shape becomes smoother, indicating a smaller estimation error. As the noise level $\tau$ grows, the recovered circle turns into a band, illustrating greater estimation error at each data point.

\begin{figure}
    \centering
    \begin{tikzpicture}[scale=0.7]
     \node[inner sep=0] at (0,0) {
\includegraphics[width=0.4\textwidth]{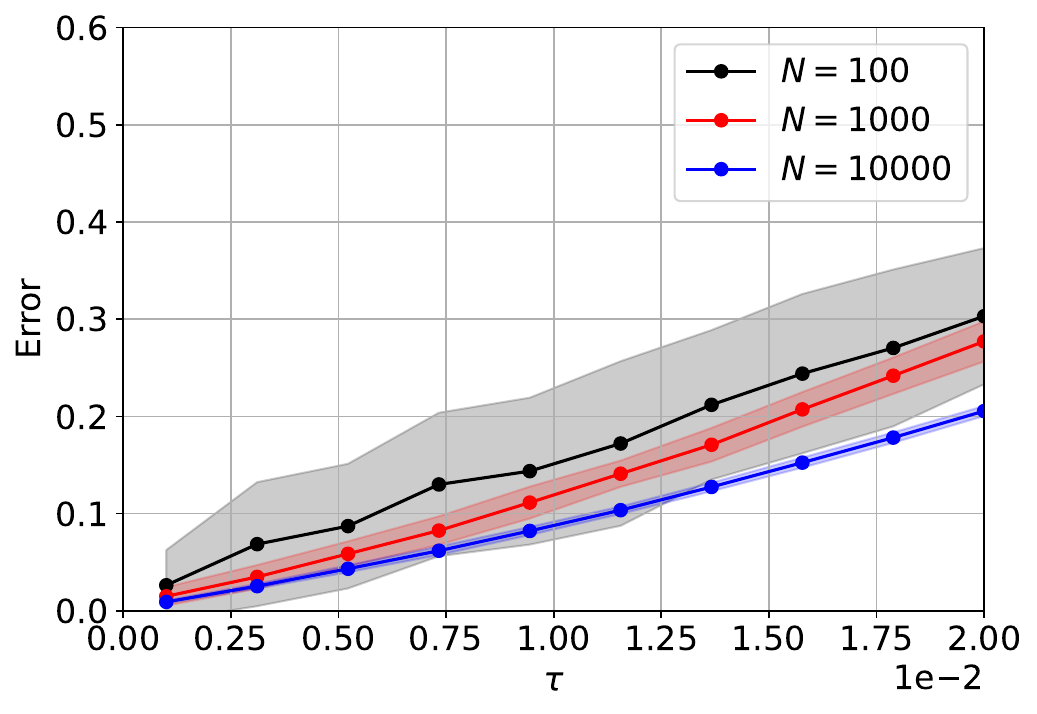}};
    \node[inner sep=0] at (10,0) {
\includegraphics[width=0.4\textwidth]{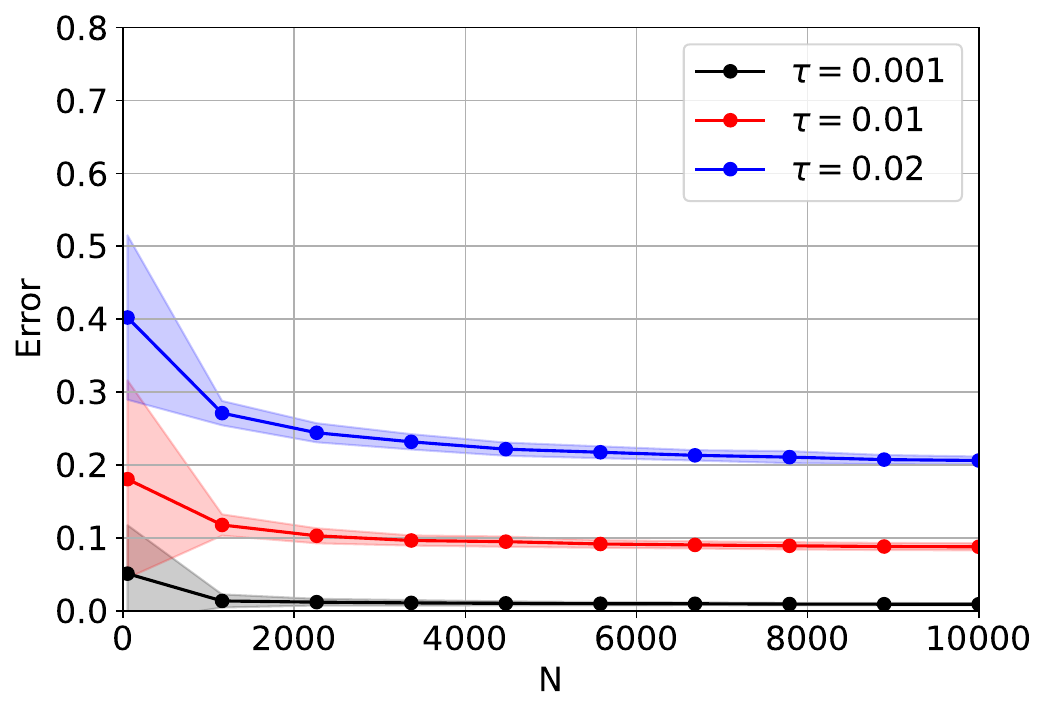}};
     \end{tikzpicture}
\caption{The permutation error on molecule data for varying noise level (left panel) and varying sample size (right panel).}
    \label{fig:estimation_errors}
   
\end{figure}

We then summarize the estimation error in Figure \ref{fig:estimation_errors} over 100 replicates. Given $N$, the estimation error decreases as $\tau$ decreases, but does not vanish due to finite sampling. For fixed $\tau$, we observe the same pattern. This shows  that the effects of sample size $N$ and noise level $\tau$ cannot replace each other.

\section*{Acknowledgement}
Yuehaw Khoo's research is partially supported by DMS-2111563, DMS-2339439, DE-SC0022232.
Xin T. Tong's research is supported by Singapore MOE grant Tier-1-A-8000459-00-00. Wanjie Wang's research is supported by Singapore MOE grant Tier-1-A-8001451-00-00. 
Yuguan Wang's research is supported by the Department of Statistics, University of Chicago. The authors thank John Harlim and Xiuyuan Cheng for their valuable comments and discussions. The authors also thank the anonymous reviewers and the associate editor for their insightful comments.

\appendix

\section{Algorithms in bandwidth selection and dimension reduction}
\label{sec:appA}
\subsection{Data-adaptive method for tuning the bandwidth}
We define the unnormalized Gaussian kernel  as 
\begin{align}
    \tilde k(x,y;\sigma)=\exp(-(2\sigma^2)^{-1}\|x-y\|^2).
\end{align}
To choose an appropriate bandwidth $\sigma>0$,
we use the method proposed in~\cite{4623181}, where it was observed that  when the bandwidth $\sigma$ is well tuned,  the following summation can be approximated by the mean-value integral 
\begin{align}
    S(\sigma)&=\frac{1}{N^2}\sum_{i,j=1}^N \tilde k(X_i,X_j; \sigma)\approx \frac{1}{\text{vol}(\mathcal{M})^2}\int_{\mathcal{M}} \int_{\mathcal{M}}   \tilde k(x_1,x_2;\sigma) \,\mathrm{d}x_1\mathrm{d}x_2.
\end{align}
For small $\sigma$, the narrow Gaussian integral can be approximated  as 
\begin{align}
    \int_{\mathcal{M}}   \tilde k(x_1,x_2;\sigma) \,\mathrm{d}x_1\approx 
     \int_{\mathbb{R}^d}   \tilde k(x_1,x_2;\sigma) \,\mathrm{d}x_1=(2\pi\sigma)^{d/2}
\end{align}
because $\mathcal{M}$ is locally well-approximated by its tangent space $\mathbb{R}^d$, and here $d=1$. From that we can obtain $\log S(\sigma)\approx (1/2) \log \sigma + (1/2)\log (2\pi)$.   This suggests that we should choose $\sigma$ in a region where  $\log S(\sigma)$ is approximately a linear function of $\log \sigma$. 

Algorithmically, we discretize $\log \sigma \in [-8,1]$ into $300$ equispaced points.  We then maintain a sliding window of size $30$ over these  points. Within each sliding window, we regress  the computed values of $\log S(\sigma)$  against $\log \sigma$ and compute the  $R^2$ value.  We choose $\sigma$ as the left endpoint of the sliding window having the largest $R^2$. 

When choosing the threshold $\sigma_0$, if the sample size $N\le 500$,  we let all points to be connected and keep all weights.  Otherwise, for each point, we find the distance between itself and its $500$-th nearest neighbor, and we average these distances over all points to be our $\sigma_0.$

\subsection{PCA matrix denoising for high-dimensional data}\label{subsec:app}

In challenging settings, Assumption \ref{aspt:noise} that $\|Z_i - X(t_i)\| \leq \epsilon$ may not hold. 
For example, for diverging $d$, a high-dimensional noise $e_i \sim N(0, \epsilon^2 I_d)$ has a uniform bound $\|e_i\| \approx \sqrt{d \log N} \epsilon $ and hence ruin the convergence of eigenvectors. 
In this section, we introduce a denoising algorithm for $Z$. A full discussion on the uniform error control of this denoising algorithm can be found in \cite{tong2025uniform}. 

A popular denoising method is to project the data into the eigen-space. 
Suppose $X$ has a low rank of $rank(X) = r \ll d$. Let $U \in \reals^{d \times r}$ be the matrix formed by the first $r$ left singular vectors of $X$ and $\mathrm{col}(U)$ be the eigen-space spanned by columns of $U$. 
The projection of $X$ into $\mathrm{col}(U)$ remains $X$, i.e. $UU^T X = X$, while $UU^T \mathcal{E}$ largely reduces the magnitude of noise. 
Without the access of $X$ and hence $U$, we use the singular value decomposition of $Z$ to get $\tilde{U}$ and denoising data $\tilde{Z} = \tilde{U}\tilde{U}^T Z$, as below. 
\begin{algorithm}
\caption{Denoise and compress noisy high-dimensional data.}
\label{alg:svd}
\begin{algorithmic}[1]
    \Require Noisy data $Z \in \mathbb{R}^{d \times N}$, rank $r \ll d$
    \State Let $\tilde{U} = (u_1, \ldots, u_r) \in \mathbb{R}^{d \times r}$ be the matrix formed by the left singular vectors of $Z$ corresponding to the $r$ largest singular values.
    \State Denoise the data by $\tilde{Z}_i = \tilde{U}\tilde{U}^T Z_i$ for each $i$, and hence $\tilde{Z} = \tilde{U}\tilde{U}^T Z \in \mathbb{R}^{d \times N}.$
    \Ensure Denoised data points $\tilde{Z}_i \in \mathbb{R}^{d}$, $i \in [N]$
\end{algorithmic}
\end{algorithm}
When $rank(X) = r \ll d$, the denoised data has a uniform error bound that 
$\max_i\|\tilde{Z}_i - X(t_i)\|=\tilde{O}(\epsilon \sqrt{r})$ when $e_i \sim \mathcal{N}(0,\epsilon^2 I_d)$ by \cite{tong2025uniform}. It is a significant improvement compared to the uniform error bound for the original data 
$\max_i\|Z_i - X(t_i)\|=\tilde{O}(\epsilon\sqrt{d})$.

In real data, the rank of $X$ is unknown, but an upper bound $r_0 \geq r$ is usually available. 
We use a variant of the randomized range finding algorithm in  \cite{doi:10.1137/090771806} to estimate $r$ and find $\tilde U$, based on $Z$ and an over-sampling rank parameter $r_0$. 
The idea is to take the product of $Z$ and a noise matrix $G \in \reals^{N \times r_0}$, so that the eigen-space formed by left singular vectors remains the same, while the singular values are affected by the additional randomization $G$. The estimation of $r$ corresponds to the rank of the first singular value of $ZG$ that is significantly smaller compared to the largest singular value of $ZG$.
Details in Algorithm \ref{alg:rsvd}. 



\begin{algorithm}
\caption{Denoise and compress noisy high-dimensional data with over-sampling rank $r_0$.}
\label{alg:rsvd}
\begin{algorithmic}[1]
    \Require Noisy data $Z \in \mathbb{R}^{d \times N}$, over-sampling rank $r_0 \ll d$, tuning parameter $\eta$
    \State Generate a Gaussian random matrix $G \in \mathbb{R}^{N \times r_0}$ and let $Y = ZG \in \mathbb{R}^{d \times r_0}$.
    \State Compute the singular value decomposition $Y = U \Sigma V^T$.
    \State Let $\lambda_k(Y)$ denote the $k$-th largest singular value of $Y$. Estimate the rank
    \[
        \hat{r} = \min\!\left\{ 1 \leq i \leq r_0 \;\middle|\; \frac{\lambda_i(Y)}{\lambda_1(Y)} < \eta \right\}.
    \]
    \State Let $\tilde{U}$ be the $d \times \hat{r}$ matrix formed by the first $\hat{r}$ left singular vectors of $U$.
    \State Denoise the data by $\tilde{Z}_i = \tilde{U}\tilde{U}^T Z_i$ for each $i$, and hence
    \[
        \tilde{Z} = \tilde{U}\tilde{U}^T Z \in \mathbb{R}^{d \times N}.
    \]
    \Ensure Estimated rank $\hat{r}$ and denoised data matrix $\tilde{Z} \in \mathbb{R}^{d \times N}$
\end{algorithmic}
\end{algorithm}

\begin{remark}
The tuning parameter $\eta$ is a very small constant to guarantee the leftover singular vectors are mainly noise.  In numerical analysis of the molecule data, we take the over-sampling rank $r_0 = 400$ and $\eta = 10^{-3}$. The estimated rank $r$ is between 200 and 300 in repetitions. 
\end{remark}

Combining Algorithm \ref{alg:rsvd} and Algorithms \ref{alg:opencurveuniform}--\ref{alg:closedcurveuniform} gives us the outline of the temporal label recovery algorithm for high-dimensional data: first get the denoised data matrix $\tilde Z$, then apply the spectral method on $\tilde Z$ in place of $Z$. We apply this approach in Section \ref{sec:molecule} and get satisfactory results.

\section{Detailed analysis}

In the following derivations, we use 
\[
r_\sigma=6\sigma\sqrt{\log (N/\sigma)},\quad 
\delta_{c,1}=\sigma^2+\sqrt{\frac{\log N}{N\sigma^3}}
\]
To simplify the estimates, we assume that $\epsilon\leq \sigma^2\leq 1, N\geq \sigma^{-6}$.  We will aslo assume Assumptions \ref{aspt:noise} and \ref{aspt:manifold} by default, so we often omit mentioning it. Moreover, for the sake of conciseness, we tend to unify the discussion of both the open loop case and the closed case together. Note that $int_\delta$ is still well defined for closed loop case, which is simply $\calM$ (and $int_\delta^c=\emptyset$). 


Given a function $f$, we also use $\tilde{f}\in \reals^N$ to denote the vectorization of the function on $t_{[N]}$, i.e. $\tilde{f}(i)=f(t_i), i\in [N]$.

\subsection{Preparation}
In this subsection, we provide some basic estimates, mostly about the functions which will appear in our discussion. Many of these results have similar versions appear in existing works like \citep{singer2006graph,peoples2021spectral,cheng2022eigen}. We provide them here nevertheless for the same of self-completeness. 
\begin{lemma}
\label{lem:expdiff}
Let $f(t)=\exp(-t^2/2\sigma^2)$, under Assumptions \ref{aspt:noise} and \ref{aspt:manifold} with $\epsilon\leq \frac18 r_\sigma, r_\sigma<L^{-1/3}_{\calM}$, for sufficiently small $\sigma$ the following hold with high probability
\begin{enumerate}
  \item When $d(t_i,t_j)>r_\sigma$, we have $\|X_{t_i}-X_{t_j}\|<3/4 r_\sigma$,  $\|Z_i-Z_j\|\leq \frac12 r_\sigma$, and
    \[
    f(d(t_i,t_j))\leq \sigma^8, \quad f(\|X_{t_i}-X_{t_j}\|)\leq \sigma^8,\quad f(\|Z_{i}-Z_{j}\|)\leq \sigma^8. 
    \]
\item When $d(t_i,t_j)<r_\sigma$, 
    \[
    |f(d(t_i,t_j))-f(\|X_{t_i}-X_{t_j}\|)|  =\Otilde(\sigma^2)
    \] 
    \[
    |f(\|X_{t_i}-X_{t_j}\|)-f(\|Z_{i}-Z_{j}\|)+\frac{\|X_{t_i}-X_{t_j}\|^2}{2\sigma^2}-\frac{\|Z_i-Z_j\|^2}{2\sigma^2}|  =\Otilde(\frac{\epsilon}{\sigma})
    \]
\end{enumerate}
\end{lemma}
\begin{proof}
To simplify notation, we assume $t_i\in [\pi/2, 3/2\pi]$ in the closed loop case. This does not sacrifice generality since we can always rotate the angle coordinate so the assumption holds. Then $|t_i-t_j|>r_\sigma$ is equivalent to $d(t_i,t_j)>r_{\sigma}$, and we will use the former since it is easier to interpret. No treatment is needed for open loop. 

Assumption \ref{aspt:manifold} indicates that 
\[
0\leq |t_i-t_j|-\|X_{t_i}-X_{t_j}\|\leq L_{\calM} \|X_{t_i}-X_{t_j}\|^3. 
\]
When $|t_i-t_j|>r_\sigma$, we can conclude that
\[
\|X_{t_i}-X_{t_j}\|\geq \frac{3}{4}r_\sigma,\quad \|Z_{i}-Z_{j}\|\geq \frac{1}{2}r_\sigma.
\]
Plug these estimate into the formulation of $f(t)$, we have claim 1). 

If $|t_i-t_j|<r_\sigma$, then  
\[
0\leq |t_i-t_j|-\|X_{t_i}-X_{t_j}\|\leq L_{\calM} \|X_{t_i}-X_{t_j}\|^3\lesssim \sigma^2, \quad
|\|X_{t_i}-X_{t_j}\|-\|Z_{i}-Z_{j}\||\leq 2\epsilon.
\]
Let $f(t)=\exp(-t^2/2\sigma^2)$, then $f'(t)=-\frac{t}{\sigma^2}f(t)$. So between $[a,b]\subset \reals^+$, there is a $c\in [a,b]$ so that 
\[
|f(a)-f(b)|\leq \frac{|a-b||c|}{\sigma^2} f(c)\leq \frac{|a^2-b^2|}{\sigma^2} \max\{f(a),f(b)\}\leq \frac{|a-b||a+b|}{\sigma^2}. 
\]
Let $a=\|X_{t_i}-X_{t_j}\|=O(r_\sigma)$ and $b=|t_i-t_j|$ or $b=\|Z_{i}-Z_{j}\|$, we have our claim 2). 

\end{proof}

\begin{lemma}
\label{lem:cos}
The following holds, if $t\in int_{r_\sigma}$, for any fixed integer $n$
\[
|\int^{2\pi}_0 (\cos(nt/2)-\cos(ns/2)) \frac{1}{\sqrt{2\pi}\sigma}\exp(-\frac{|t-s|^2}{2\sigma^2})ds-\frac{n^2}4\sigma^2 \cos(2t)|
\leq O(\sigma^4). 
\]
If $t\in int^c_{r_\sigma}$,
\[
|\int^{2\pi}_0 (\cos(nt/2)-\cos(ns/2)) \frac{1}{\sqrt{2\pi}\sigma}\exp(-\frac{|t-s|^2}{2\sigma^2})ds|\leq \Otilde(\sigma^2). 
\]
\end{lemma}
\begin{proof}
By Lemma 8 of \citep{coifman2006diffusion}, we have
\[
|\int^{\infty}_{-\infty} (\cos(nt/2)-\cos(ns/2)) \frac{1}{\sqrt{2\pi}\sigma}\exp(-\frac{|t-s|^2}{2\sigma^2})ds-\frac{n^2}4\sigma^2 \cos(2t)|
\leq O(\sigma^4). 
\]
Then note that by Lemma \ref{lem:expdiff},
\begin{equation}
\label{tmp:smallint}
\int_{|s-t|>r_\sigma}(\psi(s)-\psi(t))\frac{1}{\sqrt{2\pi}\sigma}\exp(-\frac{\|s-t\|^2}{2\sigma^2})ds
\lesssim  4\sigma^4\int\frac{1}{2\sqrt{\pi}\sigma}\exp(-\frac{\|s-t\|^2}{4\sigma^2})ds\lesssim\sigma^4,
\end{equation}
holds for any bounded $|\psi|\leq 1$, we have our first claim. 

As for the second claim, note that when $|t-s|<r_\sigma, t\in int^c_{r_\sigma}$, for some point $u$ between $t$ and $s$, we have
\[
|\cos(nt/2)-\cos(ns/2)|=\frac{n(t-s)}{2}|\sin (un/2)|\leq \frac{nr_\sigma}{2}\cdot \frac{|3r_\sigma n|}{2}=O(r_\sigma^2). 
\]
Also seeing that the integral for $|s-t|>r_\sigma$ is insignificant, by \eqref{tmp:smallint},
\[
|\int^{2\pi}_0 (\cos(nt/2)-\cos(ns/2)) \frac{1}{\sqrt{2\pi}\sigma}\exp(-\frac{|t-s|^2}{2\sigma^2})ds|\lesssim
r_\sigma^2|\int^{+\infty}_{-\infty}  \frac{1}{\sqrt{2\pi}\sigma}\exp(-\frac{|t-s|^2}{2\sigma^2})ds\leq \Otilde(r_\sigma^2). 
\]
\end{proof}

\begin{lemma}
\label{lem:anglelength}
Suppose $|a-\cos\theta|\leq \delta ,|b-\sin \theta|\leq \delta$, and $\zeta$ satisfies $\cos \zeta =\frac{a}{\sqrt{a^2+b^2}}, \sin \zeta =\frac{b}{\sqrt{a^2+b^2}}$, then $\min\{|\theta-\zeta|,|2\pi+\theta-\zeta|, |-2\pi+\theta-\zeta|\}<\delta \pi$. 
\end{lemma}
\begin{proof}
Consider the projection of a  vector $[a,b]$ onto the unit circle, which will be $z=[\cos\zeta, \sin \zeta]$. Due to the projection property, the $R^2$ distance from $z$ to $[\cos \theta, \sin \theta]$ will be less than $2\delta$. 

Note that on the unit circle, if length of a curve is $0\leq s\leq \pi$, the $R^2$ distance of its two end points is $\sqrt{2-2\cos s}=2\sin(s/2)$. Note that $f(s)=\sin s/s$ satisfies $f'(x)=\frac{\cos s s-\sin s}{s^2}<0$ for $s\leq \pi$, so 
\[
\frac{2}{\pi}\leq \frac{2\sin s/2}{s}\leq 1. 
\]
So $|\theta-\zeta|\leq \delta \pi$. 
\end{proof}

\subsection{Solving a Poisson equation}
One of the key steps of our derivation is solving a Poisson equation. Here we provide some estimates regarding the solution.

Recall that $\phi_k$ is a function and $\phitilde_k$ is a vector with entries
\[
\phi_k(t)=\frac{1}{\sqrt{2\pi}\sigma}\exp(-\frac{\|t_k-t\|^2}{2\sigma^2}),\quad 
\phitilde_k(i)=\phi_k(t_i). 
\]
Our first step is finding a function $\psi_k$ so that $L^X \psitilde_k\approx c\phitilde_k$ for some constant $c$. Recall that $\psitilde_k$ is the vectorization of $\psi_k$ on $t_{[N]}$. To do so, we first note that when $p$ is uniform, $d^X(i)$ is approximately a fixed constant $Nd_0$ (see Lemma \ref{lem:concentrateX} latter), so 
\begin{align*}
L^X(i,\cdot)\psitilde_k&\approx \frac{1}{d_0N} \sum_{j=1}^N k(X_{t_i}, X_{t_j})(\psi_k(t_i)-\psi_k(t_j))\\
&\approx \frac{1}{d_0}\int (\psi_k(t_i)-\psi_k(x))\frac{1}{\sqrt{2\pi\sigma^2}}\exp(-\frac{|x-t_i|^2}{2\sigma^2})dx
\end{align*}
by the law of large numbers. Since this needs to hold for all $t_i$, essentially we are seeking a function  $\psi_k(t)$ so that 
\begin{equation}
\label{eqn:Poisson}
\int (\psi_k(y)-\psi_k(x))\frac{1}{\sqrt{2\pi\sigma^2}}\exp(-\frac{|x-y|^2}{2\sigma^2})dx
=\phi_k(y).
\end{equation}
Now consider a stochastic model where $x=y+\sigma\xi$ with $\xi$ being a standard Gaussian. Then \eqref{eqn:Poisson} can be written as 
\[
\psi_k(y)-\E[\psi_k(x)]=\phi_k(y).
\]
This is known as a Poisson equation. Nice thing about Poisson equations, is that their solutions can be written  down explicitly as an infinite sum, which further allows us to derive explicit bounds. 

We will start with the case where $\sigma=1$:
\begin{lemma}
\label{lem:ufunction}
Suppose we let 
\[
u_z(x)=\sum_{n=1}^\infty \frac{1}{\sqrt{2\pi n}}(1-\exp(-\frac{|x-z|^2}{2n})).
\]
Then $u_z$ satisfies the following for certain $C_0$
\begin{enumerate}
    \item $|u_z(x)|\leq C_0\min\{|x-z|^2, 1+\frac{1}{|x-z|^3}\}.$
    \item $|u'_z(x)|\leq C_0\min\{|x-z|, 1+\frac{1}{|x-z|}\}\leq C_0^2$
    \item $\int (u_z(y)-u_z(x))\frac{1}{\sqrt{2\pi}}\exp(-\frac{|x-y|^2}{2})dx=\frac{1}{\sqrt{2\pi}}\exp(-\frac{|y-z|^2}{2}).$
\end{enumerate}
\end{lemma}

\begin{proof}
For the first claim, we note that $0\leq 1-\exp(-a)\leq a$. So 
\[
0\leq u_z(x)\leq \sum_{n=1}^\infty \frac{1}{\sqrt{2\pi n}}\frac{|x-z|^2}{2n}=c_0 |x-z|^2
\]
for some constant $c_0$. Let $f(u)=\frac{1}{\sqrt{u}}(1-\exp(-\frac{|x-z|^2}{2u}))$, we check that it is decreasing
\[
\dot{f}(u)=-\frac{1}{2\sqrt{u}^3}(1-\exp(-\frac{|x-z|^2}{2u}))-
\frac{|x-z|^2}{2\sqrt{u}^5}(1-\exp(-\frac{|x-z|^2}{2u}))<0.
\]
Therefore we see that for some constants $C$ and $C'$, 
\begin{align*}
&\sum_{n=1}^\infty \frac{1}{\sqrt{n}}(1-\exp(-\frac{|x-z|^2}{2n}))\\
&\leq 1+C\int^\infty_1 (1-\exp(-\frac{|x-z|^2}{2u})\frac{1}{\sqrt{u}} du \quad\text{(infinite sum is bounded by integral)}\\
&=1+C\int^1_0 (1-\exp(-\frac{|x-z|^2}{2}v)\sqrt{v} dv\quad \text{(let $v=1/u$)}\\
&\leq 1+\frac{C}{|x-z|^3}\int^{|x-z|^2}_0 \exp(-w/2)\sqrt{w} dw \quad \text{(let $w=|x-z|^2v$)}\\
&\leq 1+\frac{C'}{|x-z|^3}.
\end{align*}

For the second claim, note that  for some constant $c_1$
\[
|u'_z(x)|=\sum_{n=1}^\infty \frac{|x-z|}{\sqrt{2\pi n}n}\exp(-\frac{|x-z|^2}{2n})\leq |x-z|\sum_{n=1}^\infty \frac{1}{\sqrt{2\pi n}n}=c_1|x-z|. 
\]
Let $f(u)=\frac{1}{u^{1.5}}\exp(-\frac{|x-z|^2}{2u})$, we check its derivative
\[
\dot{f}(u)=(\frac{|x-z|^2}{2u^{3.5}}-\frac{3}{2u^{2.5}})\exp(-\frac{|x-z|^2}{2u}).
\]
So if $u>\frac13|x-z|^2$, then $\dot{f}(u)<0$, else $\dot{f}(u)>0$. So 
\begin{align*}
&\sum_{n=1}^\infty \frac{|x-z|}{\sqrt{n}n}\exp(-\frac{|x-z|^2}{2n})\\
&\leq |x-z|f(\frac13|x-z|^2)+2|x-z|\int^\infty_1 \exp(-\frac{|x-z|^2}{2u})\frac{1}{u^{1.5}} du\\
&=\frac{9}{|x-z|^2}+2|x-z|^2\int^1_0 \exp(-\frac{|x-z|^2}{2}v)\sqrt{1/|x-z|^2v} dv\\
&\leq \frac{9}{|x-z|^2}+2\int^{\infty}_0 \exp(-\frac12 w)\frac{1}{\sqrt{w}} dw=O(1+\frac{1}{|x-z|^2}).
\end{align*}

For the third claim, note that 
\begin{align*}
&\int \frac{1}{\sqrt{2\pi}}\frac{1}{\sqrt{2\pi n}}\exp(-\frac{|x-z|^2}{2n})\exp(-\frac{|y-x|^2}{2})dx
=\frac{1}{\sqrt{2\pi(n+1)}}\exp(-\frac{|y-z|^2}{2(n+1)}),
\end{align*}
this is because we can interpret the left hand side of the equation as the transition probability from $y$ to $z$, where $x\sim y+z_1$, and $z\sim x+\sqrt{n}z_2$, and  $z_1, z_2$ are two independent standard Gaussian. Clearly the transition probability from $y$ to $z$ will be $\mathcal{N}(y,(n+1))$, which is the density described by the right hand side of the equation. Therefore
\begin{align*}
&\int u_z(x)\frac{1}{\sqrt{2\pi}}\exp(-\frac{|x-y|^2}{2})dx\\
&=
\sum_{n=1}^\infty \left(\frac{1}{\sqrt{2\pi n}}-\int \frac{1}{\sqrt{2\pi n}} \frac{1}{\sqrt{2\pi}}\exp(-\frac{|x-y|^2}{2})\exp(-\frac{|x-z|^2}{2n})dx \right)\\
&=\sum_{n=1}^\infty \left(\frac{1}{\sqrt{2\pi n}}-\frac{1}{\sqrt{2\pi(n+1)}}\exp(-\frac{|x-z|^2}{2(n+1)})\right)\\
&=u_z(x)-\frac{1}{\sqrt{2\pi}}\exp(-\frac{|x-z|^2}{2}).
\end{align*} 
\end{proof}

Then the solution of \eqref{eqn:Poisson} is given by a rescaling of $u_z$ as below
\begin{lemma}
\label{lem:psolve}
The solution of \eqref{eqn:Poisson} is given by 
\[
\psi_{k}(x)=\sum_{n=1}^\infty \frac{1}{\sqrt{2\pi \sigma^2 n}}(1-\exp(-\frac{|x-t_k|^2}{2\sigma^2 n}))
\]
Moreover, it satisfies the following estimate for some universal constant
\[
0\leq \psi_k(x)\leq \frac{C_0}{\sigma}\max\{\frac{|x-t_k|^2}{\sigma^2},\frac{\sigma^3}{1+|x-t_k|^3}\}\leq \frac{C_0}{\sigma},
\]
\[
 \psi_k'(x)\leq \frac{C_0}{\sigma^2}\min\{\frac{|x-t_k|}{\sigma},1+\frac{\sigma^2}{|x-t_k|^2}\}\leq \frac{C_0}{\sigma^2}. 
\]
\end{lemma}
\begin{proof}
    Simply note that $\psi_k(x)=\frac{1}{\sigma}u_{t_k/\sigma}(x/\sigma)$, and then apply Lemma \ref{lem:ufunction}. 
\end{proof}

\subsection{Concentration results for clean data $X$}
In this subsection we provide some concentration results for quantities related to the clean data $X$. Some versions of them have already appeared in \citep{cheng2022eigen,peoples2021spectral}. 

\begin{lemma}
\label{lem:concentrateX}
Suppose $4\sigma\sqrt{1/\log \sigma}\leq 1/{4L_{\calM}}, N\geq C\log N/\sigma^6$.
Given each $i$, we split $[N]$ into two sets. We say $j\in I_i$ if  $d(t_j,t_i)\leq r_\sigma$. Otherwise, we denote $j\in I_i^c$. Then there is a constant $C$ so that with high probability $1-\frac{1}{N^3}$, the following hold for all $i\in [N]$:
\begin{enumerate}
    \item $\max_{i\in [N]} |I_i|\leq C Nr_\sigma$. For any $\delta\geq N^{-b}$ with $0<b<1$, $\max_{i\in [N]} |j\in \{d(t_i,t_j)\leq \delta\}|\leq C N\delta$ 
    \item $\max_{i\in [N]}\sum_{j\in I_i^c} k(X_{t_i},X_{t_j})\leq C N\sigma^5$.
    \item $\max_{i\in [N]}\sum_{j\in I_i^c} k(X_{t_i},X_{t_j})^2\leq C N\sigma^5$.
    \item $N/C<d^X(i)<CN,$ and $N/C<\sum_{j\in [n]}k(X_{t_i},X_{t_j})^2<CN$.
    \item
    $\max_{i: t_i\in int_{r_\sigma}}|d^X(i)/N-\frac 1{2\pi}|<C(\sigma^2+\sqrt{\log N/N\sigma}),$
    \item $\max\{|\frac{1}{N}\sum sin( k t_i)^2-\frac12|,
    |\frac1N\sum \cos(k t_i)^2-\frac12|,
    |\frac{1}{N}\sum sin(k t_i)\cos(k t_i)|\}\leq \frac{C\sqrt{\log N}}{\sqrt{N}}.$
    \item In the open curve case, both $i^*=\arg\max(t_i)$ and $\arg\min(t_i)$,
    $d^X(i^*)/N\leq \frac 1{4\pi}+C(\sigma^2+\sqrt{\log N/N\sigma}).$
\end{enumerate}

\end{lemma}
\begin{proof}
To simplify notation, we assume $t_i\in [\pi/2, 3/2\pi]$ in the closed loop case. This does not sacrifice generality since we can always rotate the angle coordinate so the assumption holds. Then $|t_i-t_j|>r_\sigma$ is equivalent to $d(t_i,t_j)>r_{\sigma}$, and we will use the former since it is easier to interpret. 

For claim 1, let  fixed $t_i$ is fixed, let $I_{i,j}=1_{\{\|X_{t_i}-X_{t_j}\|\leq r_\sigma\}}$. We will show the claims hold for each fixed $i$ with probability $1-1/N^4$, then a union bound will yield our result. 

Note that  $I_{i,i}\equiv 1$, and $I_{i,j},j\neq 1$ are i.i.d. Bernoulli r.v.s. Note that when $I_{i,j}=1$, $|t_i-t_j|\leq r_\sigma$. So 
\begin{align*}
\E[I_{i,j}|t_i]\leq \frac1{2\pi}\int^{2\pi}_0 1_{|t_i-s|\leq r_\sigma} ds
\leq r_\sigma. 
\end{align*}
Then by Chernoff bound, for some $c$ and sufficiently large $N$ we have that  
\[
P\left[ |I_i|=\sum_{j=1}^N I_{i,j}\geq 2N r_\sigma\bigg|t_i\right]\leq \exp(-cNr_\sigma)\leq \frac{1}{N^4}. 
\]
The proof for general distance $\delta$ can be obtained similarly.

Likewise, we can let $J_{j}=1_{X_{{t}_j}\leq 4\log N/N}$. Note that $\E[J_j]=8\log N/N$. So using Bernstein inequality we can find that $P(\sum J_j=0)\leq \exp(-\frac{32(\log N)^2}{8\log N+8/3(\log N)})\leq N^{-4}$. This indicates that $P(\arg\min t_i>8\log N/n)<N^{-4}$.

For claim 2 and 3, let 
\[
S_{i,j}=k(X_{t_i},X_{t_j})1_{\|t_i-t_j\|\geq r_\sigma}=k(X_{t_i},X_{t_j})(1-I_{i,j})\leq \sigma^8(1-I_{i,j}). 
\]
Note that $S_{i,i}\equiv 0$. 
So 
\[
\sum_{j=1}^N S_{i,j}\geq N\sigma^8,\quad \sum_{j=1}^N S_{i,j}^2\geq N\sigma^8.
\]
For claim 4, 5, and 7, let $U_{i,j}=k(X_{t_i},X_{t_j})$. It  is bounded by $\frac{1}{\sigma}$ and mean being 
\begin{align*}
\E[U_{i,j}|t_i]&=\frac1{2\pi}\int^{2\pi}_0 \frac{1}{\sqrt{2\pi\sigma^2}}\exp(-\frac{\|X_{t_i}-X_s\|^2}{2\sigma^2})ds\\
&=\frac1{2\pi}\int^{2\pi}_0 \frac{1}{\sqrt{2\pi\sigma^2}}\exp(-\frac{\|X_{t_i}-X_s\|^2}{2\sigma^2})1_{\|t_i-s\|>r_\sigma}ds\\
&\quad+\frac1{2\pi}\int^{2\pi}_0 \frac{1}{\sqrt{2\pi\sigma^2}}\exp(-\frac{\|X_{t_i}-X_s\|^2}{2\sigma^2})1_{\|t_i-s\|<r_\sigma}ds\\
&\leq 2\sigma^8+
\frac{1}{2\pi}\int^{2\pi}_0 \frac{1}{\sqrt{2\pi\sigma^2}}\exp(-\frac{\|t_i-s\|^2}{2\sigma^2})ds\quad \text{by Lemma \ref{lem:expdiff}}\\
&\leq \sqrt{2\pi}\sigma^8+\frac{1}{2\pi}.
\end{align*}
And if $i^*=\arg\min t_i$, and $t_{i^*}<8\log N/N$, so the last line can be improved to be $2\sigma^8+\frac{1}{4\pi}+\frac{8}{\sigma}\log N/N\leq \frac{1}{4\pi}+\sigma^2$.

For the reverse, note that $\|X_t-X_s\|\leq |t-s|$. So if $t_i>r_\sigma$
\begin{align*}
\E[U_{i,j}|t_i]&= 
\frac1{2\pi}\int^{2\pi}_0 \frac{1}{\sqrt{2\pi\sigma^2}}\exp(-\frac{\|X_{t_i}-X_s\|^2}{2\sigma^2})ds\\
&\geq 
\frac1{2\pi}\int^{2\pi}_0 \frac{1}{\sqrt{2\pi\sigma^2}}\exp(-\frac{|t_i-s|^2}{2\sigma^2})ds\\
&\geq 
\frac1{2\pi}\int^{t_i}_{t_i-r_\sigma} \frac{1}{\sqrt{2\pi\sigma^2}}\exp(-\frac{|t_i-s|^2}{2\sigma^2})ds\\
&\geq \frac1{2\pi}(\Psi(0)-\Psi(-r_\sigma^2/\sigma^2))
\geq \frac1{4\pi}(1-\sigma^8),
\end{align*}
where $\Psi$ is the CDF of standard Gaussian. A similar lower bound can be obtained if $t_i<2\pi-r_{\sigma}$. If we also have $t_i\in int_{r_\sigma}$, then 
\begin{align*}
\E[U_{i,j}|t_i]&= 
\frac1{2\pi}\int^{2\pi}_0 \frac{1}{\sqrt{2\pi\sigma^2}}\exp(-\frac{\|X_{t_i}-X_s\|^2}{2\sigma^2})ds\\
&\geq\frac1{2\pi}\int^{2\pi}_0 \frac{1}{\sqrt{2\pi\sigma^2}}\exp(-\frac{|t_i-s|^2}{2\sigma^2})ds\\
&\geq 
\frac1{2\pi}\int^{t_i+r_\sigma}_{t_i-r_\sigma} \frac{1}{\sqrt{2\pi\sigma^2}}\exp(-\frac{|t_i-s|^2}{2\sigma^2})ds\\
&\geq \frac1{2\pi}-\sigma^8.
\end{align*}

Moreover, $U_{i,j}^2=k(X_{t_i},X_{t_j})^2\leq 1/\sigma^2$, so
\[
\E[U^2_{i,j}|t_i]=\frac1{2\pi}\int \frac{1}{2\pi\sigma^2}\exp(-\frac{\|X_s-X_{t_i}\|^2}{\sigma^2})ds\leq \frac{1}{\sigma}.
\]
Recall 
\[
\frac{d^X(i)}{N}=\frac{1}{\sqrt{2\pi}\sigma N}+\frac1N\sum_{j\neq i} U_{i,j},\quad \E[d^X(i)/N|t_i]=\frac{1}{\sqrt{2\pi}\sigma N}+\frac1N\sum_{j\neq i} \E[U_{i,j}|t_i]=:\bar{U}_i. 
\]
which is in $[2/C,\frac{1}{2\pi}+C\sigma^8]$
for some constant $C$. And if $t_i\in int_{r_\sigma}$, then $\E[d^X(i)/N|t_i]\in [\frac1{2\pi}-C\sigma^8,\frac1{2\pi}+C\sigma^8]$. If $t_i=\arg\min t_i$ in open curve, then $\bar{U}_i\leq \frac{1}{4\pi}+\sigma^2$.
Then by Bernstein inequality, for any $\delta$
\[
P(|d^X(i)/N-\bar{U}_i|>\delta|t_i)=\exp(-N \delta^2 \sigma^2). 
\]
By picking a small constant $\delta=C\sigma^2$, we have our claims. The estimate for $\sum_{j\in [N]} U_{i,j}^2$ goes similar.

For claim 6, note that 
\[
\E[\sin(k t_i)^2]=\frac1{2\pi}\int^{2\pi}_0 \sin(k x)^2dx=\frac12,
\]
\[
\E[\sin(k t_i)\cos(k t_i)]=\int^{2\pi}_0 \sin(k x)\cos(k x)dx=0.
\]
Since $|\sin(k x)|\leq 1,|\cos (kx)|\leq 1$, by Bernstein inequality
\[
P(\frac1N \sum_{t=1}^N[\sin(k t_i)^2-1/2] \geq t)\leq \exp(-\frac{Nt^2}{2+2t/3})
\]
Pick  $t=\frac{2\sqrt{1/\log N}}{\sqrt{N}}$, we will have our result.

\end{proof}

\begin{lemma}
\label{lem:Laplacian}
For any function smooth function $\psi$ with bounded $O(1)$ fourth derivative, 
\[
\int^{+\infty}_{-\infty} (\psi(y)-\psi(x))\frac{1}{\sqrt{2\pi\sigma^2}}\exp(-\frac{|x-y|^2}{2\sigma^2})dx
=-\frac12\sigma^2 \Delta \psi+O(\sigma^3).
\]
\end{lemma}
\begin{proof}
Simply note that if we let $\xi\sim \mathcal{N}(0,1)$  
\begin{align*}
&\int^{+\infty}_{-\infty} (\psi(y)-\psi(x))\frac{1}{\sqrt{2\pi\sigma^2}}\exp(-\frac{|x-y|^2}{2\sigma^2})dx\\
&=\frac12\E[2\psi(y)-\psi(y+\sigma \xi)-\psi(y-\sigma \xi)]\\
&=\frac12\E[-\Delta\psi(y)\sigma^2 \xi^2+\frac{1}{24}\sigma^4(\psi^{(4)}(w_1)+\psi^{(4)}(w_2))\xi^4]=-\frac12\sigma^2 \Delta \psi+O(\sigma^4).
\end{align*}
Here $w_1, w_2$ are some random points in $[y-\sigma\xi, y+\sigma \xi]$. 
\end{proof}

\begin{lemma}
\label{lem:berstein}
Given a function $\psi(x)$ so that $|\psi(x)|\leq 1$ and $|\psi'(x)|\leq c_\psi$.  Suppose there is a function $\phi(x)$ so that
\[
\int^{+\infty}_{-\infty} (\psi(y)-\psi(x))\frac{1}{\sqrt{2\pi\sigma^2}}\exp(-\frac{|x-y|^2}{2\sigma^2})dx
=\phi(y)+O(\sigma^3).
\]
Then the following holds with high probability $1-\frac{1}{N^2}$
\[
\max_{i:t_i\in int_{2r_\sigma}}|L^Z(i,\cdot)\psitilde-\phi(t_i)|\leq \Otilde\left(r_\sigma(\sigma^2+\frac{1}{\sqrt{N\sigma}}+\frac{\epsilon}{\sigma})c_\psi +\frac{1}{\sigma N}+\frac{\epsilon}{\sigma}+\sigma^2\right)
\]
Suppose $|\psi'(x)|\leq b_\psi$ when $x\in int_{2r_\sigma}^c$, then 
\[
\max_{i:t_i\in int^c_{2r_\sigma}}(L^Z(i,\cdot)\psitilde)\leq \Otilde (b_\psi \sigma^2). 
\]
\end{lemma}

\begin{proof}
To simplify notation, we assume $t_i\in [\pi/2, 3/2\pi]$ in the closed loop case. This does not sacrifice generality since we can always rotate the angle coordinate so the assumption holds. Then $|t_i-t_j|>r_\sigma$ is equivalent to $d(t_i,t_j)>r_{\sigma}$, and we will use the former since it is easier to interpret. 

Denote
\[
(U_X\psitilde)(i)=\frac{1}{N}\sum_{j=1}^Nk(X_{t_j},X_{t_i})(\psi(t_i)-\psi(t_j)), \quad Y_{j,i}=k(X_{t_j},X_{t_i})(\psi(t_i)-\psi(t_j)).
\]
It is easy to see that $Y_{i,i}=0$.
We will study the mean and variance of $Y_{j,i}, j\neq i.$

\text{\textbf{Step 1: Find the conditional mean $Y_{j,i}$.}}
\begin{align}
\notag
&\E[k(X_{t_j},X_{t_i})(\psi(t_j)-\psi(t_i))|t_i]\\
\notag
&=\frac{1}{2\pi}\int^{2\pi}_0 (\psi(t)-\psi(t_i))\frac{1}{\sqrt{2\pi}\sigma}\exp(-\frac{\|X_t-X_{t_i}\|^2}{2\sigma^2})dt\\
\label{tmp:meansplit1}
&=\frac{1}{2\pi}\int_{t\in [0,2\pi],|t-t_i|<r_\sigma} (\psi(t)-\psi(t_i))\frac{1}{\sqrt{2\pi}\sigma}\exp(-\frac{\|X_t-X_{t_i}\|^2}{2\sigma^2})dt\\
\label{tmp:meansplit2}
&+\frac{1}{2\pi}\int_{t\in [0,2\pi],|t-t_i|>r_\sigma} (\psi(t)-\psi(t_i))\frac{1}{\sqrt{2\pi}\sigma}\exp(-\frac{\|X_t-X_{t_i}\|^2}{2\sigma^2})dt
\end{align}
We first tend \eqref{tmp:meansplit2}, because it is insignificant. Indeed, by Lemma \ref{lem:expdiff},, when $|t-t_i|>r_\sigma$, we have $\frac{1}{\sigma}\exp(-\frac{\|X_t-X_{t_i}\|^2}{2\sigma^2})\leq \sigma^6$, while $|\psi(t)|\leq 1$, so 
\[
\int_{|t-t_i|>r_\sigma} (\psi(t)-\psi(t_i))\frac{1}{\sigma}\exp(-\frac{\|X_t-X_{t_i}\|^2}{2\sigma^2})dt
\lesssim \sigma^6. 
\]
To treat \eqref{tmp:meansplit1}, we note that $\phi_i(t)\leq \sigma^8$ when $|t-t_i|\geq r_\sigma$, and $|\psi(t)|\leq 1$
\[
\int_{|t-t_i|>r_\sigma}(\psi(t)-\psi(t_i))\frac{1}{\sqrt{2\pi}\sigma}\exp(-\frac{\|t-t_i\|^2}{2\sigma^2})dt
\leq  \sigma^8. 
\]
So using the definition of $\phi$
\begin{align*}
&|\int_{|t-t_i|<r_\sigma}(\psi(t)-\psi(t_i))\frac{1}{\sqrt{2\pi}\sigma}\exp(-\frac{\|t-t_i\|^2}{2\sigma^2})dt-\phi(t_i)|\\
&=|\int_{|t-t_i|>r_\sigma}(\psi(t)-\psi(t_i))\frac{1}{\sqrt{2\pi}\sigma}\exp(-\frac{\|t-t_i\|^2}{2\sigma^2})dt|=\Otilde(\sigma^4). 
\end{align*}
See the proof of \eqref{tmp:smallint}.
We note that when $|t-t_i|\leq r_\sigma$
\begin{align*}
|\exp(-\frac{\|X_t-X_{t_i}\|^2}{2\sigma^2})-    
\exp(-\frac{|t-t_i|^2}{2\sigma^2})|
\leq \frac{r_\sigma^4}{\sigma^2}\exp(-\frac{|t-t_i|^2}{2\sigma^2}). 
\end{align*}
So because $|(\psi(t)-\psi(t_i))|\leq r_\sigma c_\psi$ and Lemma \ref{lem:expdiff}, we have
\begin{align*}
&\int^{t_i+r_\sigma}_{t_i-r_\sigma} (\psi(t)-\psi(t_i))\frac{1}{\sqrt{2\pi}\sigma}|\exp(-\frac{\|X_t-X_{t_i}\|^2}{2\sigma^2})-\exp(-\frac{|t-t_i|^2}{2\sigma^2})|dt\\
&\lesssim \frac{r_\sigma^8c_\psi}{\sigma^2} \int^{t_i+r_\sigma}_{t_i-r_\sigma} \frac{1}{\sigma\sqrt{2\pi}}=\Otilde(\sigma^3 c_\psi). 
\end{align*}
In summary:
\begin{equation}
\label{tmp:part1}
\E[k(X_{t_j},X_{t_i})(\psi(t_i)-\psi(t_j))|t_i]=\frac{\phi(t_i)}{2\pi}+\Otilde(\sigma^3c_\psi). 
\end{equation}


\text{\textbf{Step 2: Bounding the conditional 2nd moment of $Y_{j,i}$.}}
\begin{align}
\notag
\E[Y_{j,i}^2|t_i]&=\E[k(X_{t_j},X_{t_i})^2(\psi(t_j)-\psi(t_i))^2|t_i]\\
\notag
&=\frac{1}{2\pi}\int^{2\pi}_0 (\psi(t)-\psi(t_i))^2\frac{1}{2\pi\sigma^2}\exp(-\frac{\|X_t-X_{t_i}\|^2}{\sigma^2})dt\\
\label{tmp:2meansplit1}
&\leq \int^{t_i+r_\sigma}_{t_i-r_\sigma} (\psi(t)-\psi(t_i))^2\frac{1}{\sigma^2}\exp(-\frac{\|X_t-X_{t_i}\|^2}{\sigma^2})dt\\
\label{tmp:2meansplit2}
&\quad +\int_{|t-t_i|>r_\sigma} (\psi(t)-\psi(t_i))^2\frac{1}{\sigma^2}\exp(-\frac{\|X_t-X_{t_i}\|^2}{\sigma^2})dt
\end{align}
We first tend \eqref{tmp:2meansplit2}, because it is insignificant. Indeed, under our condition, when $|t-t_i|>r_\sigma$, we have $\frac{1}{\sigma^2}\exp(-\frac{\|X_t-X_{t_i}\|^2}{\sigma^2})\leq \sigma^{12}$, while $|\psi(t)|\leq 1$, so 
\[
\int_{|t-t_i|>r_\sigma} (\psi(t)-\psi(t_i))^2\frac{1}{\sigma^2}\exp(-\frac{\|X_t-X_{t_i}\|^2}{\sigma^2})dt
\lesssim \sigma^{12}. 
\]
To treat \eqref{tmp:2meansplit1}, we bound it using 
\begin{align*}
&\int^{t_i+r_\sigma}_{t_i-r_\sigma} (\psi(t)-\psi(t_i))^2\frac{1}{\sigma^2}\exp(-\frac{\|X_t-X_{t_i}\|^2}{\sigma^2})dt\\
&\leq \int^{t_i+r_\sigma}_{t_i-r_\sigma} \frac{r_\sigma^2c_\psi^2}{\sigma^2}\exp(-\frac{\|X_t-X_{t_i}\|^2}{\sigma^2})dt
\leq \int^{t_i+r_\sigma}_{t_i-r_\sigma} \frac{r_\sigma^2c_\psi^2}{\sigma^2}dt=\frac{r_\sigma^3 c_\psi^2}{\sigma^2}
\end{align*}

\text{\textbf{Step 3: Bounding fluctuation of $U_Z\psitilde$.}}
Recall that 
\[
(U_X\psitilde)(i)=\frac{1}{N}\sum_{j=1}^NY_{j,i}
\]
Previous two steps have that for $j\neq i$, $\E[Y_{j,i}|t_i]=\frac{1}{2\pi}\phi(t_i)+\Otilde(\sigma^3c_\psi)$, and $\E[(Y_{j,i})^2|t_i]=\Otilde(r_\sigma^3c^2_\psi/\sigma^2 ).$ Note also that 
$|Y_{j,i}|\lesssim \frac{1}{\sigma}$. 
So by Berstein inequality, 
\[
P(|\frac{1}{N-1}\sum_{j\neq i}Y_{j,i}-\E[Y_{j,i}]|>t)
\leq \exp(-\frac{c\sigma^2 (N-1)t^2}{r^3_\sigma c^2_\psi+t\sigma }).
\]
The upper bound will be of order $1/N^3$ when 
\begin{equation}
\label{tmp:part2}
t>4c_\psi\sqrt{\frac{r^3_\sigma\log N}{c\sigma^2 N}}+\frac{4\log N}{c\sigma(N-1)}.
\end{equation}
Note that
\[
(U_X\psitilde)(i)=\frac{1}{N}\sum_{j=1}^NY_{j,i}=\frac{1}{N}\sum_{j\neq i}Y_{j,i}.
\]
Moreover, by Lemma \ref{lem:XZperturb}, we have that with probability $1-1/N^2$, 
\begin{equation}
\label{tmp:part3}
|(U_X\psitilde)(i)-(U_Z\psitilde)(i)|
\leq \frac{1}{N}\|C^X(i,\cdot)-C^Z(i,\cdot)\|_1\|\psitilde\|_\infty\lesssim\frac{\epsilon}{\sigma}+\sigma^5,    
\end{equation}
where $C^X$ and $C^Z$ are the similarity matrix based on $X$ and $Z$ by the Gaussian kernel $C(i,j) = k(x_i, x_j;\sigma)$, respectively. 

In summary, combining \eqref{tmp:part1},\eqref{tmp:part2}, and \eqref{tmp:part3}, with probability $1-2/N^2$, we have that 
\[
\max_{i}|(U_X\psitilde)(i)-\frac{\phi(t_i)}{2\pi}|\leq \Otilde(r_\sigma^3c_\psi +
c_\psi\sqrt{\frac{\sigma}{N}}+\frac{1}{c\sigma N}+\frac{\epsilon}{\sigma}+\frac{\phi(t_i)}{N})
\]

\text{\textbf{Step 4a: The closed loop case.}}
By Lemma \ref{lem:concentrateX} claim (5) and \ref{lem:XZperturb} claim (1), we  have that with high probability $1-1/N^2$,
\[
N(d^Z(i))^{-1}=2\pi+O(\sigma^2+\sqrt{\log N/N\sigma}+\epsilon/\sigma)
\]
for all $i$. Then note that 
\begin{align*}
L^Z(i,\cdot)\psitilde&=\frac{2\pi}{N} \sum_{j=1}^N k(Z_{t_i},Z_{t_j})(\psi(t_i)-\psi(t_j))\\
&+\sum_{j=1}^N(1/\sqrt{d^Z(i)d^Z(j)}-2\pi/N)k(Z_{t_i},Z_{t_j})(\psi(t_i)-\psi(t_j)).
\end{align*}
Note that the first line is $2\pi U_Z\psitilde$, so it suffices for us to bound the second line. 
Note that when  $|t_i-t_j|>r_\sigma$, so $k(Z_{t_i},Z_{t_j})\leq \sigma^8/N$ by Lemma \ref{lem:expdiff}, so 
\[
\sum_{j: |t_i-t_j|>r_\sigma}(1/\sqrt{d^Z(i)d^Z(j)}-2\pi/N)k(Z_{t_i},Z_{t_j})(\psi(t_i)-\psi(t_j))\leq \sigma^8.
\]
Meanwhile there are at most $CNr_\sigma$ that many $t_j$ so that $|t_j-t_i|<r_\sigma$, where $\psi(t_i)-\psi(t_j)\leq c_\psi r_\sigma$. So
\[
\sum_{j: |t_i-t_j|<r_\sigma}(1/\sqrt{d^Z(i)d^Z(j)}-2\pi/N)k(Z_{t_i},Z_{t_j})(\psi(t_i)-\psi(t_j))\leq r_\sigma c_\psi\Otilde(\sigma^2+\sqrt{\log N/N\sigma}+\epsilon/\sigma).
\]
 So we would have  the same bound. 
This lead to 
\[
\|L^Z\psitilde-\phitilde\|_\infty\leq \Otilde\left((\sigma^2+\frac{1}{\sqrt{N\sigma}}+\frac{\epsilon}{\sigma})c_\psi +
c_\psi\sqrt{\frac{\sigma}{N}}+\frac{1}{\sigma N}+\frac{\epsilon}{\sigma}+\sigma^3\right).
\]

\text{\textbf{Step 4b: The open curve case.}}
Denote $j\in B$ if $t_j$ is of distance at most $r_\sigma$ from the end points $0$ and $2\pi$. $B^c$ denotes the indices that in the completement.  Note by Lemma \ref{lem:concentrateX}, 
$(d^Z(i))^{-1}=2\pi+O(\sigma^2+\sqrt{\log N/N\sigma}+\epsilon)$ if $i\in B$. Next, if $t_i$ is of distance $2r_\sigma$ from the boundary, we can decompose
\begin{align*}
L^Z(i,\cdot)\psitilde&=\frac{2\pi}{N} \sum_{j=1}^N k(Z_{t_i},Z_{t_j})(\psi(t_i)-\psi(t_j))\\
&+\sum_{j\in B}(1/d^Z(i)+1/d^Z(j)-2\pi/N)k(Z_{t_i},Z_{t_j})(\psi(t_i)-\psi(t_j))\\
&+\sum_{j\in B^c}(1/d^Z(i)+1/d^Z(j)-2\pi/N)k(Z_{t_i},Z_{t_j})(\psi(t_i)-\psi(t_j))
\end{align*}
Note that when $j\in B$, $|t_i-t_j|>r_\sigma$, so $k(Z_{t_i},Z_{t_j})\leq \sigma^8/N$ by Lemma \ref{lem:expdiff}, so 
\[
\sum_{j\in Bd}(1/d^Z(i)+1/d^Z(j)-2\pi/N)k(Z_{t_i},Z_{t_j})(\psi(t_i)-\psi(t_j))\leq \sigma^8.
\]
And when $j\in B^c$, $1/d^Z(j)=2\pi/N+O(\sigma^2+\sqrt{\log N/N\sigma}+\epsilon)$. So we would have  the same bound using proof similar to the one in step 4a. 

When $t_i\in int_{2r_\sigma}^c$, $N/d^Z(i)$ is bounded above and below by constants with high probability by Lemma \ref{lem:Xinfty} and \ref{lem:XZperturb}. So there are some constant $C$ so that 
\begin{align*}
&\sum_{j}(1/d^Z(i)+1/d^Z(j))k(Z_{t_i},Z_{t_j})(\psi(t_i)-\psi(t_j))\\
&\leq \frac{C}{N}\sum_{j}k(Z_{i},Z_{j})|\psi(t_i)-\psi(t_j)|\\
 &\leq \frac{C}{N}(\sum_{j: |t_i-t_j|<r_\sigma}(k(t_i,t_j)+\sigma^2+\epsilon/\sigma^2)b_\psi r_\sigma +2N\sigma^7)\\
 &=\Otilde (b_\psi \sigma^2)
\end{align*}

which can further be bounded using the method in Step 4a. This  leads to our second claim.  





\end{proof}

The next Lemma shows that for two ways of normalizing Laplacians, they yield similar result
\begin{lemma}
\label{lem:tworw}
When $\calM$ is a closed loop, consider $\Ltilde^X=(I-(d^X)^{-1}C^X)$ and $L^X=(I-(d^X)^{-1/2}C^X (d^X)^{-1/2})$, then suppose $\Ltilde^X$ and $L^X$ share the same eigenvalues. $\phitilde$ is an eigenvector of $L^X$  if and only if $(d^X)^{-1/2} \phitilde$ is an eigenvector of $\Ltilde^X$. 
\end{lemma}
\begin{proof}
Simply check that 
\[
(d^X)^{-1/2}(I-(d^X)^{-1/2}C^X (d^X)^{-1/2})\phitilde=\lambda (d^X)^{-1/2}\phitilde\Leftrightarrow 
(I-(d^X)^{-1}C^X)(d^X)^{-1/2}\phitilde=\lambda (d^X)^{-1/2}\phitilde. 
\]
\end{proof}

\subsection{Perturbation between X and Z}
In this subsection we provide perturbation caused by data noise.

\begin{lemma}
\label{lem:XZperturb}
Under Assumptions \ref{aspt:manifold} and \ref{aspt:noise}, $\sigma^7 N\leq 1$, the following hold with $1-O(1/N)$,  
\begin{enumerate}
    \item For all $i\in [N]$, $|d^Z(i)-d^X(i)|/N\leq \|C^X(i,\cdot)-C^Z(i,\cdot)\|_1/N= \Otilde(\epsilon/\sigma+\sigma^6)$
    \item $\max_{i\in [N]}\|C^X(i,\cdot)-C^Z(i,\cdot)\|/N= \Otilde((\epsilon/\sigma+\sigma^6)/\sqrt{N\sigma})$
     \item $\max_{i\in [N]}\|L^X(i,\cdot)-L^Z(i,\cdot)\|= \Otilde((\epsilon/\sigma+\sigma^6)/\sqrt{N\sigma})$
     \item $\|L^X-L^Z\|\leq \max_{i\in [N]}\|L^X(i,\cdot)-L^Z(i,\cdot)\|_1=  \Otilde(\epsilon/\sigma+\sigma^6).$
\end{enumerate}
\end{lemma}
\begin{proof}
Denote $I_i=\{j\in [N], d(t_i,t_j)\leq r_\sigma\}$.
 By the Lemma \ref{lem:expdiff} and \ref{lem:concentrateX}, with high probability the following holds for some constant $C$,
\[
|I_i|\leq CN r_\sigma,\quad 
\text{for all } i\in [N].
\]

So 
\begin{align*}
   &\sum_{j=1}^N |\exp(-
   \frac{\|X_{t_i}-X_{t_j}\|^2}{2\sigma^2})-\exp(-\frac{\|Z_{i}-Z_{j}\|^2}{2\sigma^2})|1_{j\in I_i}\leq C|I_i|\epsilon/\sigma\lesssim CN\epsilon. 
\end{align*}
\begin{align*}
   &\sum_{j=1}^N |\exp(-
   \frac{\|X_{t_i}-X_{t_j}\|^2}{2\sigma^2})-\exp(-\frac{\|Z_{i}-Z_{j}\|^2}{2\sigma^2})|1_{j\notin I_i}\leq N\sigma^8, 
\end{align*}
In summation, we have
\[
\sum_{j\neq i}|C^X(i,j)-C^Z(i,j)|/N=\sum_{j\neq i} \frac{1}{\sqrt{2\pi}\sigma N}|\exp(-\frac{\|X_{t_i}-X_{t_j}\|^2}{2\sigma^2})-\exp(-\frac{\|Z_{i}-Z_{j}\|^2}{2\sigma^2})|
=\Otilde(\epsilon/\sigma+\sigma^6). 
\]
This leads to
\[
|d^X(i,i)-d^Z(i,i)|=|\sum_{j\neq i}C^X(i,j)-\sum_{j\neq i}C^Z(i,j)|\leq \sum_{j\neq i}|C^X(i,j)-C^Z(i,j)|=N\Otilde(\epsilon/\sigma+\sigma^6) 
\]
This concludes our first claim.

For the second claim,
given each $i$, using Lemma \ref{lem:expdiff} and \ref{lem:concentrateX},  
\[
\sum_{j=1}^N |\exp(-\frac{\|X_{t_i}-X_{t_j}\|^2}{2\sigma^2})-\exp(-\frac{\|Z_{i}-Z_{j}\|^2}{2\sigma^2})|^21_{j\in I_i}\leq CN(\sigma^{5}+\epsilon/\sigma)^2r_\sigma, 
\]
\[
\sum_{j=1}^N |\exp(-\frac{\|X_{t_i}-X_{t_j}\|^2}{2\sigma^2})-\exp(-\frac{\|Z_{i}-Z_{j}\|^2}{2\sigma^2})|^21_{j\notin I_i}\leq N\sigma^{12}. 
\]
In summation, we have
\[
\sum_{j\neq i}|C^X(i,j)-C^Z(i,j)|^2/N^2=\sum_{j\neq i} \frac{1}{\sqrt{2\pi}\sigma^2 N^2}|\exp(-\frac{\|X_{t_i}-X_{t_j}\|^2}{2\sigma^2})-\exp(-\frac{\|Z_{i}-Z_{j}\|^2}{2\sigma^2})|^2
=\Otilde( \frac{1}{N\sigma}(\sigma^{5}+\epsilon/\sigma)^2).
\]
For the third claim, we note that for closed loop case
\[
\|L^X(i,\cdot)-L^Z(i,\cdot)\|\leq \max_j|\frac{1}{\sqrt{d^X(i)d^X(j)}}-\frac{1}{\sqrt{d^Z(i)d^Z(j)}}|\|C^X(i,\cdot)\|+\max_j|\frac{1}{d^Z(j)}|\|C^X(i,\cdot)-C^Z(i,\cdot)\|.
\]
And in the open curve case
\[
\|L^X(i,\cdot)-L^Z(i,\cdot)\|\leq \max_j|\frac{1}{d^X(i)}+\frac{1}{d^X(j)}-\frac{1}{d^Z(i)}-\frac{1}{d^Z(j)}|\|C^X(i,\cdot)\|+\max_j|\frac{1}{d^Z(j)}|\|C^X(i,\cdot)-C^Z(i,\cdot)\|.
\]
Note that claim (1) has shown that $|d^X(i)-d^Z(i)|=\Otilde(N(\epsilon/\sigma+\sigma^6))$.
Lemma \ref{lem:concentrateX} has shown that with $1-1/N$ probability, there is a constant $C$ so that $d^X(i)>cN$ for all $i$.  Plug in the estimates from the first two claims, we have the third estimate. 

For the final claim, note that $L^X,L^Z$ are symmetric, so 
\[
\|L^X-L^Z\|\leq \max_i \|L^X(i,\cdot)-L^Z(i,\cdot)\|_1. 
\]
And for closed loop case
\[
\|L^X(i,\cdot)-L^Z(i,\cdot)\|_1\leq \max_j|\frac{1}{\sqrt{d^X(i)d^X(j)}}-\frac{1}{\sqrt{d^Z(i)d^Z(j)}}|\|C^X(i,\cdot)\|_1+\max_j \frac{1}{d^Z(j)}\|C^X(i,\cdot)-C^Z(i,\cdot)\|_1.
\]
And in the open curve case
\[
\|L^X(i,\cdot)-L^Z(i,\cdot)\|_1\leq \max_j|\frac{1}{d^X(i)}+\frac{1}{d^X(j)}-\frac{1}{d^Z(i)}-\frac{1}{d^Z(j)}|\|C^X(i,\cdot)\|_1+\max_j\frac{1}{d^Z(j)}\|\|C^X(i,\cdot)-C^Z(i,\cdot)\|_1.
\]
Using claim 1 and Lemma \ref{lem:concentrateX} we will have our result.

\end{proof}

\begin{lemma}
\label{lem:LXZpert}
Suppose  $\psi\in \reals^N$ satisfies   
 $\|\psi\|_\infty\lesssim \frac{1}{\sqrt{N}}$, and for some universal  constants $\delta_2$, 
\[
\max_i|L^X(i,\cdot)\psi-\lambda^X\psi(i)|\lesssim \delta_{2}/\sqrt{N}.
\]
Then following holds with high probability, \[
\max_i|L^Z(i,\cdot)\psi-\lambda^Z\psi(i)|\leq \Otilde (\delta_{2}+\epsilon/\sigma+\sigma^6)/\sqrt{N},
\]
\end{lemma}
\begin{proof}
Note 
\begin{align*}
&|L^Z(i,\cdot)\psi-\lambda^Z\psi(i)|\\
&
\leq |L^X(i,\cdot)\psi-\lambda^X\psi(i)|
+|L^X(i,\cdot)\psi-L^Z(i,\cdot)\psi|+|\lambda^X-\lambda^Z||\psi(i)|\\
&\lesssim \frac{\delta_2}{\sqrt{N}}+|L^X(i,\cdot)\psi-L^Z(i,\cdot)\psi|+\|L^X-L^Z\||\psi(i)|.
\end{align*}
Then we use Lemma \ref{lem:XZperturb} and obtain
\[
|L^X(i,\cdot)\psi-L^Z(i,\cdot)\psi|\leq 
\frac{1}{\sqrt{N}}\|L^X(i,\cdot)-L^Z(i,\cdot)\|_1=\Otilde((\epsilon/\sigma+\sigma^6)/\sqrt{N}).
\]
 Lemma \ref{lem:XZperturb} also gives us $\|L^X(i,\cdot)-L^Z(i,\cdot)\|=\Otilde(\epsilon/\sigma+\sigma^6)$. In combine we have our claim. 
\end{proof}

\subsection{Eigenvector error projection}
The final step of the preparation is to get $\delta_1$ in Lemma \ref{lem:Xinfty}. 

\begin{lemma}
\label{lem:Czprod}
Suppose $F^Z$ is an eigenvector of $L^Z$ with eigenvalue $\lambda^Z$, $v$ is a unit norm vector and $\lambda$ is a scalar. Let $Q^Z=I-L^Z$. The following holds with high probability $1-1/N^2$: 
\begin{align*}
\sqrt{N}\sigma\max_{i\in int_{3r_\sigma}}\langle F^Z-v, Q^Z(i,\cdot)\rangle
&\leq (\lambda^Z+\epsilon+\sigma^b+\frac{1}{\sqrt{N\sigma}}+\frac{\|q_i\|}{\sqrt{N}})\|F^Z-v\|\\
&\quad+|\lambda-\lambda^Z|+\|L^Zv-\lambda v\|
\end{align*}
Where $b=1.5$ for open curve and $b=2$ for closed loop. And $q_i$ is a vector with entries:
\[
q_i(j)= N\sigma Q^Z(i,j)-{\sqrt{2\pi}} \exp(-\frac{|t_i-t_j|^2}{2\sigma^2}),\quad i,j\in [N]. 
\]
\end{lemma}
\begin{proof}
Recall the $\psi_k$ from Lemma \ref{lem:psolve}, and let
\[
\psi(x)= \sqrt{2\pi}\sigma\psi_{k}(x),\quad \phi_t(x)=\sqrt{2\pi}\exp(-\frac{\|x-t\|^2}{2\sigma^2}).
\]
Note that $\psi(x)=O(1), \psi'(x)=O(1/\sigma)$. 
Denote $r_k=L^Z\psitilde_{k}-\phitilde_{t_k}$, where $\psitilde_k$ and $\phitilde_{t_k}$ is the vecterization of $\psi(x)$ and $\phi_{t_k}(x)$ at $x_{[N]}$. We apply Lemma \ref{lem:berstein}, where 
\[
b_\psi=c_\psi=\sup_{x}|\psi'(x)|=O(\frac{1}{\sigma}).
\]
So for each fixed $k$, we have with high probability $1-1/N^2$, 
\[
\max_{j: t_j\in int_{2r_{\sigma}}}|r_k(j)|
\leq \Otilde(\delta_3),\quad 
\max_{j\in \text{int}_{2r_\sigma}^c}|r_k(j)|
\leq \exp(-\frac{\|t_j-t_k\|^2}{2\sigma^2})+\Otilde(\sigma),\quad 
\delta_3:= (\epsilon/\sigma+\sigma^2+1/\sqrt{N\sigma}).
\]
This also leads to $\|r_{k}\|\leq \sqrt{N}(\Otilde(\delta_3)+\sigma^{1.5})$. For closed loop case, this can be improved to  $\|r_{k}\|\leq \sqrt{N}(\Otilde(\delta_3))$. 

Note that 
\[
Q^Z(i,\cdot)=\frac{1}{\sigma N}\phitilde_{t_i}+\frac{1}{\sigma N}q_i
\]

So we have
\begin{align}
\notag
&\langle F^Z-v, Q^Z(i,\cdot)\rangle\\
\notag
&=\frac{1}{N\sigma}\langle F^Z-v, \phitilde_{t_i}\rangle+\frac1{ N\sigma}\langle F^Z-v, q_i\rangle\\
\notag
&\leq 
\frac{1}{N\sigma}\langle F^Z-v, L^Z\psitilde_{i}\rangle+\frac{\|r_{i}\|+\|q_i\|}{N\sigma}\|F^Z-v\|\\
\notag
&= \frac{1}{N\sigma}\langle L^Z(F^Z-v), \psitilde_{i}\rangle+\frac{\|r_{i}\|+\|q_i\|}{N\sigma}\|F^Z-v\|\\
\notag
&=\frac{\lambda^Z}{ N\sigma}\langle F^Z-v,\psitilde_{i}\rangle+\frac{(\lambda^Z-\lambda)}{ N\sigma}\langle v, \psitilde_{i}\rangle
-\frac{1}{N\sigma}\langle L^Zv-\lambda v, \hat{\psi}_{i}\rangle+\frac{\|r_{i}\|+\|q_i\|}{N\sigma}\|F^Z-v\|\\
\label{tmp:Fz-v}
&\leq \frac{1}{\sqrt{N}}(\frac{\lambda^Z}{\sigma}\|F^Z-z\|+\frac{|\lambda^Z-\lambda|}{\sigma}
+\frac{\|L^Zv-\lambda v\|}{\sigma})+\frac{\|r_{i}\|+\|q_i\|}{N\sigma}\|F^Z-v\|
\end{align}
Plug the estimates above into this upper bound we find the end result. 

\end{proof}
\begin{lemma}
\label{lem:qi}
In the open curve case, we let
\[
p_i(j)=\frac{N}{\sqrt{2\pi}}\left(\exp(-\frac{|Z_i-Z_j|^2}{2\sigma^2})-\exp(-\frac{|t_i-t_j|^2}{2\sigma^2})\right)(\frac{1}{2d^Z(i)}+\frac{1}{2d^Z(j)})
\]
\[
u_i(j)=\frac{N}{\sqrt{2\pi}}(\frac{1}{2d^Z(i)}+\frac{1}{2d^Z(j)}-\frac{2\pi}{N})\exp(-\frac{|t_i-t_j|^2}{2\sigma^2}).
\]
In the closed loop case, we let
\[
p_i(j)=\frac{N}{\sqrt{2\pi d^Z(i)d^Z(j)}}\left(\exp(-\frac{|Z_i-Z_j|^2}{2\sigma^2})-\exp(-\frac{|t_i-t_j|^2}{2\sigma^2})\right)
\]
\[
u_i(j)=(\frac{N}{\sqrt{2\pi d^Z(i)d^Z(j)}}-\sqrt{2\pi})\exp(-\frac{|t_i-t_j|^2}{2\sigma^2})
\]
Recall $q_i$ in Lemma \ref{lem:Czprod}. If $t_i\in int_{2r_\sigma}$, with high probability $1-1/N^4$
\[
\|q_i\|\leq \|u_i\|+\|p_i\|\leq
\Otilde(\sqrt{N}(\sigma^{2.5}+\frac{1}{\sqrt{N}}+\epsilon/\sqrt{\sigma})))
\]
And for $t_i\notin int_{2r_\sigma}$, we have 
\[
\|q_i\|\leq \|u_i\|+\|p_i\|\leq
\Otilde(\sqrt{N}(\sigma^{0.5}+\frac{1}{\sqrt{N}}))
\]
\end{lemma}

\begin{proof}
\text{\textbf{The closed loop case.}}
By Lemma \ref{lem:concentrateX} (5) and \ref{lem:XZperturb} (1), we  have that with high probability 
\[
(d^Z(i))^{-1}=\frac{2\pi}{N}+\frac{1}{N}O(\sigma^2+\sqrt{\log N/N\sigma}+\epsilon/\sigma)
\]
for all $i$. Then note that by Lemma \ref{lem:expdiff} and \ref{lem:concentrateX},
\begin{align*}
\|p_i\|^2&\lesssim \sum_{j=1}^N \left(\exp(-\frac{\|Z_i-Z_j\|^2}{2\sigma^2})-\exp(-\frac{|t_i-t_j|^2}{2\sigma^2})\right)^2\\
&\leq 
\sum_{j: |t_j-t_i|>r_\sigma} \left(\exp(-\frac{\|Z_i-Z_j\|^2}{2\sigma^2})-\exp(-\frac{|t_i-t_j|^2}{2\sigma^2})\right)^2\\
&\quad +\sum_{j: |t_j-t_i|<r_\sigma} \left(\exp(-\frac{\|Z_i-Z_j\|^2}{2\sigma^2})-\exp(-\frac{|t_i-t_j|^2}{2\sigma^2})\right)^2\\
&\leq N r_\sigma(\sigma^4+\epsilon^2/\sigma^2) +N\sigma^8=N\Otilde(\sigma^5+\epsilon^2/\sigma). 
\end{align*}
Note that when  $|t_i-t_j|>r_\sigma$, so $k(Z_{i},Z_{j})\leq \sigma^8$ by Lemma \ref{lem:expdiff}, so 
\begin{align*}
\|u_i\|^2=\frac1{2\pi}\sum_{j=1}^N(N/\sqrt{d^Z(i)d^Z(j)}-2\pi)^2\exp(-\frac{|t_i-t_j|^2}{\sigma^2})\leq N\Otilde(\sigma^5+\frac{1}{N}+ \epsilon^2/\sigma).    
\end{align*}

\text{\textbf{The open curve case.}}
The upper bound for $\|p_i\|$ is largely the same as the closed loop case. For the upper bound of $\|u_i\|$, note that
$(d^Z(i))^{-1}=2\pi+O(\sigma^2+\sqrt{\log N/N\sigma}+\epsilon)$ if $t_i\in int_{r_\sigma}$. Next, if $i$ is of distance $2r_\sigma$ from the boundary, we can decompose
\begin{align*}
8\pi\|u_i\|^2=&\sum_{j: |t_j-t_i|<r_\sigma}^N( N/d^Z(i)+N/d^Z(j)-4\pi)^2\exp(-\frac{|t_i-t_j|^2}{\sigma^2})\\
&+\sum_{j: |t_j-t_i|>r_\sigma}^N(N/d^Z(i)+N/d^Z(j)-4\pi)^2\exp(-\frac{|t_i-t_j|^2}{\sigma^2})
\end{align*}
Note that when  $|t_i-t_j|>r_\sigma$, so $\exp(-\frac{|t_i-t_j|^2}{\sigma^2})\leq \sigma^8$ by Lemma \ref{lem:expdiff}. But when $|t_j-t_i|<r_\sigma, t_j\in int_{r_\sigma}$ so 
\[
\sum_{|t_j-t_i|<r_\sigma}( N/d^Z(i)+N/d^Z(j)-4\pi)^2\exp(-|t_i-t_j|^2/\sigma^2)=\Otilde( |I_i| (\sigma^4+\frac{1}{N\sigma}+\epsilon^2/\sigma^2))
\]
And when $i\in int_{r_\sigma}^c$, $(N/d^Z(i)+N/d^Z(j)-4\pi)^2$ is bounded, so 
\[
4\|u_i\|^2\leq r_\sigma\Rightarrow \|u_i\|\leq \sqrt{\sigma}. 
\]
\end{proof}

\section{Proofs of the main theorems}

\subsection{Open curve}
\begin{proof}[of Theorem \ref{thm:openeigen}]

\textbf{Step 1. Eigenconvergence of $L^X$ to $\Delta$.}

For notational simplicity, we will write $F_{k+1}^X$ and $F_{k+1}^Z$ as $F^X$ and $F^Z$. Consider a vector $\phitilde\in \reals^N$ with entries $\phitilde(i)=\frac{\cos (k t_i/2)}{\sqrt{\pi N}}$. Corollary 4.1 and Theorem 4.3 of \citep{peoples2021spectral} state that with probability $1-O(\frac{1}{N^2})$, the following hold for a $c=\pm 1$
\[
\left\|\frac{c\phitilde}{\|\phitilde\|} -  F^X\right\|\lesssim \delta_1:=\sqrt{\sigma}+\frac{\log N}{N^{1/4}\sigma^3}
\]
and 
\begin{equation}
\label{tmp:eigengap}
 |\lambda^X_k/\sigma^2-k^2|\leq \delta_2:=\sigma+\frac{1}{\sqrt{N}\sigma^3},\quad k\leq  K. 
\end{equation}
(Note that  the Laplacian in \citep{peoples2021spectral} is $L^X/\sigma^2$.)

\textbf{Step 2. Eigen perturbation of $L^Z$ from $L^X$.}
By \eqref{tmp:eigengap}
the spectral gaps, $\lambda_j-\lambda_{j+1}$, of $L^X$ are of order $O(\sigma^2)$ with high probability. 
Lemma \ref{lem:XZperturb} shows that $\|L^X-L^Z\|=\Otilde(\epsilon/\sigma+\sigma^6)$, which is less than the spectral gap in our setting.
So by Davis--Khan theorem, we have $\ell_2$ perturbation of the eigenvectors:
\[
\|F^X-F^Z\|\lesssim \frac{\epsilon+\sigma^7}{\sigma^3}.  
\]
This further leads to 
\[
\left\|\frac{c\phitilde}{\|\phitilde\|} -  F^Z\right\|\leq 
\left\|\frac{c\phitilde}{\|\phitilde\|} -  F^X\right\|+\|F^X - F^Z\|\lesssim \delta_1+\frac{\epsilon+\sigma^7}{\sigma^3}. 
\]

\textbf{Step 3. Bounding $\ell_2$ distance.}
Lemma \ref{lem:concentrateX} shows that with high probability, $\|\phitilde\|=1+\Otilde(1/\sqrt{N})$. Denote  \[
t_i^R=\begin{cases}
t_i\quad &\text{if }c=1,\\
2\pi-t_i,\quad &\text{if }c=-1.
\end{cases}
\]
And let $v$ be a vector with entries $\frac{\cos (k t^R_i/2)}{\sqrt{N\pi}}$, so $v=c\phitilde$. 
Then 
\[
\|v-F^Z\|=\|c\phitilde-F^Z\|\leq \|c\phitilde/\|\phitilde\|-F^Z_k\|+|1/\|\phitilde\|-1|\lesssim \delta_1+\frac{\epsilon+\sigma^7}{\sigma^3}. 
\]

We let $v=\phitilde/\|\phitilde\|$, which is the vector form of the function $\frac{c\cos(k t/2)}{\sqrt{N\pi}}$. Also note that $\psi=\cos(k t/2)$ is an eigenfunction of $-\Delta$ with eigenvalue $\lambda_{k+1}=\frac{k^2}4$, so using  
Lemmas \ref{lem:berstein} with \ref{lem:cos}, $(\psi=c\cos(kt/2), \phi=\frac{c k^2\sigma^2}{4} \cos(kt/2),c_\psi=1,b_\psi=k r_\sigma)$ we have 
\[
\max_{i\in int_{2r_{\sigma}}}|L^Z(i,\cdot)v-\frac{k^2}4\sigma^2v(i)|\leq \frac{1}{\sqrt{N}}
\Otilde\left(
\sqrt{\frac{\sigma}{N}}+\frac{1}{\sigma N}+\frac{\epsilon}{\sigma}+\sigma^2\right)
\]
\[
\max_{i\in 
\text{int}^c_{2r_{\sigma}}}|L^Z(i,\cdot)v-\frac{k^2}4\sigma^2v(i)|\leq 
\frac{1}{\sqrt{N}}\Otilde\left(
\sigma^3\right)
\]
Then because $|int^c_{2r_\sigma}|\leq Cr_\sigma N$ with high probability 
\begin{align*}
\|L^Z v-\frac{k^2}4\sigma^2v\|&=\Otilde  \left(
\sqrt{\frac{\sigma}{N}}+\frac{1}{\sigma N}+\frac{\epsilon}{\sigma}+\sigma^2\right).
\end{align*}
Using \eqref{tmp:eigengap} and Lemma \ref{lem:XZperturb},
\[
|\lambda-\lambda^Z|\leq |\lambda-\lambda^X|+\|L^X-L^Z\|
\leq \Otilde\left(\sigma^3+\frac{1}{\sqrt{N}\sigma}+\frac{\epsilon}{\sigma}\right). 
\]

Lemma \ref{lem:qi} suggests, when $i\in int_{2r_\sigma}$, 
\[
\|q_i\|\leq \Otilde(\sqrt{N}(\sigma^{2.5}+\frac{1}{\sqrt{N}}+\epsilon/\sqrt{\sigma}))
\]
Otherwise
\[
\|q_i\|\leq \Otilde(\sqrt{N}(\sigma^{0.5}+\frac{1}{\sqrt{N}}))
\]

Then we apply Lemma \ref{lem:Czprod} with $\lambda_{k+1}=\frac{k^2}4\sigma^2$ and find 
\begin{align*}
\max_{i\in \text{int}_{3r_\sigma}}|\langle F^Z_k-v, Q^Z(i,\cdot)\rangle|
&\lesssim \frac{(\epsilon/\sigma+\sigma^{1.5}+\frac{1}{\sqrt{N\sigma}})\|F^Z_k-v\|}{\sqrt{N}\sigma}
+\frac{\epsilon/\sigma+\sigma^6}{\sqrt{N}\sigma}
+\frac{\sqrt{\frac{\sigma}{N}}+\frac{1}{\sigma N}+\epsilon/\sigma+\sigma^{2}}{\sqrt{N}\sigma}\\
&\leq \frac{(\epsilon/\sigma+\sigma^{1.5}+\frac{1}{\sqrt{N\sigma}})(\sqrt{\sigma}+\frac{1}{N^{1/4}\sigma^3}+\frac{\epsilon/\sigma+\sigma^6}{\sigma^2})}{\sqrt{N}\sigma}+\frac{\epsilon/\sigma+\sigma^{2}+\frac{1}{N\sigma}+\sqrt{\frac{\sigma}{N}}}{\sqrt{N}\sigma}\\
&\lesssim \frac{1}{\sqrt{N}\sigma}(\epsilon/\sigma+\sigma^{1.5}+\frac{1}{\sqrt{N\sigma}})(\sqrt{\sigma}+\frac{1}{N^{1/4}\sigma^3}+\frac{\epsilon}{\sigma^3})+\frac{\epsilon}{\sqrt{N}\sigma^2}=:\delta_4.
\end{align*}
Combine these two results using Lemma \ref{lem:Xinfty}, we find 
\[
\max_{i\in \text{int}_{3r_\sigma}}\|F^z_k-v\|_\infty\leq \frac{1}{\sqrt{N}}\Otilde(\delta_4).
\]

\end{proof}

\begin{proof}[Proof of Theorem \ref{thm:open}]
Using Theorem \ref{thm:openeigen} with $k=1$, we have for $t^c_i=t_i$ or $2\pi-t_i$, 
\[
\max_{t_i\in \text{int}_{\delta}}|\cos(t_i^Z/2)-\cos(t^c_i/2)|\leq \Otilde(\delta_4).
\]
Since $|\cos'(x/2)|=|-\frac12 \sin(x/2)|>|\frac14\delta|$, for $x\in int_{\delta}$ and small constant $\delta$, so we have 
\[
Err_{int_\delta}(\bft,\bft^Z)\leq \Otilde(\delta_4)\leq \Otilde (\sigma). 
\]
For seriation result, note that $\pi^c(i)=\sum_{j=1}^N 1_{t^c_j<t^c_i}$, which leads to
\[
\pi^c(i)-\pi^Z(i)=\sum_{j:t^c_j<t^c_i}^N 1_{t^Z_j>t^Z_i}-
\sum_{j:t^c_j>t^c_i}^N 1_{t^Z_j<t^Z_i}.
\]
Meanwhile if $t^c_j>t^c_i$, $t^c_j\in int_{3r_\sigma}$ and
\[
\max_{t_m\in int_{3r_\sigma}}|\cos(t_m^Z/2)-\cos(t^c_m/2)|<\frac12|\cos(t^c_j/2)-\cos(t^c_i/2)|,
\]
we find $\cos(t^Z_j/2)<\cos(t^Z_i/2)$, so
by $\cos(x/2)$ being decreasing, we find $t^Z_j>t^Z_i$. Therefore 
\begin{align*}
&\{j:t^c_j<t^c_i,t^Z_j>t^Z_i\}\\
&\subseteq \{j:|\cos(t^c_j/2)-\cos(t^c_i/2)|<|\cos(t^Z_j/2)-\cos(t^Z_i/2)|\}\\
 &\subseteq \{j: t_j\in int_{3r_\sigma}, \frac12|\cos(t^c_j/2)-\cos(t^c_i/2)|\leq \max_{t_m\in int_{3r_\sigma}}|\cos(t_m^Z/2)-\cos(t^c_m/2)|\}\cup \{t_j\in int^c_{3r_\sigma}\}\\
 &\subseteq \{t_j\in int^c_{3r_\sigma}\}\cup 
 \{|t_j-t_i|<r_\sigma\}
\end{align*}
By Lemma \ref{lem:concentrateX}, we know the cardinality of this set is at most $\Otilde(N \sigma)$. Therefore $\sum_{j:t^c_j<t^c_i}^N 1_{t^Z_j>t^Z_i}=\Otilde(N \sigma)$. The analysis for  $\sum_{j:t^c_j>t^c_i}^N 1_{t^Z_j<t^Z_i}$ is similar and is omitted. Putting them together we have the seriation result.


\end{proof}

\subsection{Closed loop}

\begin{proof}[Proof of Theorem \ref{thm:closedeigen}]
\textbf{Step 1. Eigenconvergence of $L^X$ to $\Delta$.}
The first technical issue we try to solve here is that \citep{cheng2022eigen} considers spectral properties of $\Ltilde^X$, which is different $L^X$. We show their eigenvectors are largely the same, at least for the first few. Let $\Ftilde^X_+$  be $D^{-1/2}_X F^X_+$ with normalization, it will be an eigenvector of $\Ltilde^X$ based on Lemma \ref{lem:tworw}. 
We want to show 
\[
\|F^X_+-\Ftilde^X_+\|=\Otilde(\sigma^2+ \sqrt{\log N/N\sigma}),
\]
and likewise for $\|F^X_--\Ftilde^X_-\|$. 
Recall that Lemma \ref{lem:concentrateX} showed that 
\[
\max_i|d^X(i)-\frac{1}{2\pi}|=O (\sigma^2+\sqrt{\log N/N\sigma}).
\]
Next we note $F^X_+=\frac{D^{1/2}_X\Ftilde^X_+}{\|\sqrt{d^X}\Ftilde^X_+\|}$, where 
\[
\|\sqrt{d^X} \Ftilde^X_+-\tfrac{1}{\sqrt{2\pi}}\Ftilde^X_+\|=O (\sigma^2+\sqrt{\log N/N\sigma}).
\]
So 
\[
\|F^X_+-\Ftilde^X_+\|\leq \frac{\|D^{1/2}_X\Ftilde^X_+-\frac{1}{\sqrt{2\pi}}\Ftilde^X_+\|}{\|\sqrt{d^X}\Ftilde^X_+\|}
+\frac{1}{\sqrt{2\pi}}|\frac{1}{\|\sqrt{d^X}\Ftilde^X_+\|}-\sqrt{2\pi}|=O (\sigma^2+\sqrt{\log N/N\sigma}).
\]
Consider two vectors $\phi_+$ and $\phi_-$ with entries $\phi_+(i)=\frac{\sqrt{2}}{\sqrt{N}}\cos (kt_i)$, $\phi_-(i)=\frac{\sqrt{2}}{\sqrt{N}}\sin (kt_i)$. 
Consider the projection of 
 $\phi_+$ and $\phi_-$ onto the subspace spanned by $\Ftilde^X_+$ and $\Ftilde^X_-$, and $F^X_+$ and $F^X_-$ 
\begin{equation}
\label{tmp:Rdefn}
\tilde{R}_1=\begin{bmatrix}
\langle \Ftilde_+^X,\phi_+\rangle \quad \langle \Ftilde_-^X,\phi_+\rangle\\
\langle \Ftilde_+^X,\phi_-\rangle \quad \langle \Ftilde_-^X,\phi_-\rangle
\end{bmatrix}, \quad R_1=\begin{bmatrix}
\langle F_+^X,\phi_+\rangle \quad \langle F_-^X,\phi_+\rangle\\
\langle F_+^X,\phi_-\rangle \quad \langle F_-^X,\phi_-\rangle
\end{bmatrix}
\end{equation}
\citep{cheng2022eigen} Theorem 5.5 and remark 5 shows that
\[
\|[\phi_+,\phi_-] - [\Ftilde^X_+, \Ftilde^X_-]\tilde R_1\|_F=\Otilde(\sigma^2+ \sqrt{\log N/N\sigma^3}). 
\]
Then because $\|F^X_{\pm}-F^X_{\pm}\|=O(\sigma^2+\sqrt{\log N/N\sigma})$ which are of lower order, we have 
\[
\|[\phi_+,\phi_-] - [F^X_+, F^X_-] R_1\|_F=\Otilde(\sigma^2+ \sqrt{\log N/N\sigma^3}). 
\]

\textbf{Step 2. Eigen perturbation of $L^Z$ from $L^X$.}
By Lemma \ref{lem:concentrateX}, we have that $\|L^X-L^Z\|\lesssim \epsilon/\sigma$. So by Davis--Khan theorem, there is an orthogonormal matrix $R_2$ so that 
\[
\|[F_+^X,F_-^X]-[F_+^Z,F_-^Z]R_2\|_F\lesssim \frac{\epsilon}{\sigma^3}. 
\]
This leads to 
\[
\|[\phi_+,\phi_-] - [F^Z_+, F^Z_-]R\|_F\lesssim \delta_1:=\sigma^2+ \sqrt{\log N/N\sigma^3}+\frac{\epsilon}{\sigma^3}
\]
where $R=R_1R_2$. 

\textbf{Step 3. $R$ is approximately a rotation or rotation-reflection.}
Lemma \ref{lem:Rrotation} shows that with high probability, $R_1=R'_1+r(\delta_1)$, where $R'_1$ is a $2\times 2$ orthogonormal matrix, and $r(\delta_1)$ denotes a $2\times 2$ matrix with all entries being of order $\delta_1$. Therefore $R^{-1}=R_2^{-1}R_1^{-1}$ will also be close to an orthogonormal matrix. In particular, there will be a $\theta_0\in [0,2\pi]$ and $c\in \{-1,+1\}$  so that
\[
R^{-1}=\begin{bmatrix}
    \cos \theta_0 & -c\sin \theta_0\\
    \sin \theta_0 & c\cos \theta_0
\end{bmatrix}+r'(\delta_1)
\]
$r'(\delta_1)$ denotes a $2\times 2$ matrix with all entries being of order $\delta_1$. 
Note that when $c=1$, $R^{-1}$ is approximately a rotation matrix. When $c=-1$, it is approximately a rotation-reflection.

\textbf{Step 4. Bounding $\ell_2$ distance.}
Denote $[\psi_+,\psi_-]=[\phi_+,\phi_-]R^{-1}$ and \[
t_i^R=ct_i+\theta_0\quad \text{mod}\quad 2\pi. 
\]
And let $\psi'_+$ be a vector with entries $\frac{\sqrt{2}\cos (k t^R_i)}{\sqrt{N}}$, $\psi'_-$ be a vector with entries $\frac{\sqrt{2}\sin (kt^R_i)}{\sqrt{N}}$. Then $\|\psi'_\pm-\psi_\pm\|\lesssim \delta_1, \|\psi'_\pm-\psi_\pm\|_\infty\lesssim \delta_1/\sqrt{N}$. 
 Note that by Lemma \ref{lem:concentrateX}, their norm are $1+O(\delta)$. Meanwhile, because $\ell_2$ norm is preserved under orthogonormal transformation, we would have 
\[
\|[\psi'_+,\psi'_-] - [F^Z_+, F^Z_-]\|_F\leq
\|[\psi_+,\psi_-] - [F^Z_+, F^Z_-]R\|_F+
\delta_1\leq 2\delta_1. 
\]
\textbf{Step 5. Bounding $\ell_\infty$ distance.}
Note first that in this case, the interior is simply the whole set. 
We will just bound the error of $F^Z_+$ with $\psi'_+$. The error for $F^Z_-$ is similar. We let $v=\psi'_+$, which is the vector form of the function $\frac{\cos(t^R)}{\sqrt{N\pi}}$. Also note that $\psi=\cos(k t^R)$ is an eigenfunction of $-\Delta$ with eigenvalue $\lambda=k^2$, so using  
Lemma \ref{lem:Laplacian}, \ref{lem:berstein} with \ref{lem:cos}, $(\psi=\cos(kt^R), \phi=\frac{k^2}2\sigma^2 \cos(kt^R))$ we have 
\[
\|L^Z(i,\cdot)v-\frac{k^2}2\sigma^2v\|_\infty\leq \Otilde(\sigma^2+\epsilon/\sigma+\sqrt{\frac{\sigma}{N}}+\frac{1}{\sigma N})
\]
Also we have Theorem 5.4 from \citep{cheng2022eigen}, which shows
\[
|\lambda-\lambda^Z|\leq |\lambda-\lambda^X|+\|L^X-L^Z\|
\leq \Otilde\left(\sigma^4+\frac{\sigma^{1.5}}{\sqrt{N}}+\epsilon/\sigma\right). 
\]
(Note that  the Laplacian in \citep{cheng2022eigen} is $L^X/\sigma^2$.)

Lemma \ref{lem:qi} suggests
\[
\|q_i\|\leq \Otilde(\sqrt{N}(\sigma^{2.5}+\frac{1}{\sqrt{N}}+\epsilon/\sqrt{\sigma}))
\]
Put all these information into Lemma \ref{lem:Czprod} with $\lambda_2=\sigma^2$ and find 
\begin{align*}
\max_{i}|\langle F^Z-v, Q^Z(i,\cdot)\rangle|
&\leq \frac{(\epsilon/\sigma+\sigma^2+1/\sqrt{N\sigma})\|F_z-v\|}{\sqrt{N}\sigma}+\frac{\sigma^{2.5}+\sqrt{\sigma/N}+\frac{1}{\sigma N}+\epsilon/\sigma}{\sqrt{N}\sigma}\\
&\leq \frac{(\epsilon/\sigma+\sigma^2+1/\sqrt{N\sigma})(\sigma^2+\sqrt{1/N\sigma^3}+\epsilon/\sigma^3)}{\sqrt{N}\sigma}+\frac{\sigma^{2}+\sqrt{\sigma/N}+\frac{1}{\sigma N}+\epsilon/\sigma}{\sqrt{N}\sigma}\\
&\leq \frac{1}{\sqrt{N}}(\frac{\epsilon}{\sigma^2}+\sigma^2+\frac{\epsilon^2}{\sigma^4}+\frac{\epsilon}{\sqrt{N}\sigma^{3.5}}+\frac{1}{\sqrt{N\sigma^3}}). 
\end{align*}
Combine these two results using Lemma \ref{lem:Xinfty}, we find 
\[
\|F^Z-v\|_\infty\leq \frac{1}{\sqrt{N}}\Otilde(\frac{\epsilon}{\sigma^2}+\sigma^2+\frac{1}{\sqrt{N\sigma^3}}+\frac{\epsilon}{\sqrt{N}\sigma^{3.5}}).
\]






\end{proof}

\begin{lemma}
\label{lem:Rrotation}
Consider $R=\begin{bmatrix}
\langle F_+^X,\phitilde_+\rangle \quad \langle F_-^X,\phitilde_+\rangle\\
\langle F_+^X,\phitilde_-\rangle \quad \langle F_-^X,\phitilde_-\rangle
\end{bmatrix}$, then $R$ is approximatley a rotation with high probability: with probability $1-O(1/N)$, there is a $\theta_+$ and $b=\pm1$, so that 
\begin{align*}
R=\begin{bmatrix}
\cos\theta_+ \quad \sin \theta_+\\
-b\sin \theta_+ \quad b\cos\theta_+
\end{bmatrix}+O(\delta_{c,1}),\quad \delta_{c,1}=\sigma^2+\sqrt{\frac{\log N}{N\sigma^3}}.  
\end{align*}
\end{lemma}

\begin{proof}
Note that $\|F_+^X\|=\|F_-^X\|=1,$ and $F_+^X\bot F_-^X$. Consider the projection of $\phitilde_\pm$ onto the subspace spanned by $F_+^X$ and $F_-^X$:
\[
\phi'_+=\langle F_+^X,\phitilde_+\rangle F^X_++
\langle F_-^X,\phitilde_+\rangle F^X_-
\]
and
\[
\phi'_-=\langle F_+^X,\phitilde_-\rangle F^X_++
\langle F_-^X,\phitilde_-\rangle F^X_-
\]
From \citep{cheng2022eigen} Theorem 5.5 and remark 5, we know that 
\[
\|\phi'_+-\phitilde_+\|+\|\phi'_--\phitilde_-\|\lesssim \delta_{c,1}
\]
Meanwhile, we can find a $\theta_+$ and $\theta_-$ so that 
\[
\langle F_+^X,\phi'_\pm\rangle=\cos \theta_\pm \|\phi'_\pm\|,\quad
\langle F_-^X,\phi'_\pm\rangle=\sin \theta_\pm \|\phi'_\pm\|
\]
Then from Lemma \ref{lem:concentrateX} claim 6, we know 
\[
|\|\phitilde_\pm\|^2-\frac12|=O(\sqrt{\log N/N\sigma})=O(\delta_{c,1}),\quad
|\langle\phitilde_-,\phitilde_+\rangle|=O(\sqrt{\log N/N\sigma})=O(\delta_{c,1}).
\]
with high probability. This leads to 
\begin{align*}
\cos (\theta_+-\theta_-)\|\phi'_+\|\|\phi'_-\|=\langle \phi'_+,\phi'_-\rangle
&=\langle \phitilde_+,\phitilde_-\rangle+O(\delta_{c,1})=O(\delta_{c,1}).
\end{align*}
Since 
\[
|\|\phi'_\pm\|^2-\frac12|=|\|\phitilde_\pm\|^2-\frac12|+O(\delta_{c,1})=O(\delta_{c,1})
\]
we have  $\theta_+-\theta_-=\frac{\pi}{2}+O(\delta_1)$ or $\theta_+-\theta_-=-\frac{\pi}{2}+O(\delta_1)$. We let $b=1$ in the first case, and $b=-1$ in the second. Note then
\begin{align*}
R=\begin{bmatrix}
\langle F_+^X,\phi'_+\rangle \quad \langle F_-^X,\phi'_+\rangle\\
\langle F_+^X,\phi'_-\rangle \quad \langle F_-^X,\phi'_-\rangle
\end{bmatrix}=
\begin{bmatrix}
\|\phi'_+\|\cos\theta_+ \quad \|\phi'_+\|\sin \theta_+\\
\|\phi'_-\|\cos\theta_- \quad \|\phi'_-\|\sin \theta_-
\end{bmatrix}
=\begin{bmatrix}
\cos\theta_+ \quad \sin \theta_+\\
-b\sin \theta_+ \quad b\cos\theta_+
\end{bmatrix}+O(\delta_{c,1}). 
\end{align*}
\end{proof}

\it{Proof of Theorem \ref{thm:closed}.}
Denote
\[
\delta_4=(\frac{\epsilon}{\sigma^2}+\sigma^2+\frac{1}{\sqrt{N\sigma^3}}+\frac{\epsilon}{\sqrt{N}\sigma^{3.5}})
\]
Fix a time point $t_i$. For simplicity of discussion, we assume that $\pi/2<t_i\leq 3/2\pi$. This helps us to avoid $t_i$ near $0$ and $2\pi$, where the distance is cumbersome to discuss. And if $t_i$ is not in this range, note that we can just set a different angle to be $0$. By Lemma \ref{lem:anglelength},
\[
Err(\bft^c,\bft^Z)\leq \Otilde(\delta_4). 
\]
For seriation result, note that $\pi^c(i)=\sum_{j=1}^N 1_{t^c_j<t^c_i}$, which leads to
\[
\pi^c(i)-\pi^Z(i)=\sum_{j:t^c_j<t^c_i}^N 1_{t^Z_j>t^Z_i}-
\sum_{j:t^c_j>t^c_i}^N 1_{t^Z_j<t^Z_i}.
\]
Meanwhile if $t^c_j>t^c_i$,  and $\delta_4<\frac12|t^c_j/2-t^c_i/2|,$
 we find $t^Z_j>t^Z_i$. Therefore 
\begin{align*}
\sum_{j:t^c_j<t^c_i}^N 1_{t^Z_j>t^Z_i}\leq |\{j:t^c_j<t^c_i,t^Z_j>t^Z_i\}|
\end{align*}
By Lemma \ref{lem:concentrateX} (1), we know the cardinality of this set is at most $\Otilde(N \delta_4)$ with high probability. Therefore $\sum_{j:t^c_j<t^c_i}^N 1_{t^Z_j>t^Z_i}=\Otilde(N \sigma)$. The analysis for  $\sum_{j:t^c_j>t^c_i}^N 1_{t^Z_j<t^Z_i}$ is similar and is omitted. Putting them together we have the seriation result. 
\[
Err(\bft^c,\bft^Z)=N\Otilde(\delta_4)
\]


\qed

\subsection{Proof for the graphical methods}

\begin{proof}
\textbf{Step 1: we will find an end point with high probability.}
By Lemma \ref{lem:concentrateX} claim 7 and Lemma \ref{lem:XZperturb}, we see that with high probability
\[
d^Z(i^*)/N\leq \frac{1}{4\pi}+\sigma^2+\epsilon/\sigma,
\max_{i: t_i\in int_{r_\sigma}}d^Z(i)/N\geq \frac{1}{2\pi}-\sigma^2-\epsilon/\sigma
\]
for $i^*=\arg\max t_i$ and $i^*=\arg\min t_i$. So if $\sigma^2+\epsilon/\sigma\ll 1$, then $d^Z(i^*)<d^Z(i)$ for all $i\in int_{r_\sigma}$. In other words, the maximizer $d^Z(j^*)$ will correspond to a $t_{j^*}\in int_{r_\sigma}^c$.

\textbf{Step 2: the graphical distance based method will roughly recover the correct ordering.}
 Next we show that if $r_\sigma<t_i<t_j-3\sigma_0$, then $Z_i$ will be ranked in front of $Z_j$. Let $Z_{j^*}=Z_{(0)},Z_{(1)},\ldots, Z_{(k)}=Z_j$ be the shortest graphical path connecting $Z_j$ and $Z_{j^*}$, while 
 $\|Z_{(i)}-Z_{(i+1)}\|\leq \sigma_0$. 
 Note that $\|Z_{(i)}-Z_{(i+2)}\|\geq \sigma_0$, else $Z_{(i+1)}$ can be removed to obtain a shorter path.

 By Assumptions \ref{aspt:manifold} and \ref{aspt:noise}, this indicates that $|t_{(i)}-t_{(i+1)}|\leq \sigma_0+2\epsilon+0.1\sigma_0\leq 1.2\sigma_0$. Then there is an $s\in [k]$,  $|t_{(s)}-t_i|\leq 0.6\sigma_0$. Clearly $s<k-2$, since otherwise $|t_j-t_i|\leq |t_j-t_{(k-1)}|+|t_{(k-1)}-t_{(k-2)}|+|t_i-t_{(k-2)}|\leq 3\sigma_0$. Also because  $\|Z_{(i)}-Z_{(i+2)}\|\geq \sigma_0$, $d_G(Z_{(s)},Z_{(k)})\geq (\frac{1}{2}(k-s-1))\sigma_0=\sigma_0$, while $d_G(Z_{(s)},Z_j)\leq 0.6\sigma+0.1\sigma\leq 0.7\sigma$. Therefore 
 \[
d_G(Z_{j^*},Z_i)=d_G(Z_{j^*},Z_{(s)})+d_G(Z_{i},Z_{(s)})
>d_G(Z_{j^*},Z_{(s)})+d_G(Z_{j},Z_{(s)})\geq d_G(Z_{j^*},Z_j).
\]
 So $\hat{\pi}^G_i<\hat{\pi}^G_j$.

\textbf{Step 3: finding the time stamp.} By step 2, we note that
\[
|\{j: t_j<t_i-3\sigma_0\}|\leq \hat{\pi}^G_i\leq N+1-|\{j: t_j>t_i+3\sigma_0\}|. 
\]
By Bernstein's inequality, 
\[
P(|\{j: t_j<t_i-3\sigma_0\}|\leq \frac{N}{2\pi}(t_i-4\sigma_0))
\leq \exp(-\frac{1}{10}\sigma_0^2 N)<N^{-4}. 
\]
\[
P(|\{j: t_j>t_i+3\sigma_0\}|\leq \frac{N}{2\pi}(2\pi-t_i-4\sigma_0))
\leq \exp(-\frac{1}{10}\sigma_0^2 N)<N^{-4}. 
\] 
So with high probability
\[
\frac{t_i-4\sigma_0}{2\pi}\leq \frac{\hat{\pi}^G_i}{N}\leq \frac{t_i+4\sigma_0}{2\pi}. 
\]
Therefore $|\that^G_i-t_i|\leq 4\sigma_0$ for all $i$ with high probability.

\textbf{Step 4: Combining the two sorting results.} Let $i$ and $j$ be the $\delta N$-th and $(1-\delta) N$-th data points in the $\hat{\pi}^G$ ordering. Then Theorem \ref{thm:open} indicates that $|\that_i-t_i|=o(1), |\that_j-t_j|=o(1)$,  or $|2\pi-\that_i-t_i|=o(1), |2\pi-\that_j-t_j|=o(1)$. In either case, step 8 of Algorithm \ref{alg:opencurvepatch} will match the ordering of $\hat{\pi}^G$ with $\hat{\pi}$. From step 3, we can see that all points in $int^c_{r\sigma}$ will be stamped by the $\that^G$. So the error of the final time stamp will be the same the one in Theorem \ref{thm:open} plus $\sigma_0$. 
\end{proof}


\bibliographystyle{imsart-number}
\bibliography{ref}
\end{document}